\documentclass[twocolumn,twocolappendix]{aastex631}
\let\splitbox\relax

\usepackage{xspace}
\usepackage{hyperref}
\usepackage{enumitem}
\usepackage{amssymb}
\usepackage{amsmath}
\usepackage{pifont}

\graphicspath{{./}{figures/}}
\usepackage{lipsum}
\usepackage{graphicx}
\usepackage{subfigure}
\usepackage{comment}
\usepackage{booktabs}
\usepackage{amsmath}
\usepackage{etoolbox}
\usepackage{IEEEtrantools}
\usepackage{parskip}
\usepackage{ragged2e}
\usepackage{textcomp,gensymb}
\usepackage{physics}
\usepackage{multirow}
\let\tablenum\relax
\usepackage{siunitx} % For alignment of numbers
\usepackage{adjustbox}

\usepackage{tabularx, booktabs,array}
\begin{document}
\UseRawInputEncoding
%\linenumbers
%\switchlinenumbers

\title{Revisiting
galaxy cluster scaling relations through dark matter-gas coherence: scatter dependence on dynamical state.}

\correspondingauthor{Giulia Cerini}
\email{giulia.cerini@jpl.nasa.gov}

\author[0000-0002-5273-4634]{Giulia Cerini}
\affiliation{Department of Physics, University of Miami, 1320 S Dixie Hway, Coral Gables, FL 33146, USA}
\affiliation{Jet Propulsion Laboratory, California Institute of Technology, 4800 Oak Grove Dr, Pasadena, CA 91109, USA}

\author[0000-0001-6411-3686]{Elena Bellomi}
\affiliation{Center for Astrophysics $\vert$ Harvard $\&$ Smithsonian, 60 Garden St., Cambridge, MA 02138, USA}

\author[0000-0002-1697-186X]{Nico Cappelluti} 
\affiliation{Department of Physics, University of Miami, 1320 S Dixie Hway, Coral Gables, FL 33146, USA}

\author{Sabina Khizroev}
\affiliation{Department of Physics, University of Miami, 1320 S Dixie Hway, Coral Gables, FL 33146, USA}

\author[0000-0001-8914-8885]{Erwin T. Lau}
\affiliation{Nara Women’s University, Kitauoyanishi-machi, Nara, Nara 630-8506, Japan}
%\affiliation{Department of Physics, University of Miami, 1320 S Dixie Hway, Coral Gables, FL 33146, USA}

\author[0000-0002-5554-8896]{Priyamvada Natarajan}
\affiliation{Department of Astronomy, Yale University, 219 Prospect Street, New Haven, CT 06511, USA}
\affiliation{Department of Physics, Yale University, P.O. Box 208121, New Haven, CT 06520, USA}
\affiliation{Black Hole Initiative, Harvard University, 20 Garden Street, Cambridge, MA 02138, USA}

\author[0000-0003-3175-2347]{John ZuHone}
\affiliation{Center for Astrophysics $\vert$ Harvard $\&$ Smithsonian, 60 Garden St., Cambridge, MA 02138, USA}

\begin{abstract} \label{abstract}

Galaxy clusters, the most massive, dark matter dominated and most recently assembled structures in the Universe are key tools for probing cosmology. However, uncertainties in scaling relations that connect cluster mass to observables like X-ray luminosity and temperature remain a significant challenge. In this paper, we present the results of an extensive investigation of 329 simulated clusters from IllustrisTNG300 cosmological simulations. Our analysis involves cross-correlating dark matter and the hot X-ray emitting gas, considering both the 3D and 2D projected distributions to account for projection effects. We demonstrate that this approach is highly effective in evaluating the dynamical state of these systems and validating the often-utilized assumption of hydrostatic equilibrium, which is key for inferring cluster masses and constructing scaling relations. Our study revisits both the X-ray luminosity - mass and X-ray temperature - mass scaling relations and demonstrates how the scatter in these relations correlates with the clusters' dynamical state. We demonstrate that matter - gas coherence enables the identification of an optimal set of relaxed clusters, reducing scatter in scaling relations by up to $40\%$. This innovative approach, which integrates higher-dimensional insights into scaling relations, might offer a new path to further reduce uncertainties in determining cosmological parameters from galaxy clusters.
\end{abstract}


\section{Introduction} \label{sec:intro}
Galaxy clusters are the most significant cosmic structures, not only for their masses and scales but also for the deeper understanding of cosmology and astrophysics that they offer. They are the largest repositories of dark matter (hereafter DM), with masses in the range $10^{14-15}M_{\odot}$. Although their gravitational potential is dominated by the non-baryonic DM component, they contain X-ray emitting hot gas with typical temperatures of $10^{7-8}K$, the Intra-Cluster Medium (ICM), that accounts for $10\text{-}20\%$ of the total mass of the cluster.

 \begin{figure*}
\centering
      %\subfigure[]{\includegraphics[width=0.32\textwidth]{good_figures/halo5_relaxed.png}}
      %\subfigure[]{\includegraphics[width=0.32\textwidth]{good_figures/halo6_par_relaxed.png}}
      %\subfigure[]{\includegraphics[width=0.32\textwidth]{good_figures/halo1_nonrelaxed.png}}
      %\subfigure[]       
    % Subfigure 1
    \includegraphics[width=0.32\textwidth]{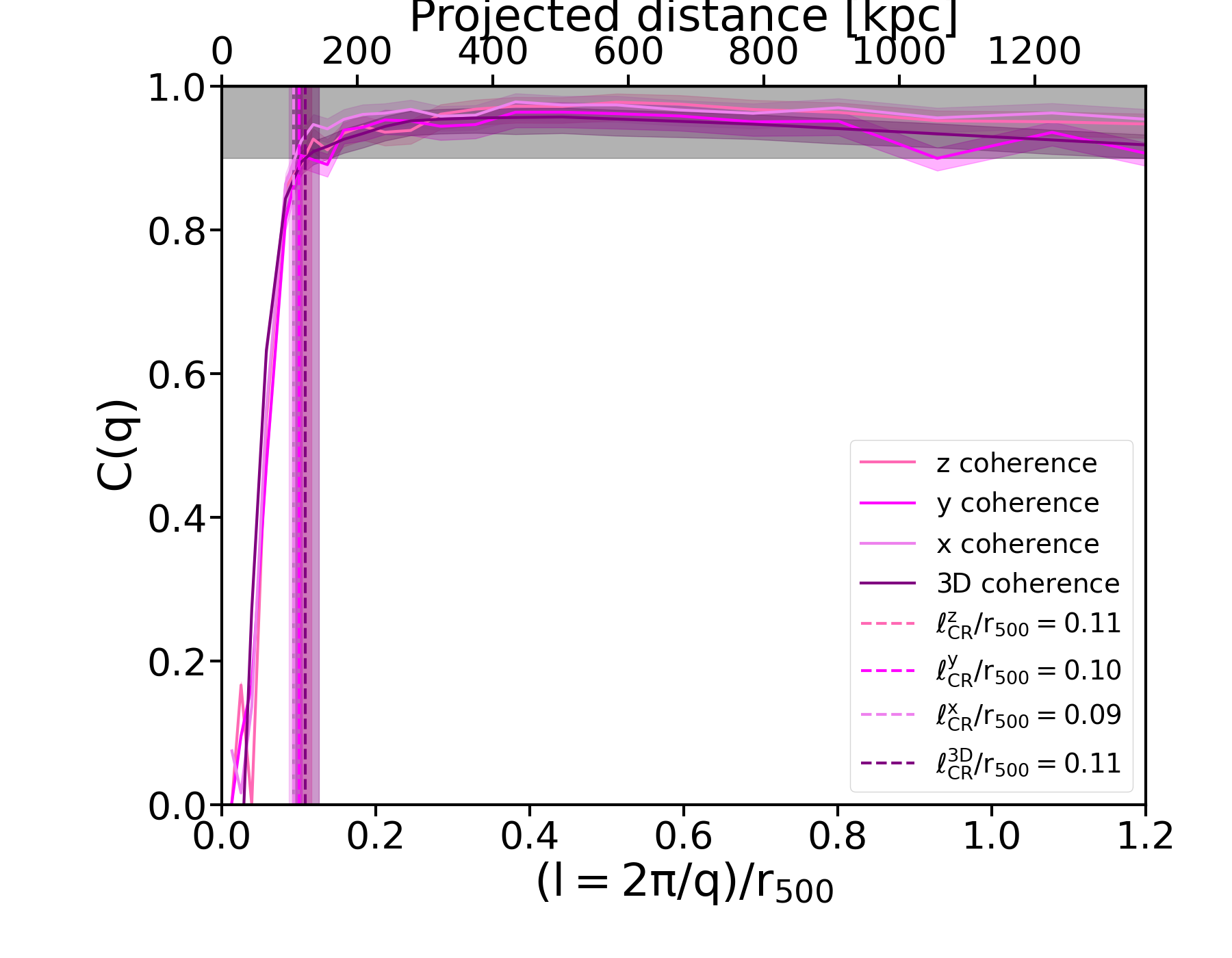}
    % Subfigure 2
    \includegraphics[width=0.32\textwidth]{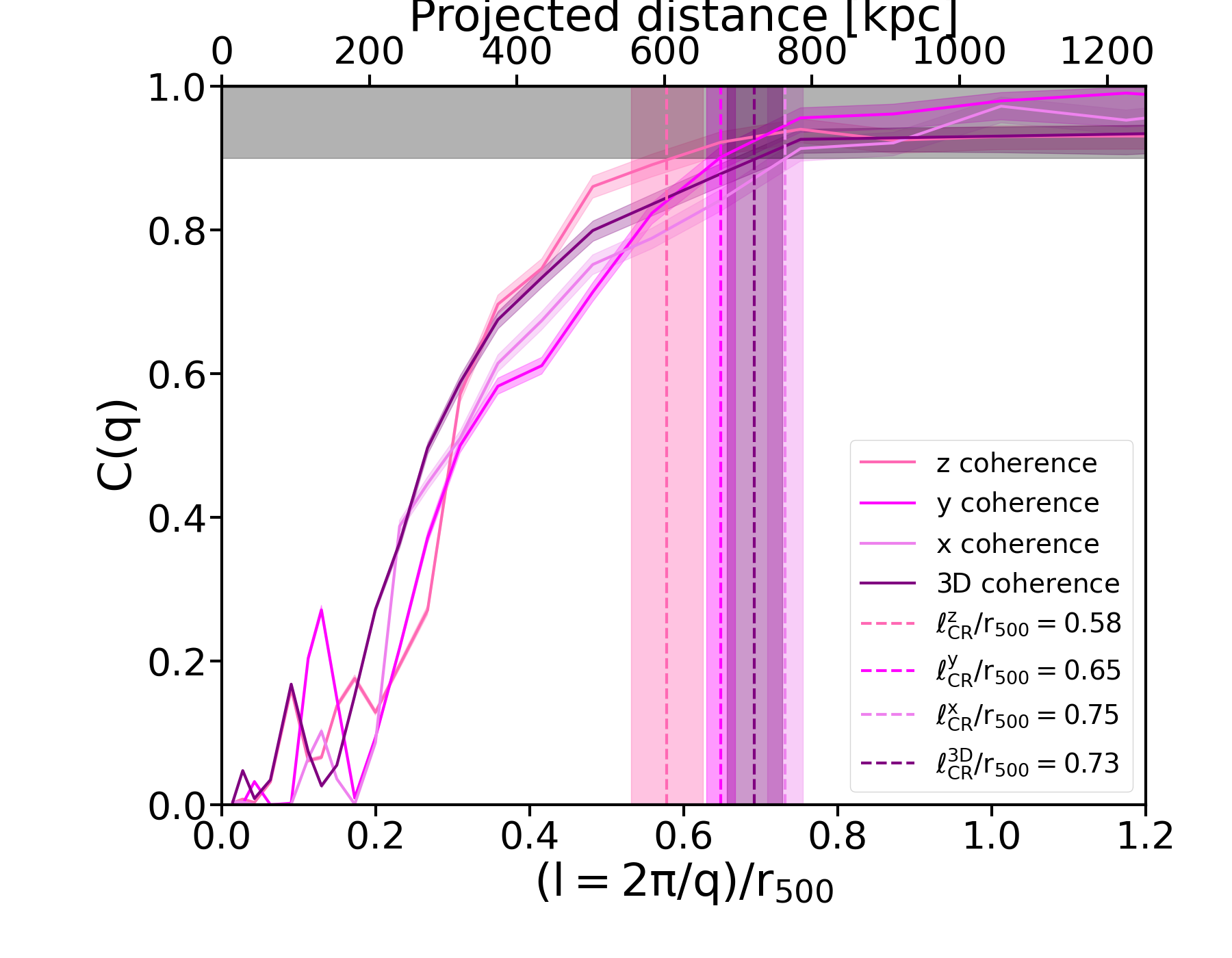}
    % Subfigure 3
    \includegraphics[width=0.32\textwidth]{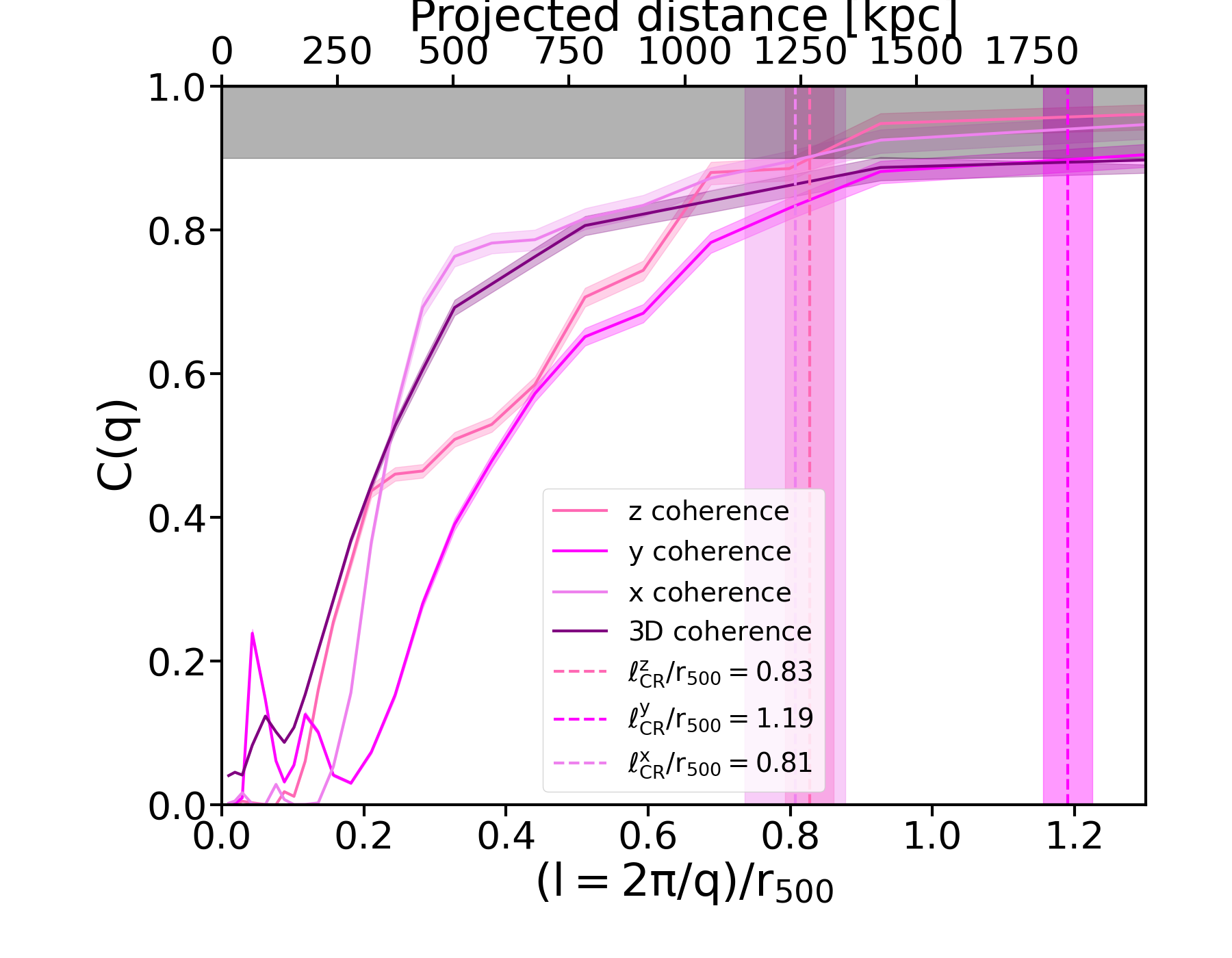}
    
      \caption{Examples of coherence analysis for clusters in different dynamical states, according to the value of the \textit{coherence length} (vertical dashed straight lines), normalized to $r_{500}$: relaxed cluster (panel (a)), partially unrelaxed cluster (panel (b)), unrelaxed cluster (panel (c)). For the unrelaxed cluster, the \textit{coherence length} in 3D is not definable. In all panels, we show in pink, magenta, violet and purple the results obtained from the 3 projections - z, y and x - and the 3D analysis, respectively. Shaded areas represent 1$\sigma$ errors, computed as explained in Section~\ref{coherence} .}
 \label{example_c}
 \end{figure*}

In the current bottom-up scenario for the formation of cosmic structures, small initial density fluctuations in the DM are amplified by gravity, leading to the creation of the massive, DM-dominated structures we observe today (\citealt{Press1974}). Thus, to a first approximation, clusters of galaxies are simple objects whose properties are defined only by their mass and redshift ($z$). This leads to simple power-law correlations between different observable properties, such as temperature and luminosity, referred to as scaling relations, that correlate these observable properties of the clusters with their masses. These scaling relations arise as a consequence of the ``self-similar'' assembly process (see, for instance, \citealt{Kaiser1986}, \citealt{Battaglia2012}, \citealt{B_hringer_2012}, \citealt{Giodini2013}). The study of the scaling relations between the key parameters characterizing galaxy clusters has provided a unique way of constraining cosmological parameters, that complements results from other observational cosmic probes. Indeed, observations of structure formation and assembly over a range of physical (Gpc to pc) and temporal ($10^9$ to $10^{13}$ yr) scales have been successfully accounted for by the concordance cosmological model that comprises Cold Dark Matter (CDM) and a cosmological constant ($\Lambda$). The $\Lambda$CDM paradigm is well studied and observationally supported on a range of scales, by the Cosmic Microwave Background (CMB) data, high-$z$ supernovae, galaxy surveys, and scaling relations of galaxy clusters (e.g., \citealt{Bahcall2002}, \citealt{Tonry2003}, \citealt{Vik2009}, \citealt{Allen11}, \citealt{Pill2012}, \citealt{Mantz2015}, \citealt{Schell2017}, \citealt{Pacaud_2018}, \citealt{Planck2020}, \citealt{Pandey22}, \citealt{2022Riess}, \citealt{2023DES}, \citealt{ghirardini2024srgerosita}). However, tensions between CMB early-time measurements and large-scale structure late-time experiments have emerged in both the expansion history and the growth of structure via the $H_0$-tension (\citealt{2021DiValentino}) and the $S_8$-tension (\citealt{Heymans2021}), respectively. Additionally, significant sources of error persist in the cluster-based estimations of cosmological parameters (e.g., \citealt{Vikhlinin_2009}, \citealt{2010Rozo}, \citealt{Allen11}, \citealt{WEINBERG201387}).

The derivation of cosmological constraints from cluster abundance depends primarily on accurate determinations of cluster masses; this remains a major challenge for cosmology, even for clusters in the local Universe. Indeed, the mass of a cluster is not directly measurable. For instance, \citealt{Stopyra_2021} found large uncertainties in the mass estimation of clusters in the local region at $<135\ \textrm{Mpc}~h^{-1}$ (approximately $z<0.046$), even when different methods were used to infer these masses: weak lensing (\citealt{Bonnet94},\citealt{Fahlman_1994}), the Sunyaev-Zel$^{\prime}$dovich (SZ) method (\citealt{S1970}, \citealt{S1980}), dynamical estimates using the virial theorem
(\citealt{Merritt1987}), and mass estimates from the X-ray emission (\citealt{Evrard1996}). In particular, the latter three methods are based on the assumption that the gas is in hydrostatic equilibrium in the cluster potential. Indeed, it is typically assumed that both the galaxy distribution and the gas responsible for the X-ray emission efficiently trace the gravitational potential that is dominated by the dark matter. However, the assumption of hydrostatic equilibrium may not always hold true, as deviations are expected in assembling clusters undergoing merging activity. These mergers can induce significant non-thermal pressure from gas motions, which is considered the primary source of departure from equilibrium (e.g., \citealt{Lau_2009}). Additionally, non-gravitational processes, such as feedback from active galactic nuclei or gas turbulence, can further disrupt the equilibrium, leaving distinct spatial and velocity imprints on the cluster's structure (e.g., \citealt{Churazov_2012}, \citealt{zhuravleva17}). In addition, the observed scaling relations, derived in many recent works, point to deviations from the ’self-similar’ scenario, suggesting that non-gravitational effects are likely to play an important role in shaping the observed physical state of clusters (see, for instance, (e.g., \citealt{Bhat2008}, \citealt{McCarthy2010}, \citealt{Fabjan2011}, \citealt{Giodini2013}, \citealt{Bulbul_2019}, \citealt{Lovisari2020})). Studies of ongoing merging clusters (e.g., \citealt{Markevitch_2002}, \citealt{Clowe_2006}, \citealt{Emery_2017}) have also clearly revealed that, in these extreme cases, there is a mismatch between the (weakly) collisional gas and the (likely) collisionless dark matter distributions. However, even during later stages of mergers, when the merged cluster appears to have settled down, there is still a lag in reaching hydrostatic and virial equilibrium. These stages and their evolution are yet to be probed and characterized.

 \begin{figure*}
\centering
      \subfigure[]{\includegraphics[width=0.32\textwidth]{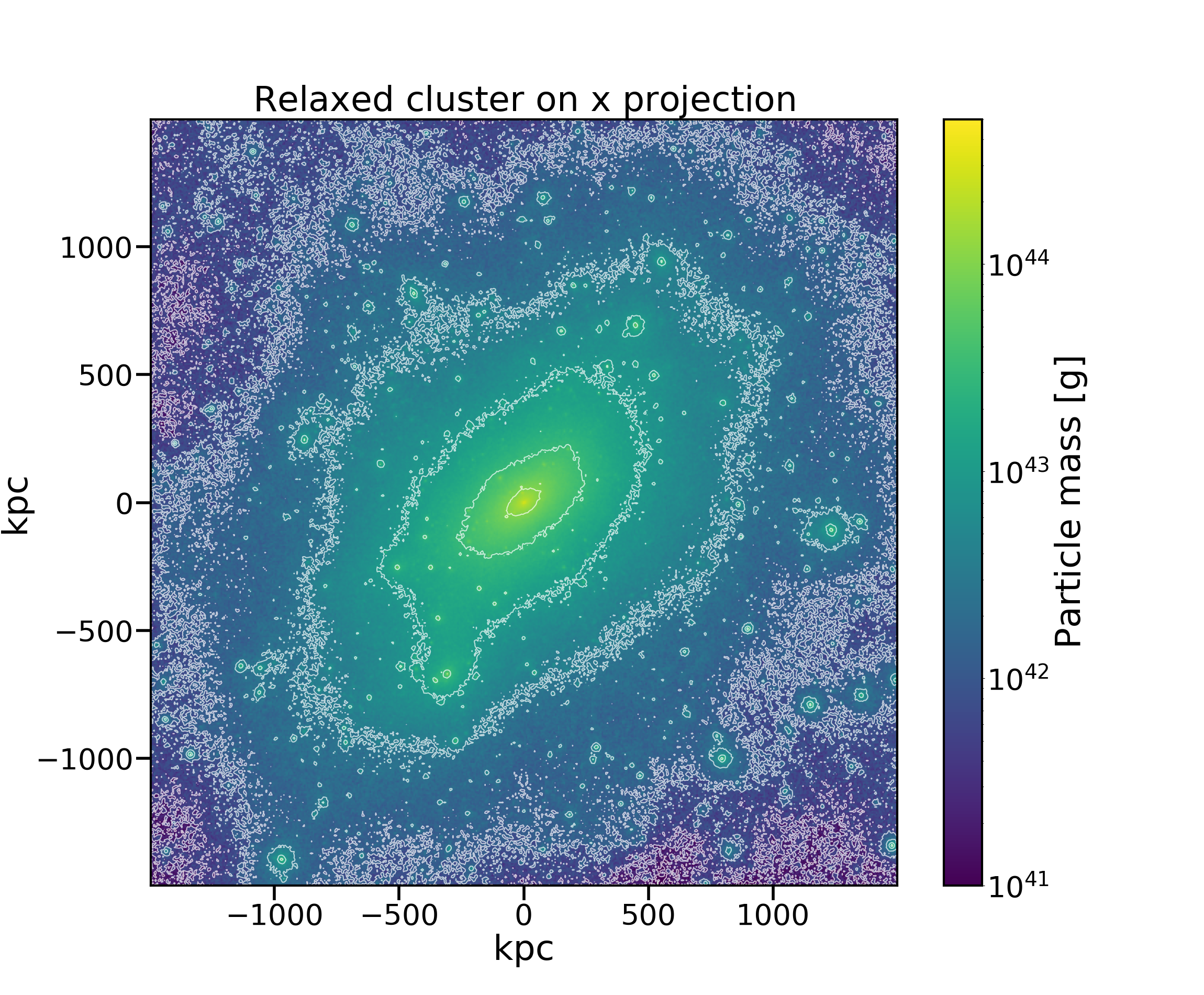}}
      \subfigure[]{\includegraphics[width=0.32\textwidth]{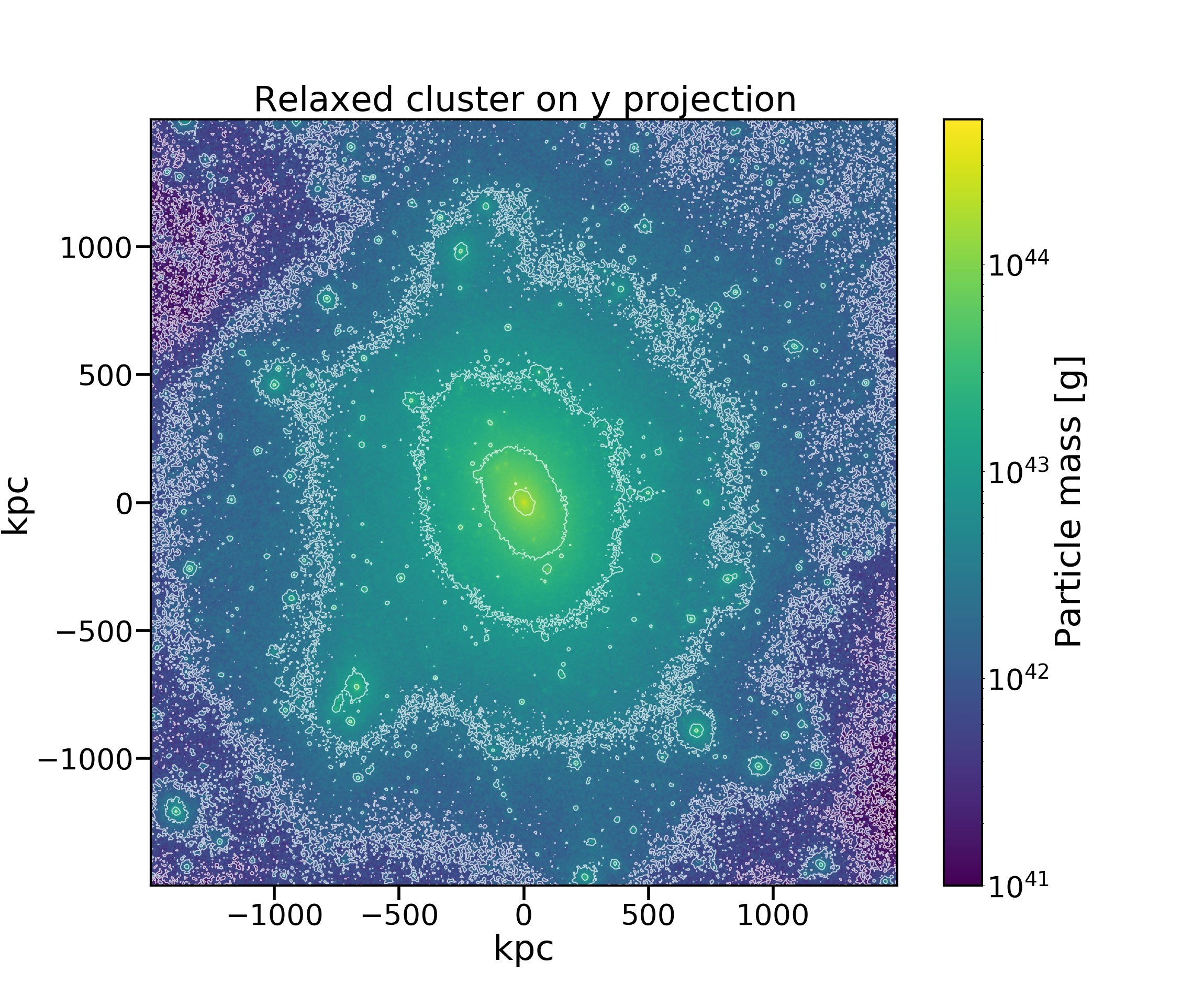}}
      \subfigure[]{\includegraphics[width=0.32\textwidth]{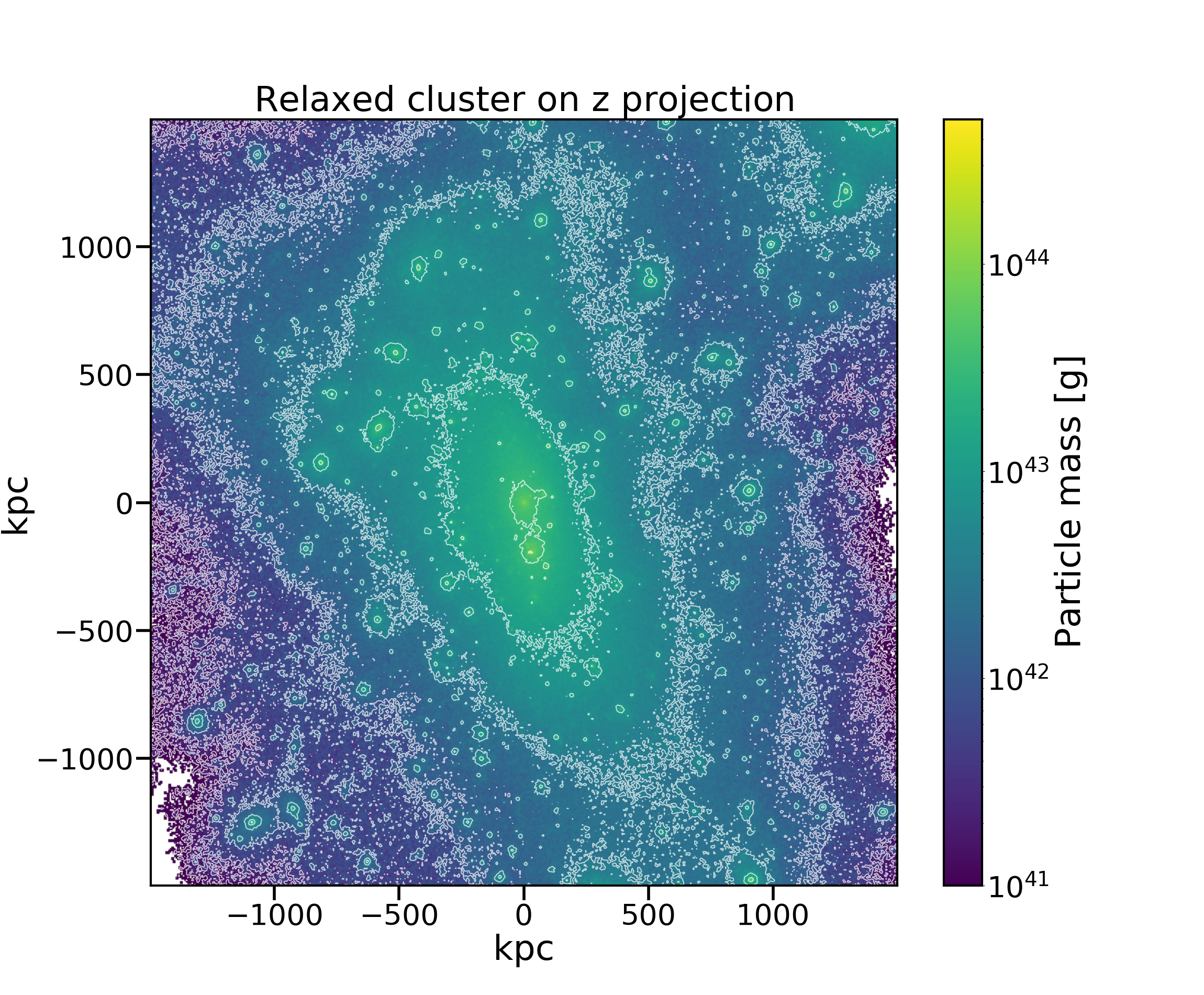}} 
      \subfigure[]{\includegraphics[width=0.32\textwidth]{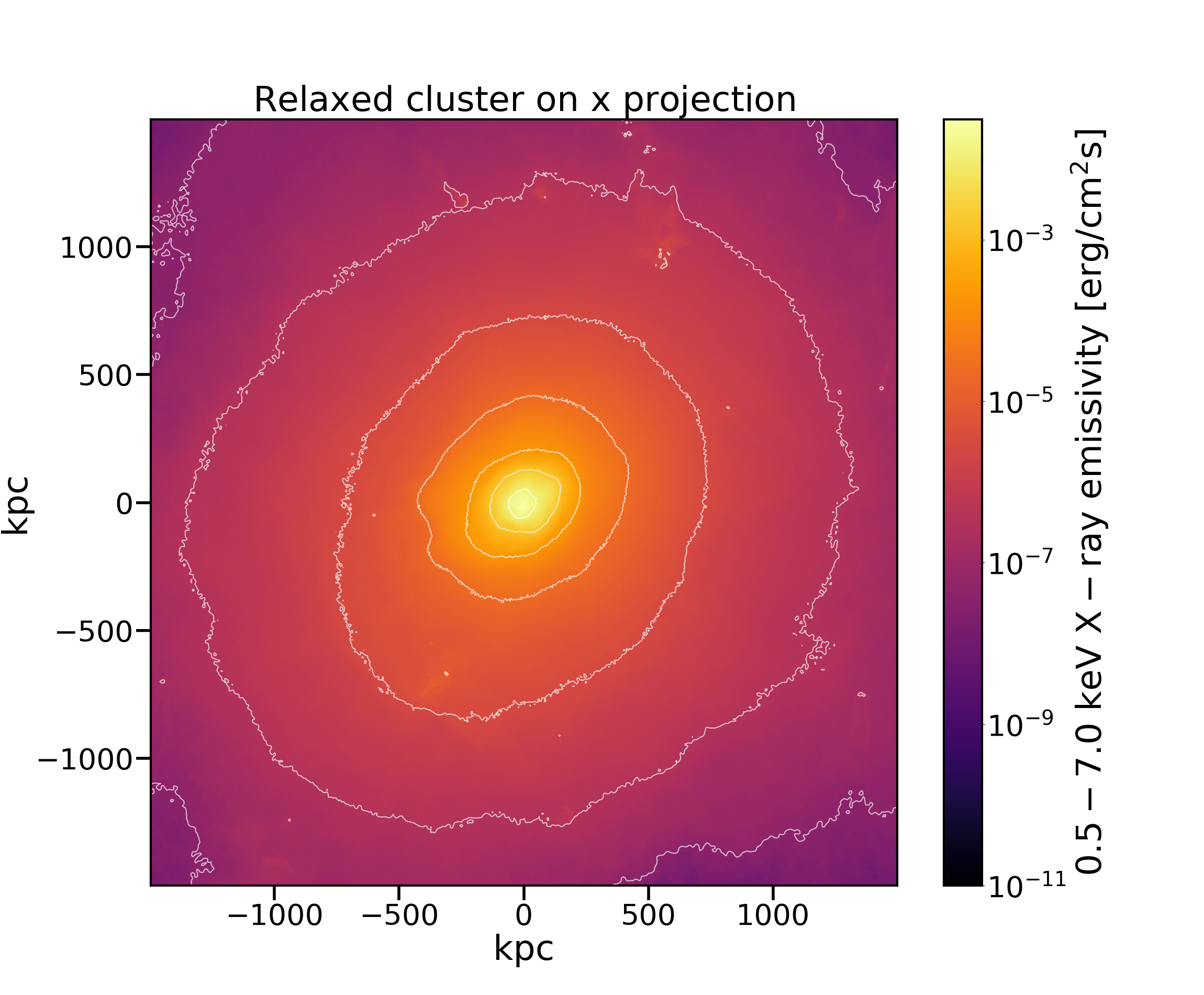}}
      \subfigure[]{\includegraphics[width=0.32\textwidth]{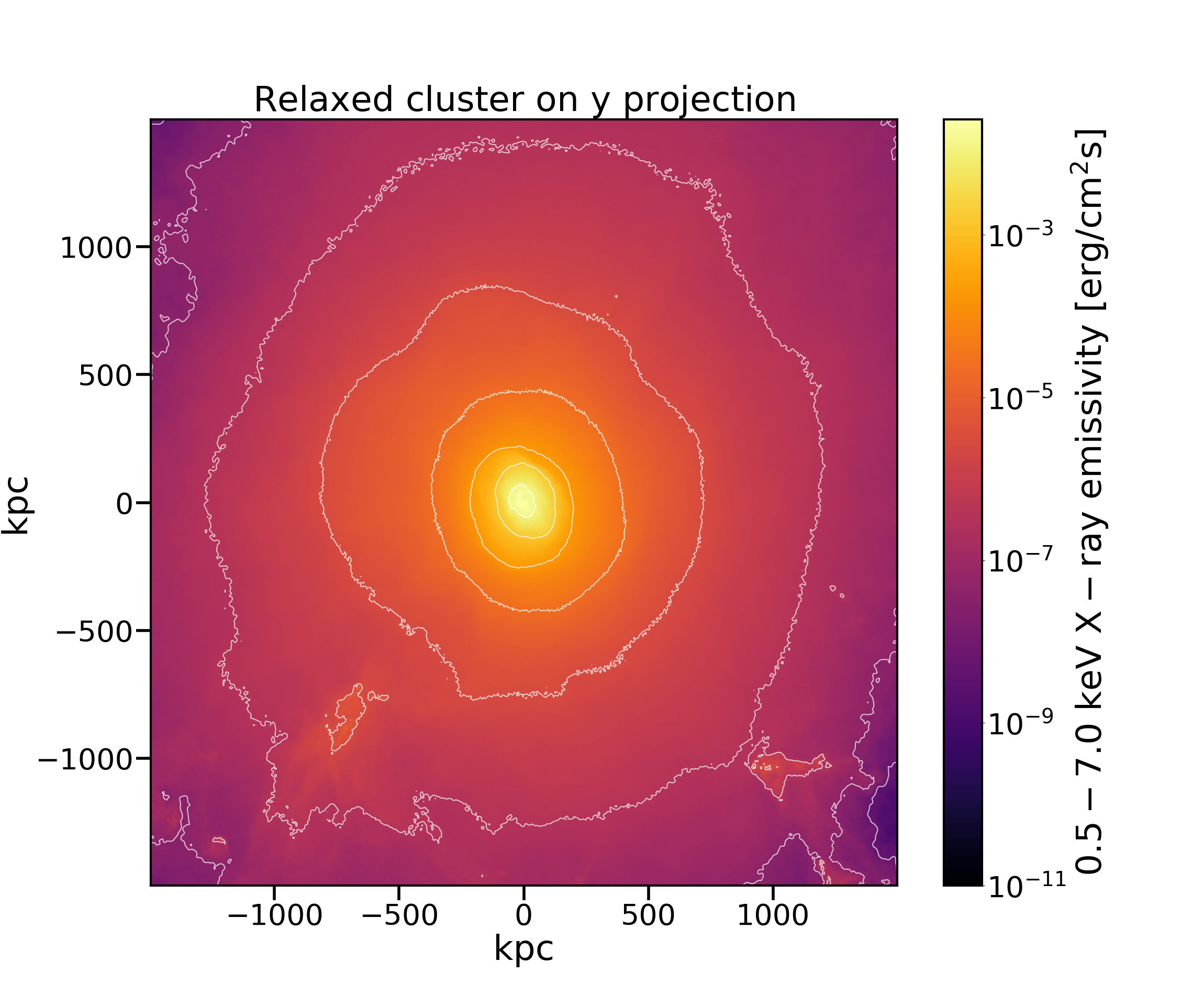}}
      \subfigure[]{\includegraphics[width=0.32\textwidth]{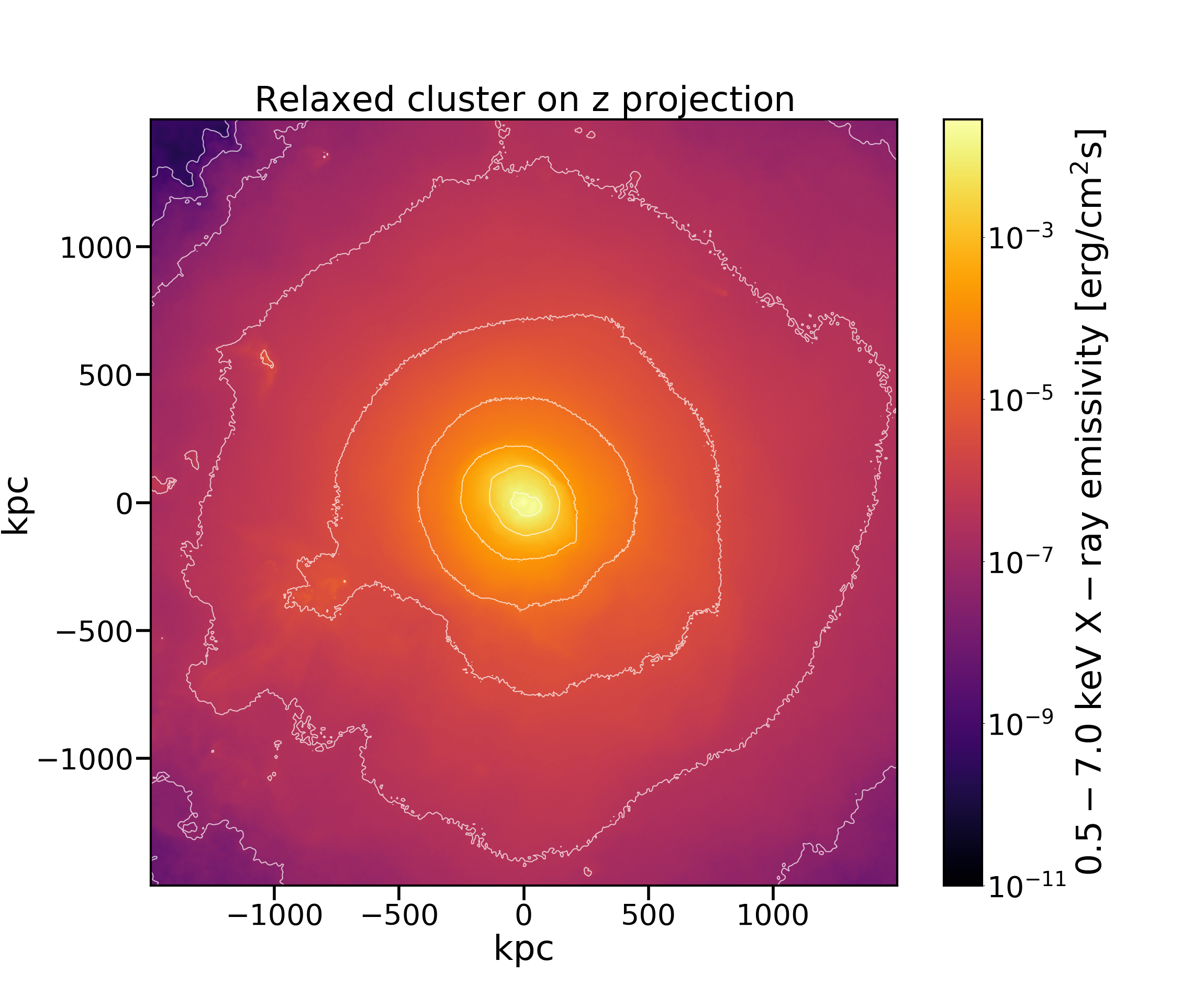}}      
    \caption{Mass (upper row) and X-ray emissivity (lower row) maps on the x, y, and z projections for the relaxed cluster whose coherence is shown in panel (c) of Fig.~\ref{example_c}. The white contours indicate logarithmic mass density and X-ray emissivity levels.}
      
 \label{rel_cluster}
 \end{figure*}

The most direct estimates of cluster masses employ measurements of the gravitational lensing distortions of background galaxies produced by foreground clusters. This technique is free of assumptions about the dynamical state of the cluster, yielding more consistent mass estimates. Even if with recent high-quality data from the Hubble Frontier Fields Initiative, mass models for some clusters have been reported at a few percent precision level \cite{Jauzac+2015}, this method remains sensitive to projection effects, which might be an additional source of uncertainty, making appearing clusters more or less relaxed than they actually are. In addition, estimating cluster masses through gravitational lensing relies on the chosen model profile, typically the Navarro-Frenk-White (NFW; \citealt{Navarro_1996}) profile, with the assumption that mass correlates with halo concentration (\citealt{Duffy2008}). \citealt{Lee2023} showed that a model bias can arise when the cluster density profile deviates from the assumed model profile, especially in merging systems. Therefore, despite the significant improvement introduced by integrating the gravitational lensing signal into the cluster mass calibration process, the observed scaling relations still deviate from the self-similar scenario. 
Constraining cosmological parameters from scaling relations using the traditional approach of analyzing all galaxy clusters, including those that may not fully adhere to the underlying assumptions, presents challenges. While larger-area sky surveys offer substantial statistical power, achieving significant improvements may require complementary strategies that account for the varying dynamical states of clusters. For instance, as we will highlight in the results and conclusions of this paper, it might be reasonable to consider clusters in various phases of their evolution separately, much like stars in different stages of their life occupy distinct regions on the Hertzsprung-Russell (HR) diagram.

Previously, the classification of disturbed cluster systems was done using several proxies: the shape of the velocity dispersion, for both galaxies and ICM (e.g., \citealt{Inogamov2003}, \citealt{Martinez12}, \citealt{2016Hitomi}, \citealt{2024Silich}) the X-ray morphology (e.g., \citealt{Mantz2015}, \citealt{Shi15}), diffuse radio emission (e.g., \citealt{Ferretti2012}), the cluster galaxy density distribution (e.g., \citealt{Wen15}), and a combination of observations at different wavelengths such as the X-ray
peak/centroid to Brightest Cluster Galaxy (BCG) offset (e.g., \citealt{MANN2012}). Although these 
methods allowed insights into understanding the dynamical
state of galaxy clusters, uncertainties persist and discrepancies are found between results obtained with different methods for the same cluster (e.g., \citealt{DeLuca21}, \citealt{Haggar2024}).

 \begin{figure*}
\centering
      \subfigure[]{\includegraphics[width=0.32\textwidth]{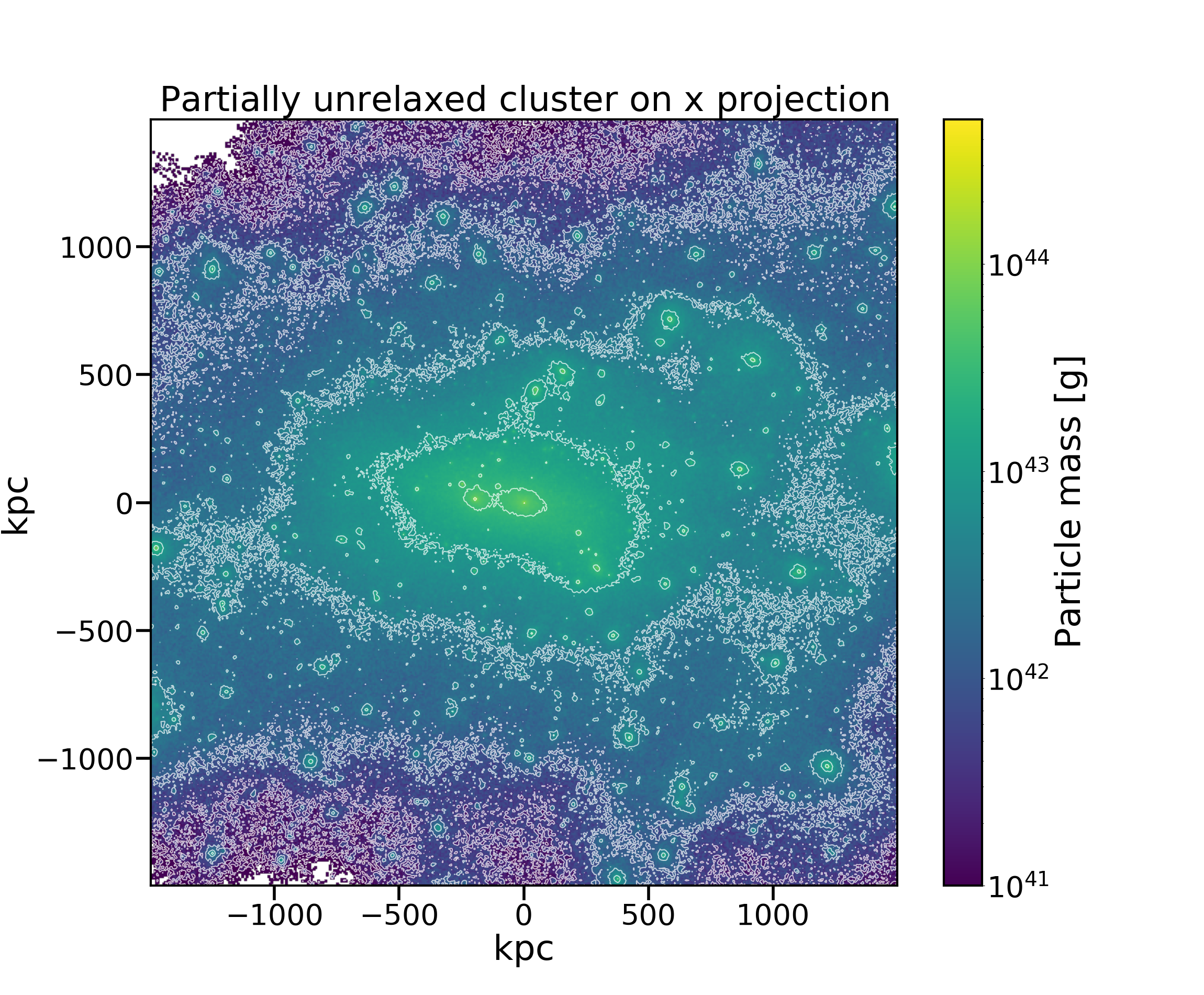}}
      \subfigure[]{\includegraphics[width=0.32\textwidth]{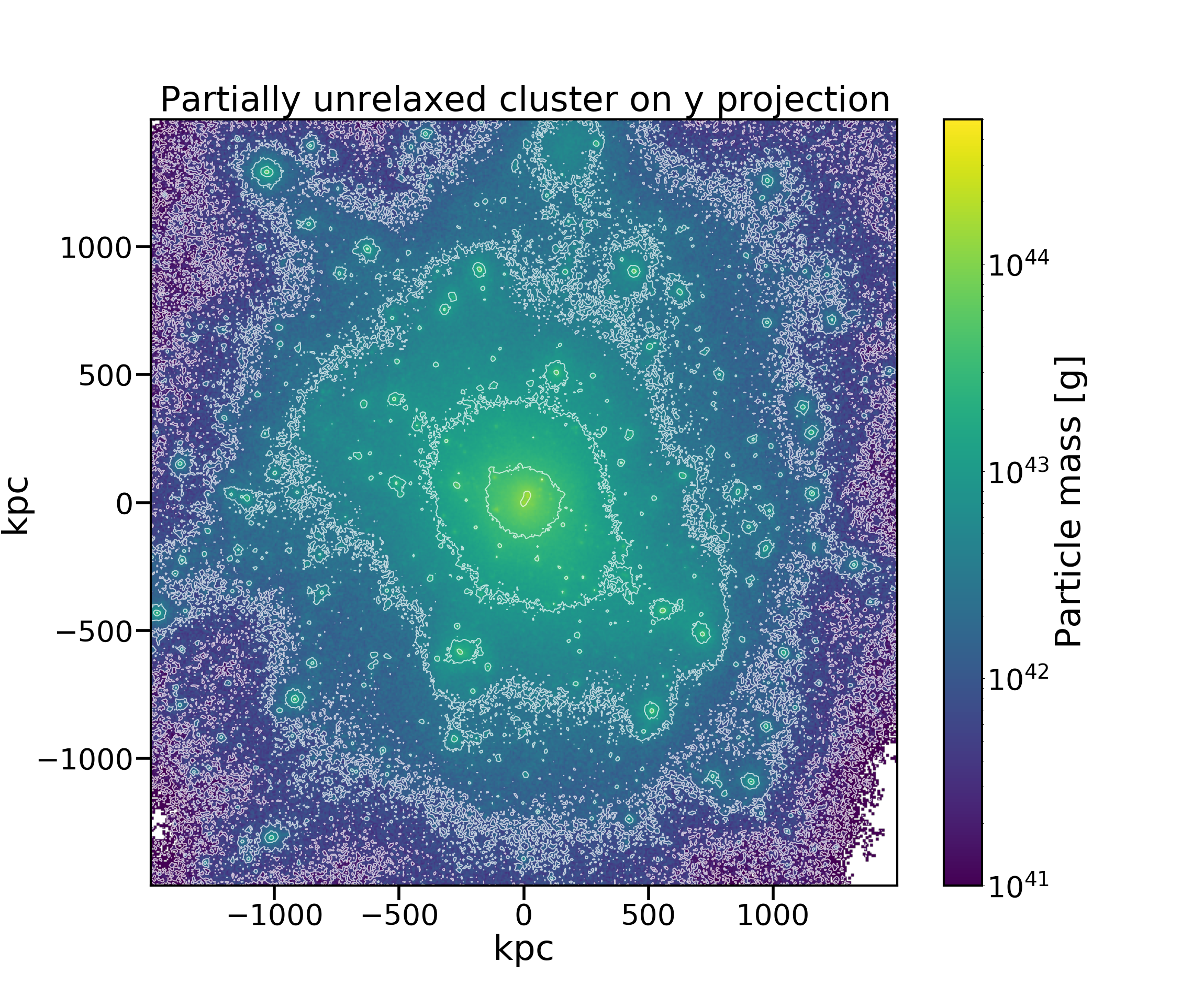}}
      \subfigure[]{\includegraphics[width=0.32\textwidth]{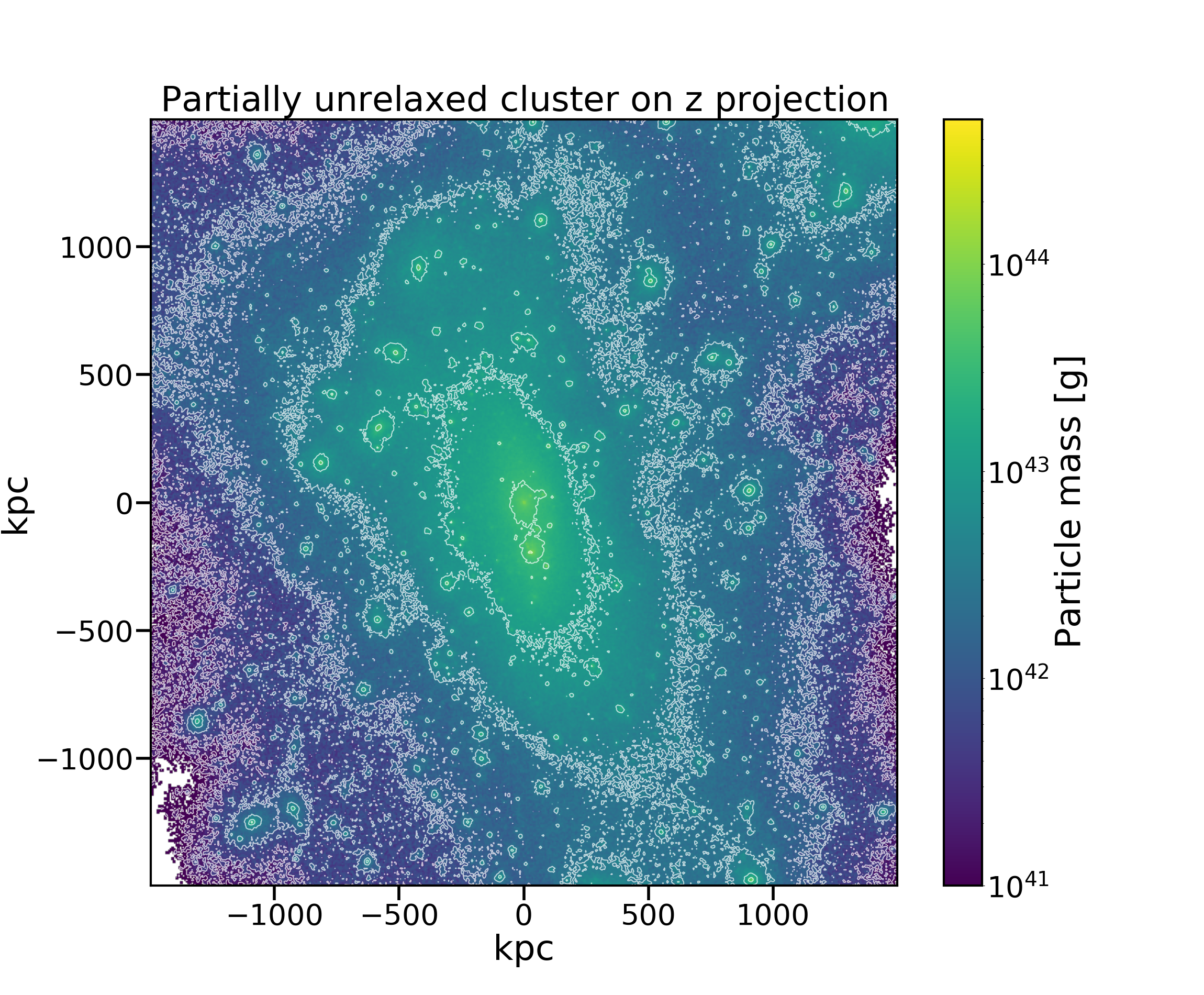}} 
      \subfigure[]{\includegraphics[width=0.32\textwidth]{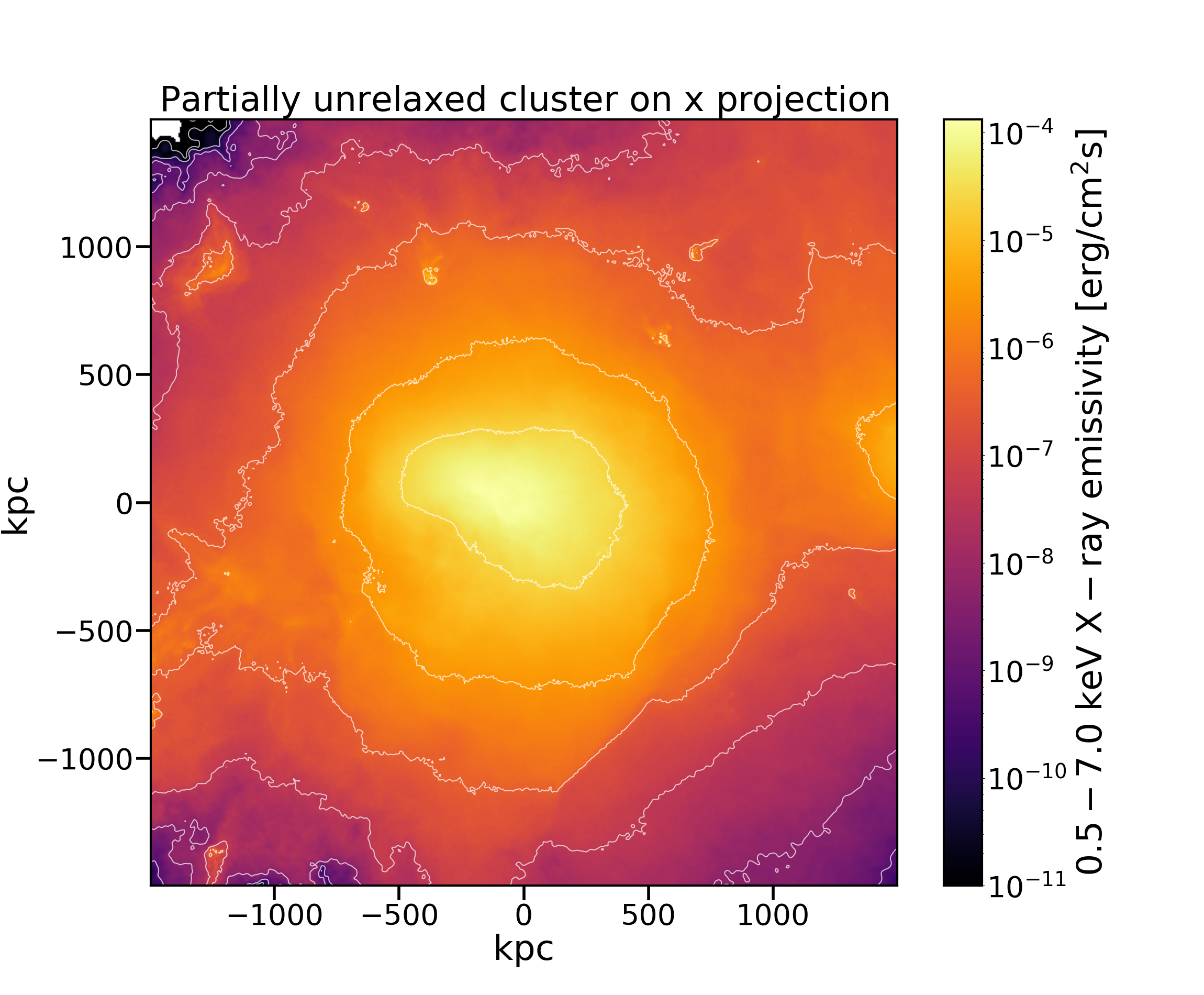}}
      \subfigure[]{\includegraphics[width=0.32\textwidth]{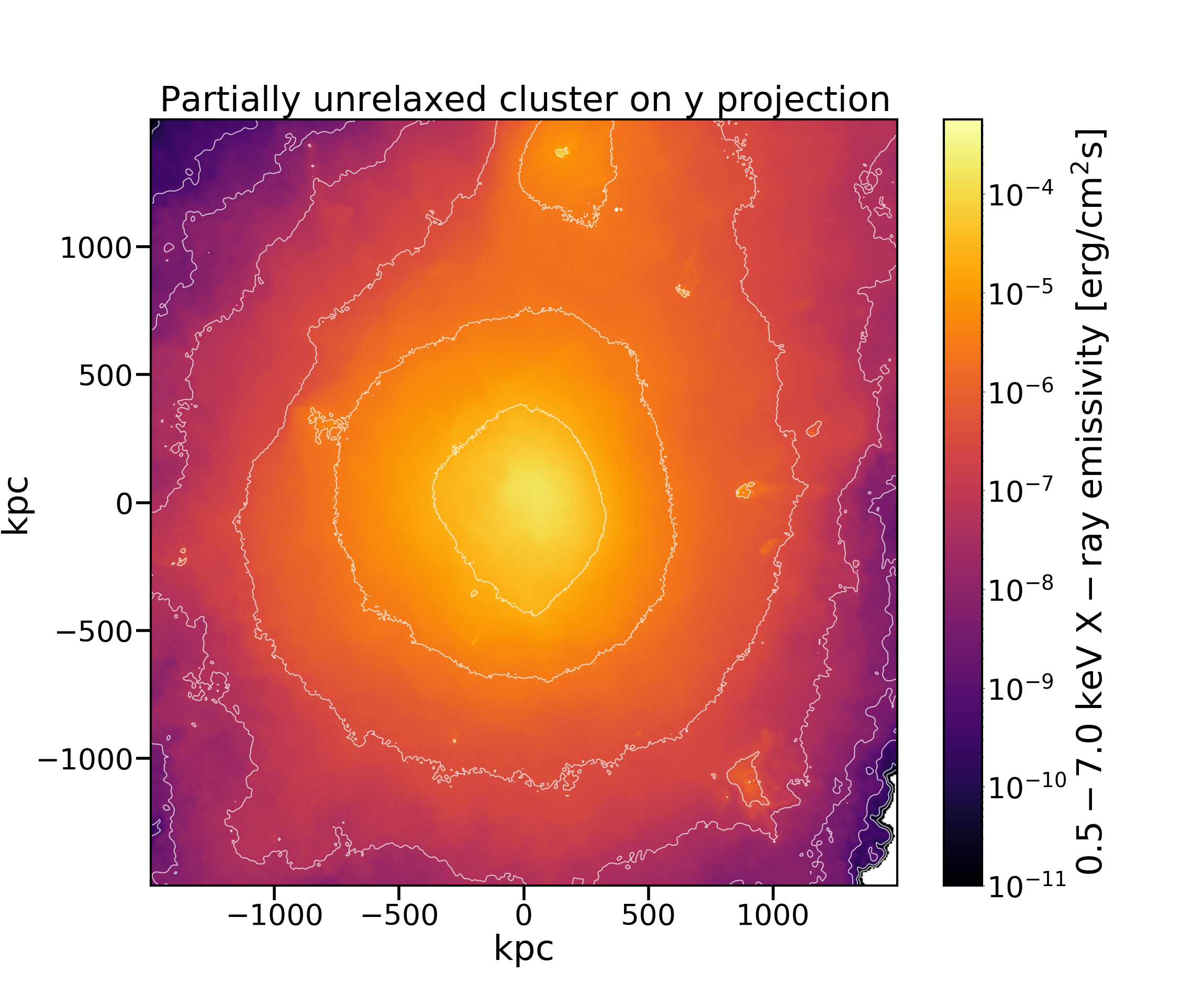}}
      \subfigure[]{\includegraphics[width=0.32\textwidth]{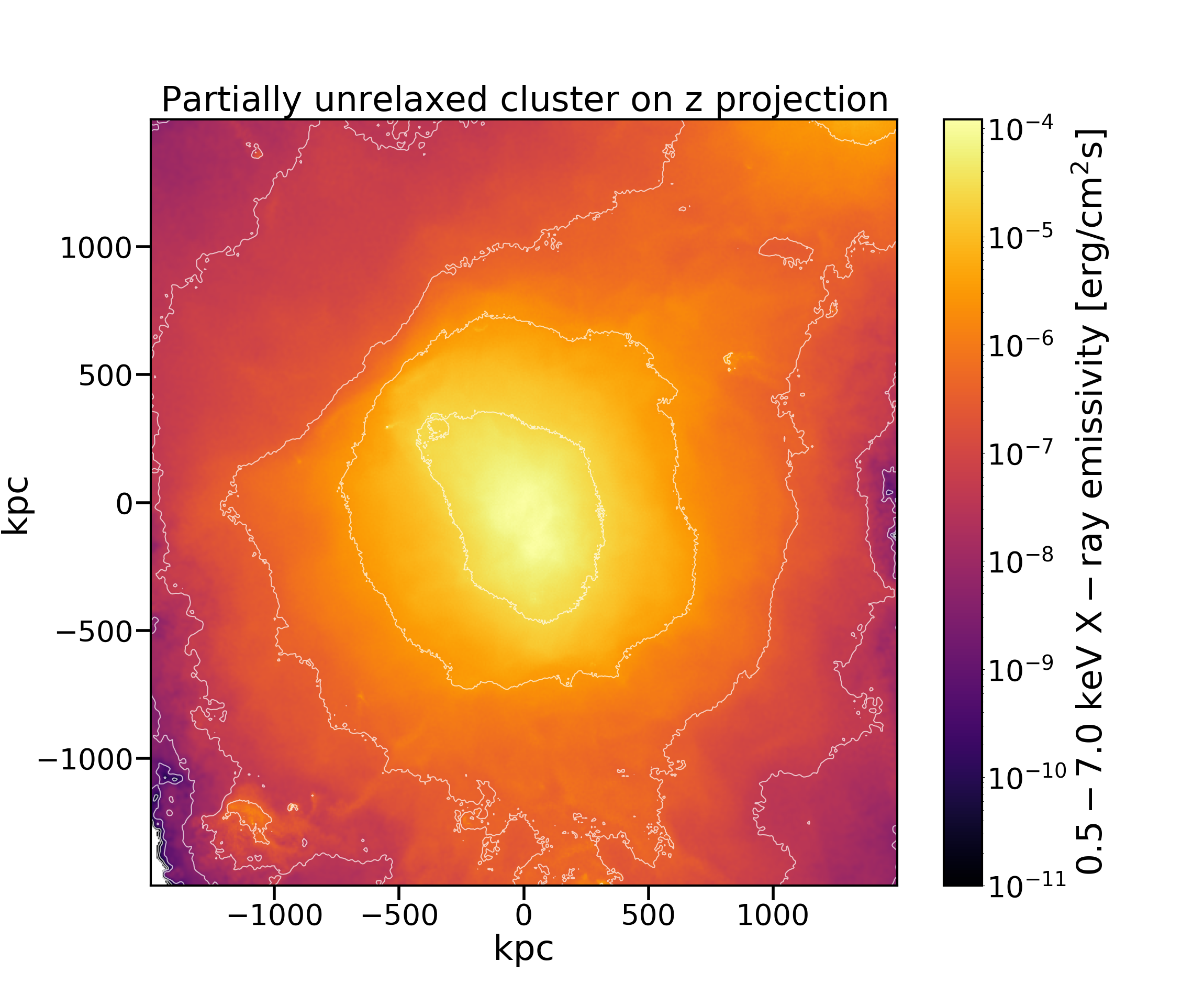}}       
    
      \caption{Mass (upper row) and X-ray emissivity (lower row) maps on the x, y and z projections for the partially unrelaxed cluster whose coherence is shown in panel (b) of Fig.~\ref{example_c}. The white contours indicate logarithmic mass density and X-ray emissivity levels.}
 \label{par_rel_cluster}
 \end{figure*}

In our previous work (\citealt{https://Cerini22}) we adopted the power spectrum and coherence analysis to assess the dynamical state of galaxy clusters. This method, already used in the past decades in many other successful studies (e.g., \citealt{Kashlinsky_2005}, \citealt{Churazov_2012}, \citealt{Cappelluti_2012}, \citealt{Kashlinsky_2012}, \citealt{Cappelluti_2013}, \citealt{Helgason_2014}, \citealt{Cappelluti_2017}, \citealt{Eckert_2017}, \citealt{zhuravleva17}, \citealt{Li_2018}, \citealt{Kashlinsky_2018}) in other contexts, was tested for two real clusters - Abell 2744 and Abell 383 - and four additional simulated clusters from the OMEGA500 simulation suite. The approach involved cross-correlating the fluctuations in the matter and X-ray surface brightness distributions (the first one inferred from high-resolution mass maps derived from gravitational lensing) to quantify how well the ICM gas traces the underlying dark matter potential. This study led to the proposal of a new quantity, the \textsl{Coherence Radius} $R_{\rm CR}$, renamed \textsl{galaxy cluster coherence length} $\ell_{\rm CR}$ in this paper. The updated terminology provides a clearer understanding of its true meaning and prevents it from being misinterpreted as deriving from a radial profile. While innovative within the framework of galaxy cluster studies, this definition is distinct from previous uses of ``coherence length'' in large-scale structure analyses. For example, in \citealt{Kash1992}, the term describes the scale over which peculiar velocities — deviations from the uniform Hubble flow — remain correlated. Instead, the \textsl{galaxy cluster coherence length} $\ell_{\rm CR}$ introduced here specifically quantifies the scale above which fluctuations in the mass distribution and the X-ray surface brightness of galaxy clusters are coherent at the 90$\%$ level. This novel application ensures consistency with established terminology in power spectrum analysis but represents a fundamentally different concept tailored to the internal properties of galaxy clusters. Importantly, this definition differs from previous uses of "coherence length" in cosmology, as it applies to localized structures and examines the coherence between dark matter and gas distributions within individual clusters. 
For simplicity, throughout the remainder of this paper, we will refer to the \textsl{galaxy cluster coherence length} $\ell_{\rm CR}$ as the \textsl{coherence length}, while maintaining the specificity of its definition in the galaxy cluster context.

The \textsl{coherence length} was shown to be particularly useful in assessing the equilibrium state of clusters and identifying deviations from it. In this work, we apply the coherence analysis to the study of simulated clusters from the cosmological TNG300 simulation suite. We analyze both the 3D distributions of dark matter and hot X-ray emitting gas and projections on 3 perpendicular planes, in order to address how projection effects impact the determination of the dynamical state of the clusters and the scatter in scaling relations. We revisit $L_X-M$ and $T_X-M$ scaling relations - with $L_X$, $T_X$ and $M$ X-ray luminosity, X-ray temperature and mass of the cluster, respectively - incorporating the \textsl{coherence length} $\ell_{\rm CR}$ obtained from the power spectrum analysis of fluctuations to relate the scatter in these relations to the dynamical state of the clusters.

This paper is structured as follows: the data are described in Section~\ref{data}, the coherence analysis is described in Section~\ref{coherence}, the scaling relations modeling and fitting are treated in Section~\ref{scale_rel}, results and conclusions are presented in Section
~\ref{results} and \ref{conclusions}, respectively. Throughout the paper, errors are quoted at 1$\sigma$ level unless otherwise specified. 

 \begin{figure*}
\centering
      \subfigure[]{\includegraphics[width=0.32\textwidth]{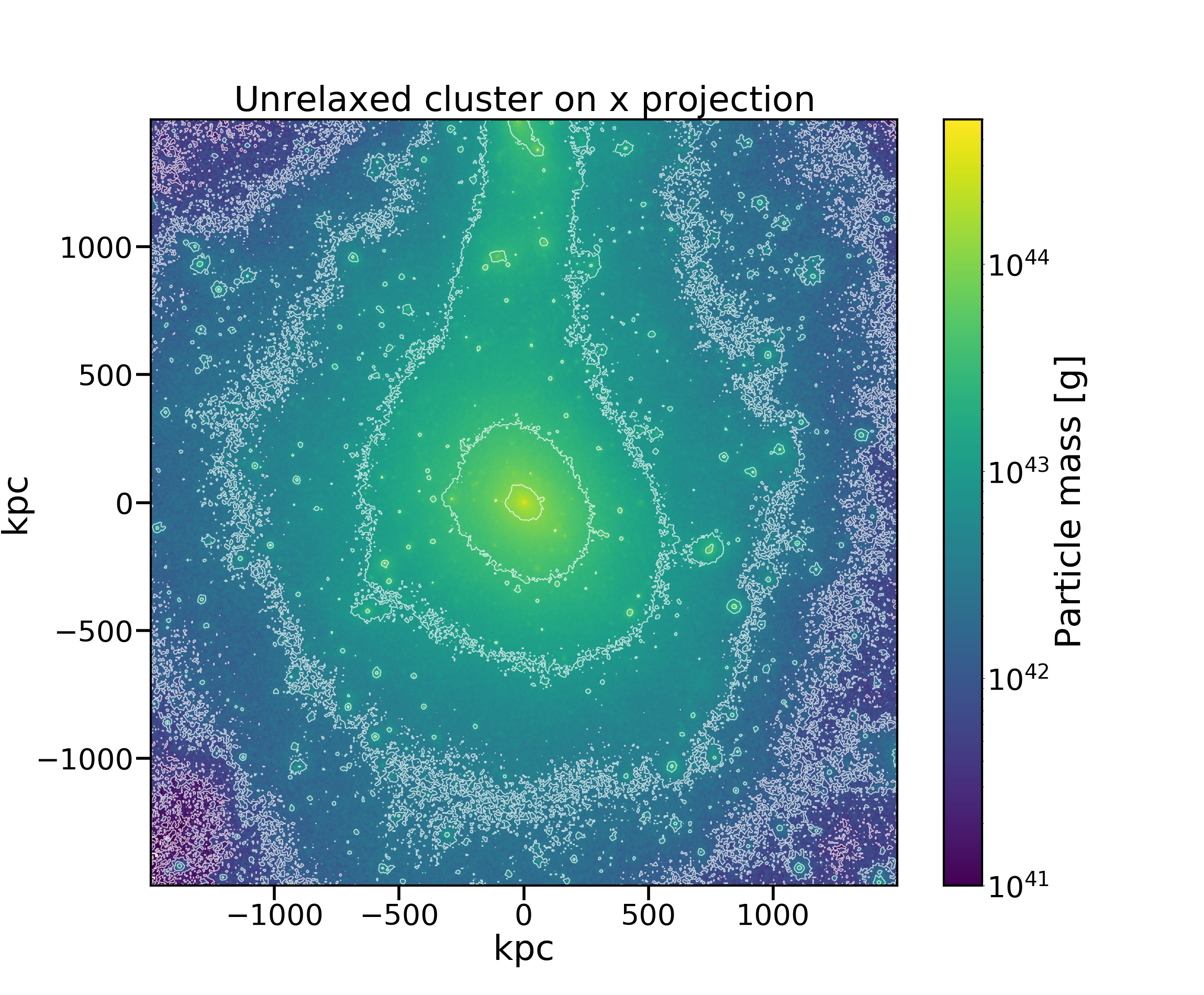}}
      \subfigure[]{\includegraphics[width=0.32\textwidth]{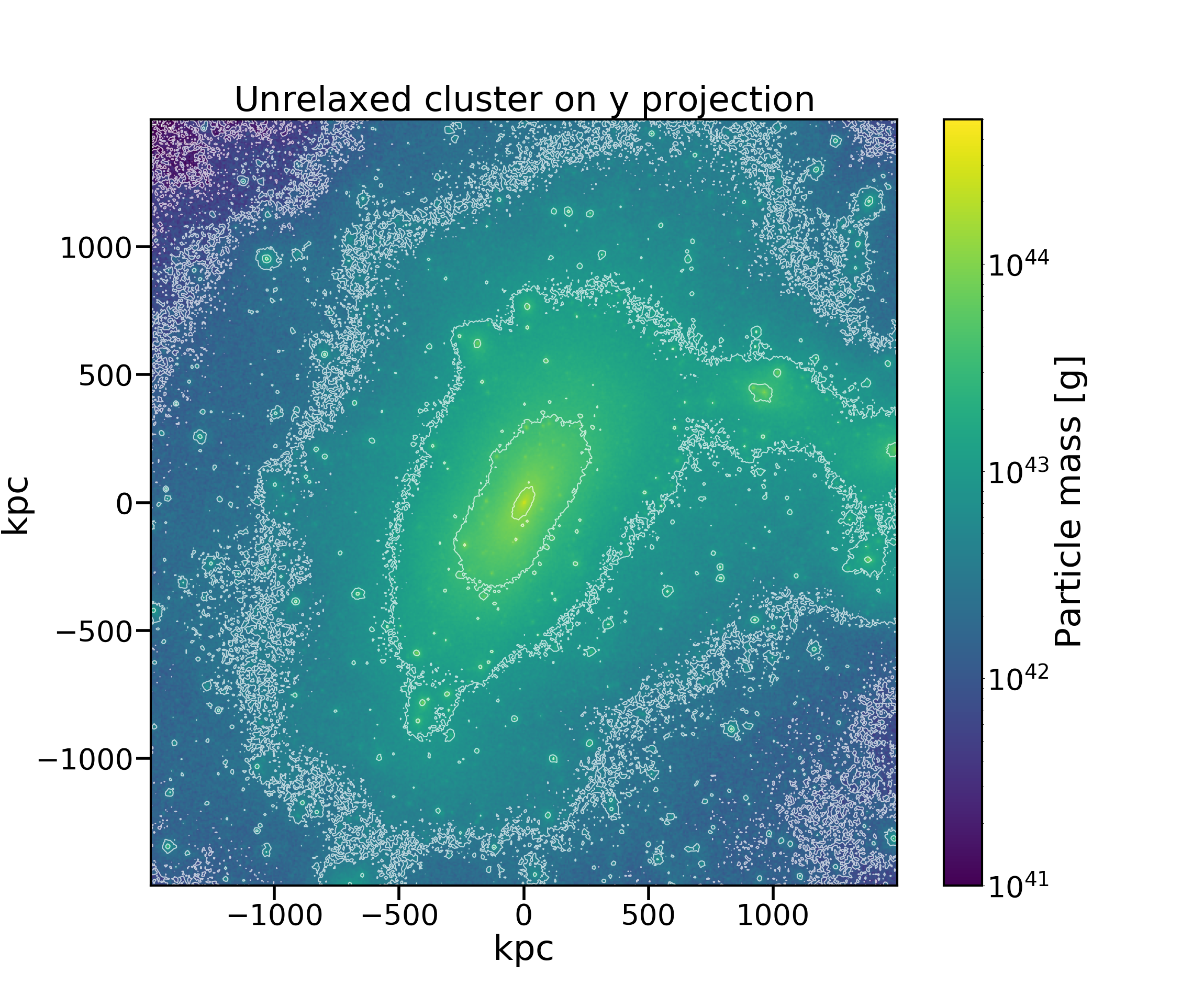}}
      \subfigure[]{\includegraphics[width=0.32\textwidth]{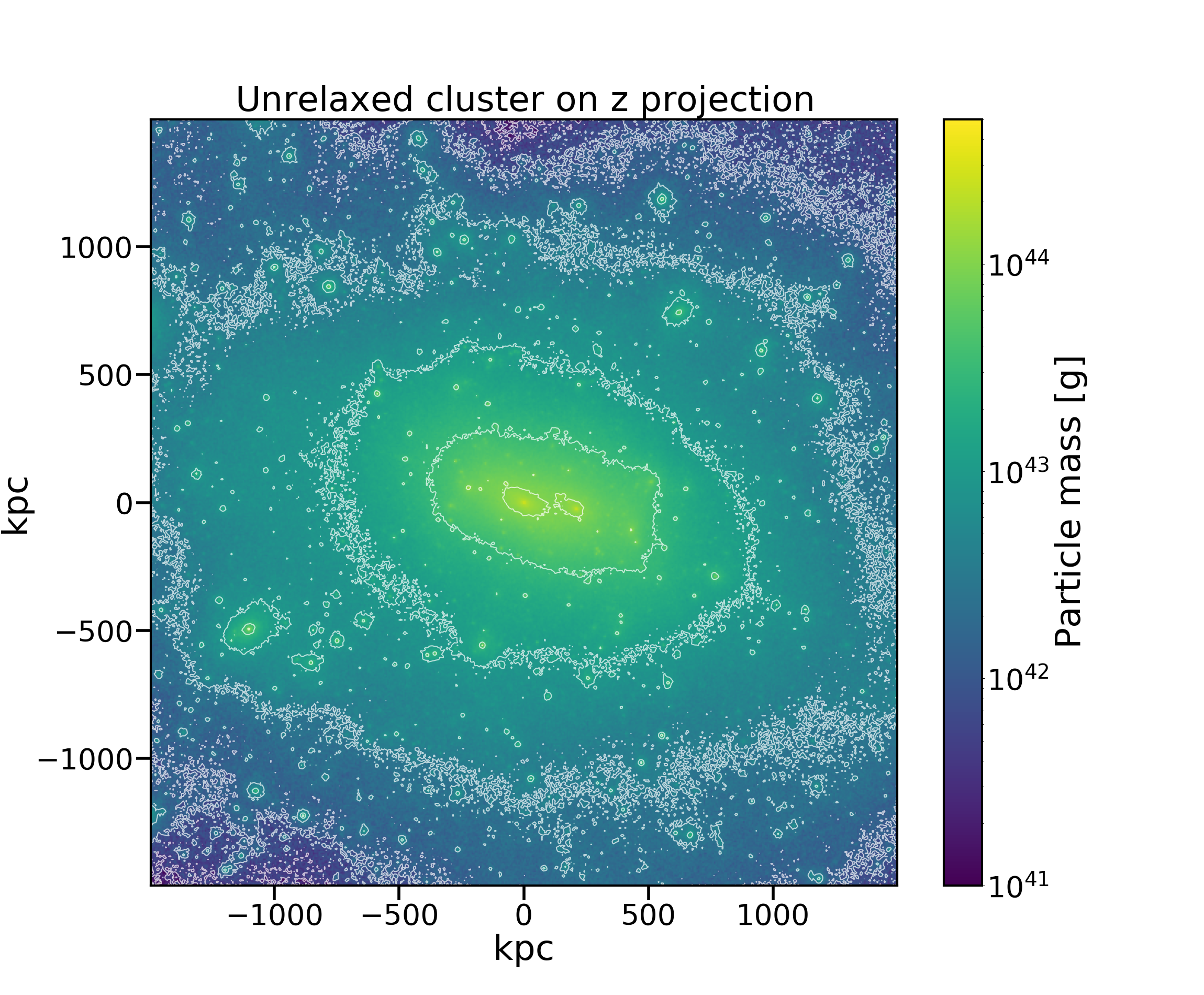}} 
      \subfigure[]{\includegraphics[width=0.32\textwidth]{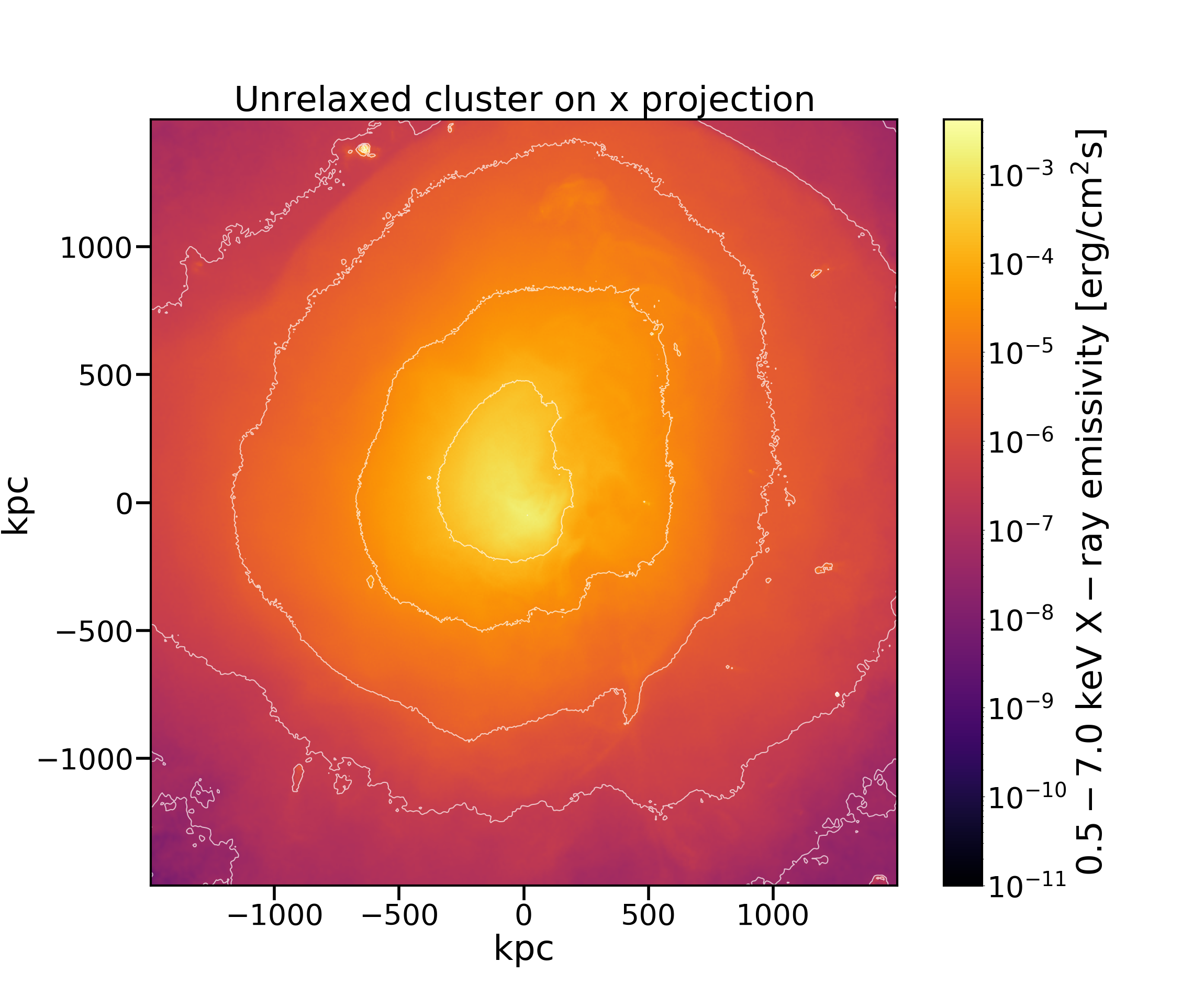}}
      \subfigure[]{\includegraphics[width=0.32\textwidth]{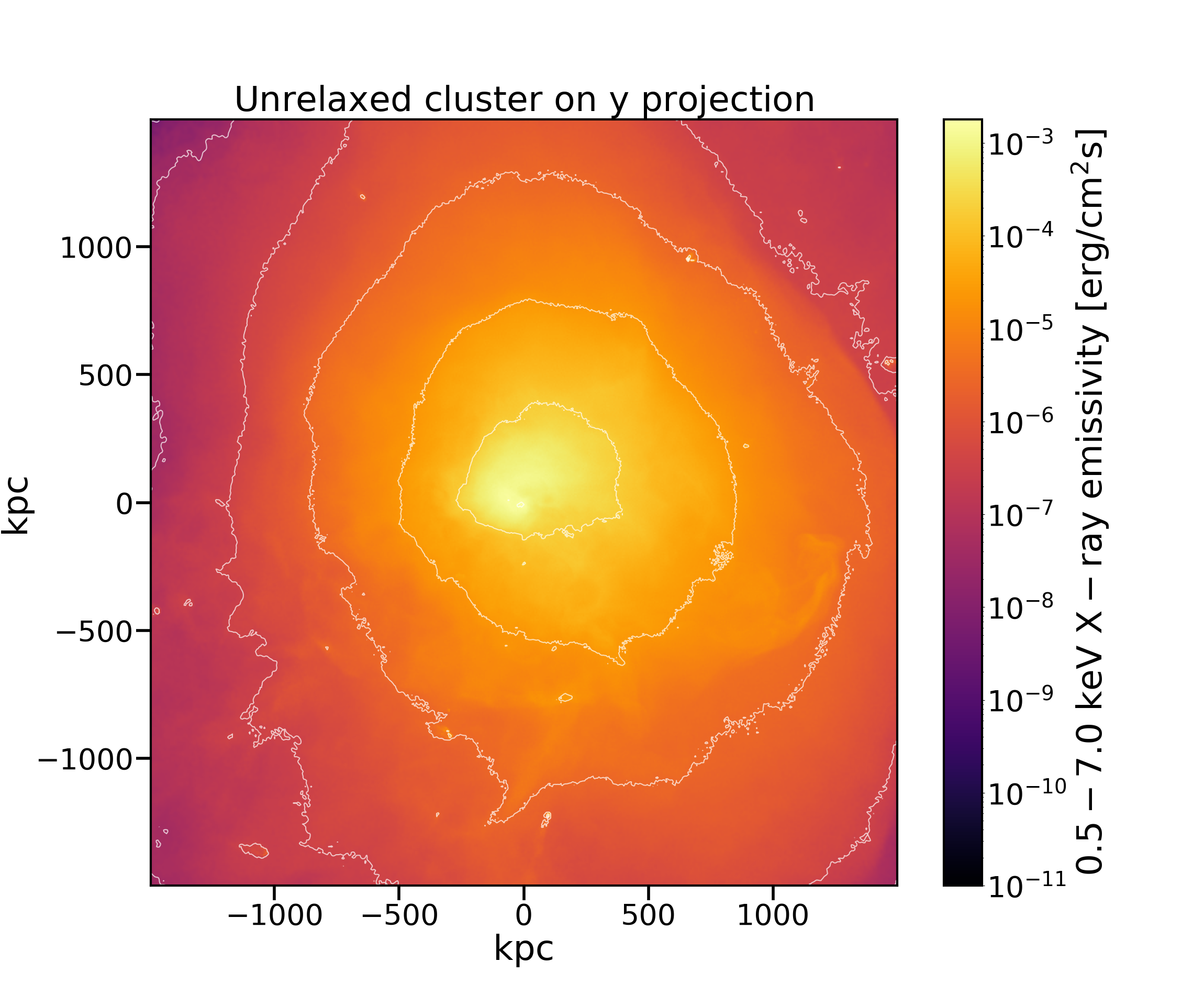}}
      \subfigure[]{\includegraphics[width=0.32\textwidth]{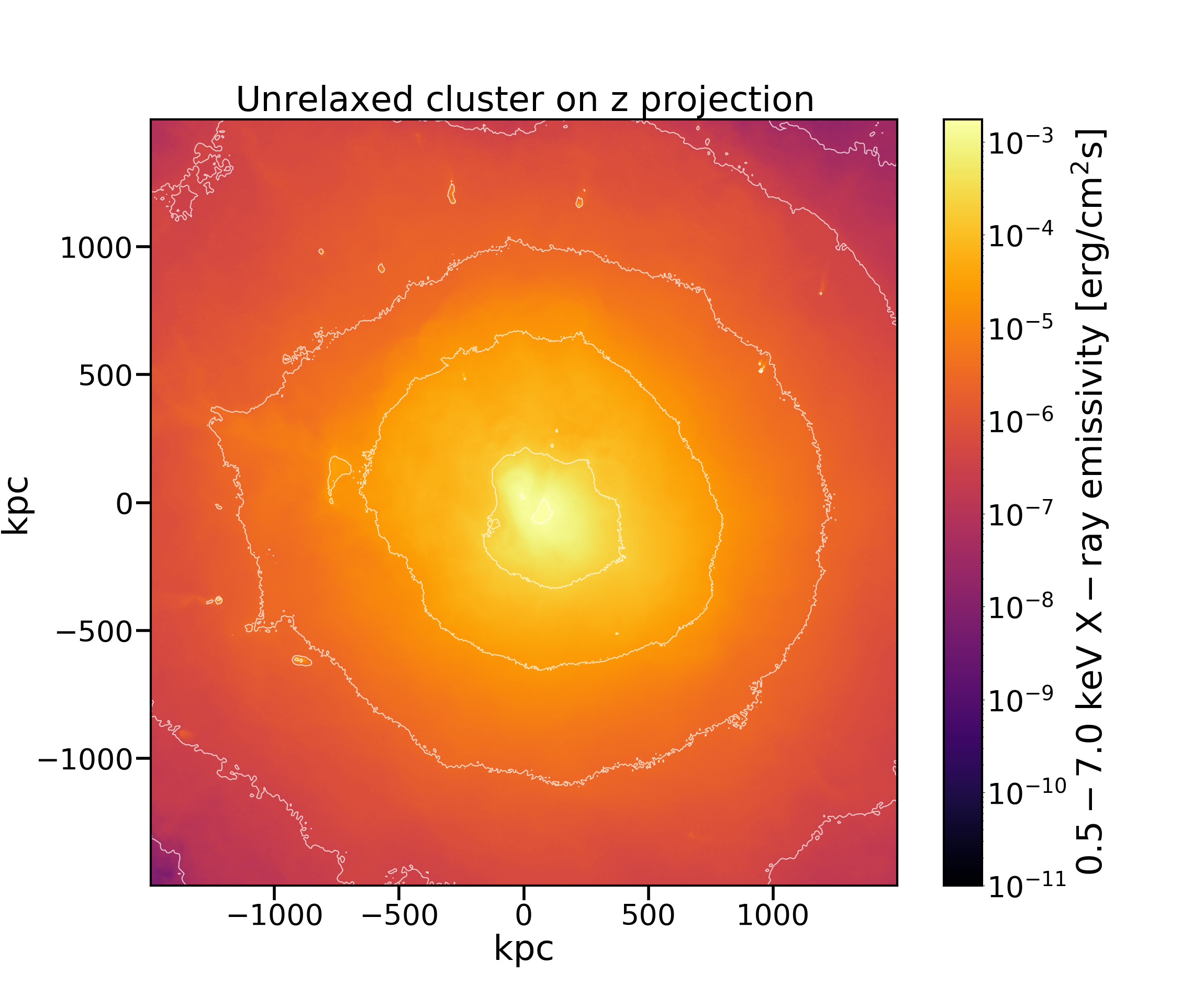}}      
    
      \caption{Mass (upper row) and X-ray emissivity (lower row) maps on the x, y, and z projections for the unrelaxed cluster whose coherence is shown in panel (c) of Fig.~\ref{example_c}. The white contours indicate logarithmic mass density and X-ray emissivity levels.}
 \label{unrel_cluster}
 \end{figure*}

\section{The Dataset: TNG300 simulations} \label{data}

The TNG300 simulation suite is one of the cosmological magnetohydrodynamical simulations of galaxy formation within the IllustrisTNG project \footnote{https://www.tng-project.org} (\citealt{2018Pillepich}, \citealt{2018Martinacci}, \citealt{2018Nelson}, \citealt{2018Springel}, \citealt{2018Naiman}, \citealt{Nelson:2018uso}, \citealt{2019Pillepich}). 

IllustrisTNG presents a new galaxy formation model within the $\Lambda$CDM paradigm (\citealt{2017Weinberger},\citealt{2018Pillepich}) that includes radiative cooling of the gas, star formation and evolution, chemical enrichment, stellar feedback, and supermassive black hole (BH) growth and feedback. These simulations reproduce galaxies with realistic stellar masses and sizes, as well as galaxy groups and clusters with gas fractions in better agreement with observations. The TNG300 simulations follow the co-evolution of DM, gas, stars, and supermassive BHs within a cubical volume of approximately 300 Mpc on a side. 

The adopted cosmological parameters are given by a matter density $\Omega_m=\Omega_{DM}+\Omega_b=0.3089$, baryonic density $\Omega_b=0.0486$, cosmological constant $\Omega_{\Lambda}=0.6911$, Hubble constant $H_0=100~h~$km~s$^{-1}$Mpc$^{-1}$, with $h=0.6774$, normalization $\sigma_8=0.8159$ and spectral index $n_s=0.9667$ (taken from Planck, \citealt{2016Planck}). Halos, subhalos, and their basic properties are obtained with the friends-of-friends and SUBFIND algorithms (\citealt{Davis:1985rj}, \citealt{2001Springel}, \citealt{2009Dolag}). Here we use 329 $z=0$ galaxy clusters, adopting the spherical overdensity $M_{500}$, obtained by summing the mass of all particles and cells enclosed within $r_{500}$ (the radius of the sphere around the cluster enclosing an average density 500 times greater than the critical) - including dark matter, gas and stars. Our sample comprises 329 simulated clusters at redshift $z=0$ in the mass range $(0.37-11.55)\times 10^{14}M_{\odot}$, with a median of $0.97\times 10^{14}M_{\odot}$. This choice is consistent with previous simulation studies using TNG300 that include group-scale halos to expand sample sizes and investigate a wider dynamical range (e.g. \citealt{2018Pillepich}) and aligns well with recent observational selections (e.g. \citealt{ghirardini2024srgerosita}). For every cluster, we obtained 3D mass, gas density, temperature, and X-ray surface brightness data cubes (the latter in the energy range $0.5-7.0$ keV) with size $3\times3\times3$~Mpc, as well as projected $3\times3$~Mpc maps along the three perpendicular axes - x, y and z. We adopted a grid of $512\times 512\times 512$ ($512\times 512$) pixels over the $3\times3\times3$~Mpc ($3\times3$~Mpc) field of view, corresponding to a pixel scale of approximately 5.9 kpc. The choice of $0.5-7.0$ keV band ensures consistency with some common practices in both simulation-based and real observational analyses of galaxy clusters. Indeed, this range has been widely adopted in mock X-ray studies as well as in real Chandra observations of clusters (e.g., \citealt{Rasia2012}, \citealt{Henden2019}, \citealt{Ubertosi2023}). The $0.5-7.0$ keV band captures the bulk of the thermal bremsstrahlung emission from the ICM. Furthermore, our choice was guided by the need to define a band that is representative and compatible with multiple X-ray instruments we plan to use in future applications, including Chandra, XMM-Newton, and eROSITA. While eROSITA operates effectively only up to $\sim 2.3$ keV, XMM-Newton and Chandra extend to higher energies. However, Chandra's effective area declines significantly above $\sim 7.0$ keV, making this a reasonable upper limit for maintaining good sensitivity. Our adopted band, therefore, maximizes compatibility and scientific return across these instruments and aligns with some common Chandra practices.

For the X-ray emission, we adopted the APEC model 3.0.9 (\citealt{2001Smith}). To test the robustness of our X-ray luminosity maps to metallicity assumptions, we generated the luminosity maps using the APEC model with two approaches: (1) adopting a constant metallicity value of $0.3Z_{\odot}$, as commonly done in X-ray analyses of galaxy clusters (e.g., \citealt{Rasia2012}, \citealt{Nagai2007}); and (2) spatially varying metallicities directly extracted from the TNG simulation snapshots. In the second approach, each gas cell's metallicity was used to compute its individual X-ray emissivity with APEC, following procedures similar to those adopted in previous works (e.g., \citealt{Barnes2018}). In both cases, we employed the default abundance pattern from \citealt{Anders1989} as implemented in APEC. We found an average difference of 0.0013 in the final coherence measurements (representing $0.13\%$ of the maximum coherence), confirming that our results are robust against reasonable metallicity assumptions. This small difference can be explained physically by the fact that the X-ray emissivity scales as $\epsilon_X \propto n_e^2\times \Lambda(T,Z)$, where $n_e$ is the electron number density and $\Lambda(T,Z)$ is the cooling function, which depends on the gas temperature $T$ and metallicity $Z$. In clusters, the dependence on $n_e$ dominates, while metallicity variations mainly affect $\Lambda(T,Z)$ and thus have only a minor impact on large-scale morphological or coherence indicators.  This is consistent with the practice in the literature of adopting a constant metallicity approximation for synthetic X-ray analyses when focusing on global cluster properties (e.g., \citealt{Rasia2012}, \citealt{Nagai2007}). Throughout the paper, all plots and quantitative results refer to the simulations using the constant metallicity of $0.3Z_{\odot}$.

Every 3D cube was then projected along the three perpendicular axes x, y, and z. For all clusters, we computed the X-ray temperature $T_X$ and luminosity $L_X$:

\begin{eqnarray}
T_X=\frac{\sum_{i} T_i\epsilon_i\Delta S_i(\Delta V_i)}{\sum_{i} \epsilon_i\Delta S_i(\Delta V_i)},
\end{eqnarray}
\begin{eqnarray}
L_X=\frac{\sum_{i} \epsilon_i\rho_i^2\Delta S_i(\Delta V_i)}{\sum_{i} \rho_i^2},
\end{eqnarray}

where $T_i$, $\rho_i$ and $\epsilon_i$ are the temperature, density and emissivity of the cell $i$. $\Delta S_i$ and $\Delta V_i$ represent the surface areas of the cell $i$ in the projected images and the volumes of the cells in the data cubes, respectively. As explained in Section~\ref{results}, the sum is over a sphere from the center of the cluster to $r_{500}$ and over the same sphere but excluding the core region with radius $r_c=0.1r_{500}$, consistently with both theoretical and observational analyses (e.g., \citealt{2006Vikh}, \citealt{2006Kravtsov}, \citealt{Chen_2007}, \citealt{2009Vich}, \citealt{2010Hudson}, \citealt{2011Fabjan}, \citealt{2014Planelles}).

 \begin{figure*}
\centering
      \subfigure[]{\includegraphics[width=0.32\textwidth]{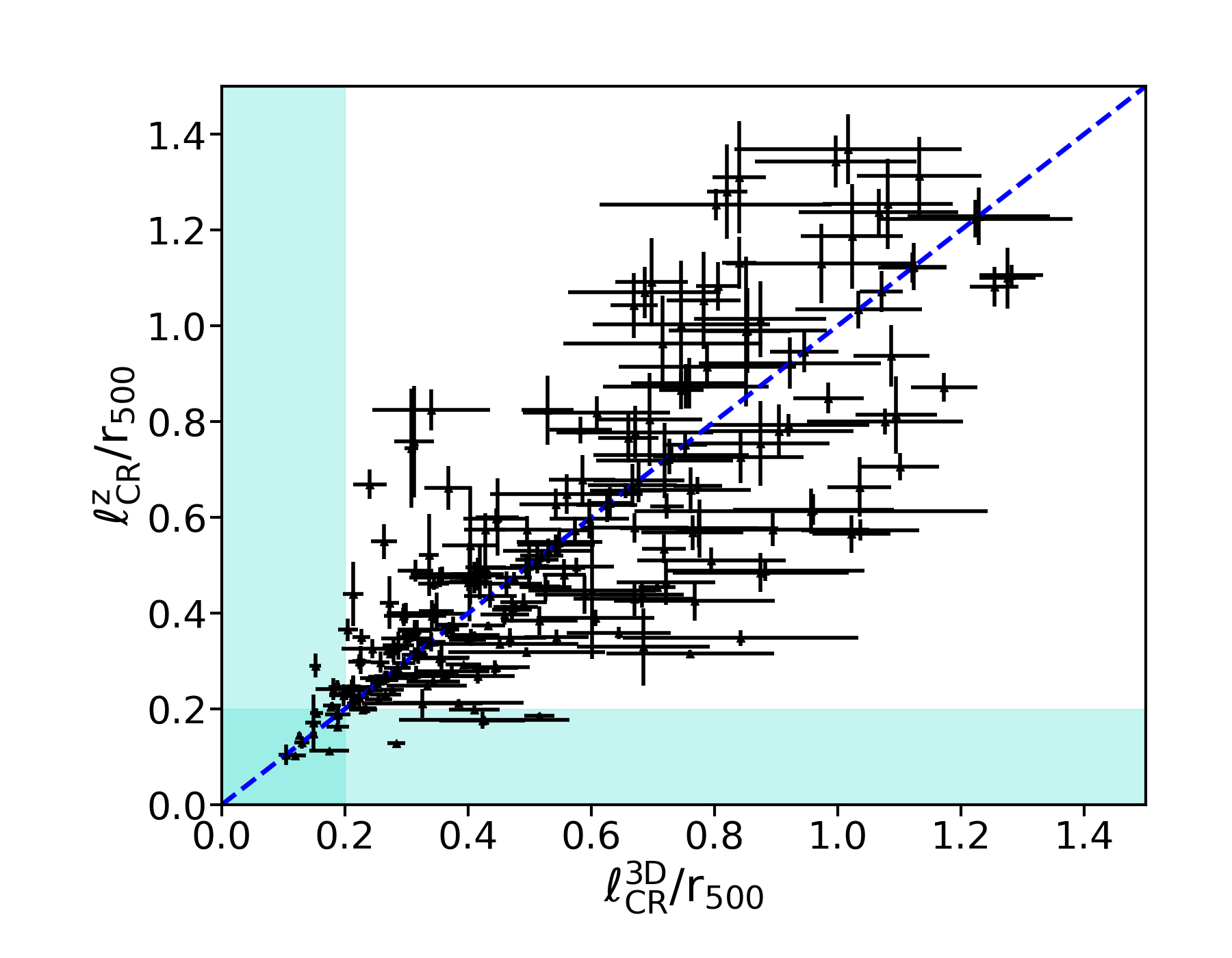}}
      \subfigure[]{\includegraphics[width=0.32\textwidth]{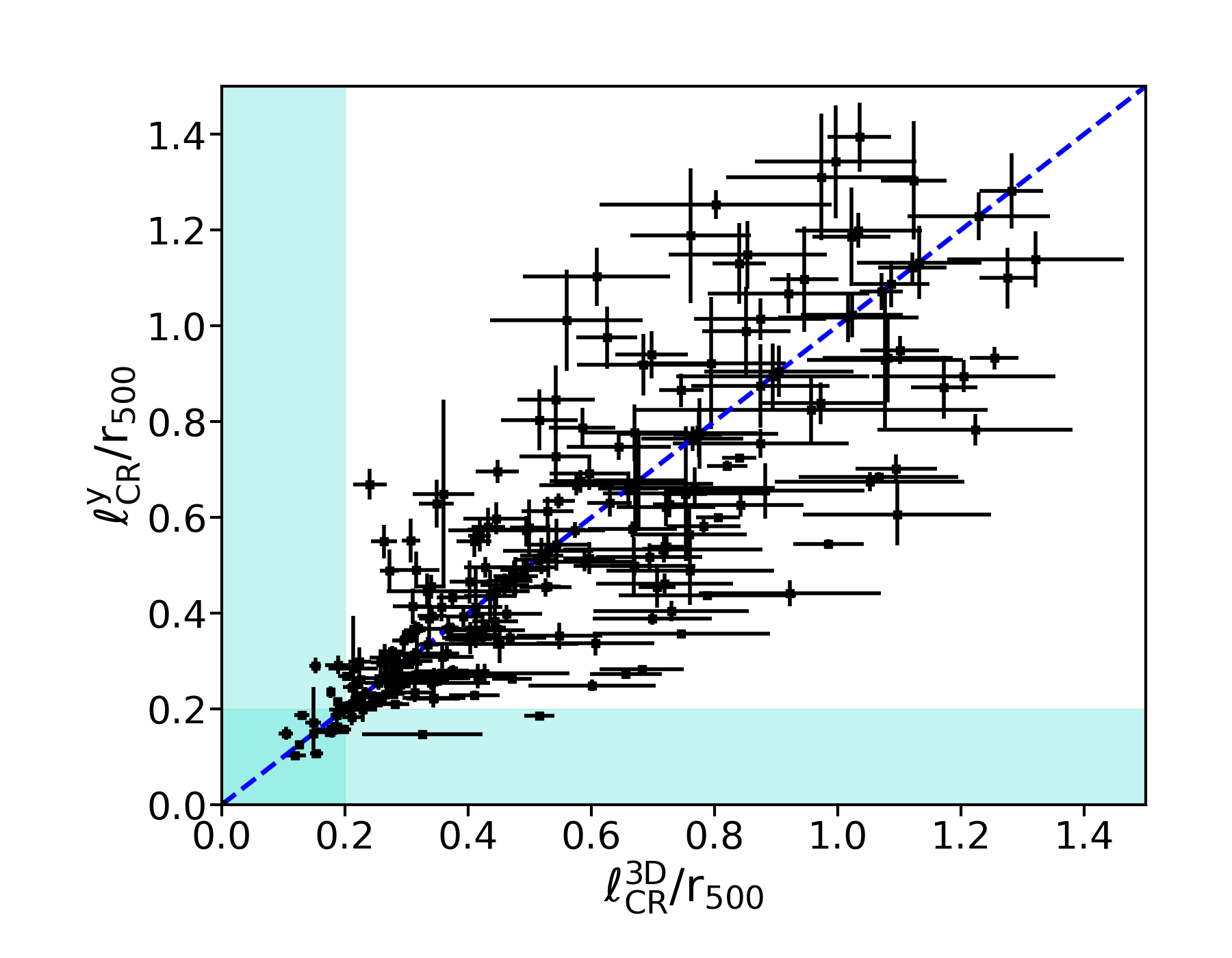}}
      \subfigure[]{\includegraphics[width=0.32\textwidth]{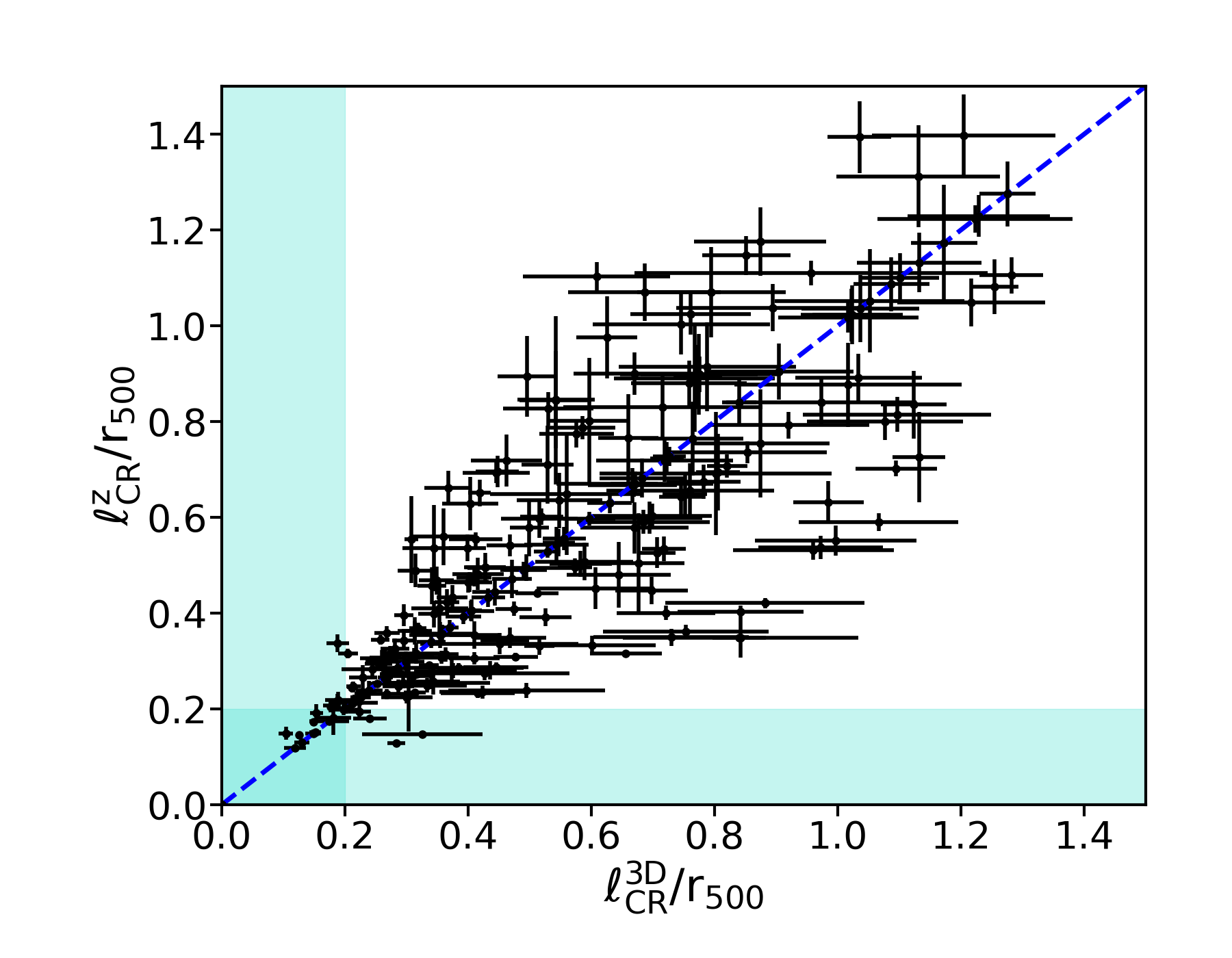}}      
    
      \caption{Relation between 2D and 3D \textit{coherence lengths}. The errors on the \textit{coherence lengths} are derived from the propagation of the errors on the coherence to the \textit{coherence length}, as explained in Section ~\ref{coherence}. The blue dashed diagonal lines represent the equality between the different \textit{coherence lengths}, while the turquoise vertical and horizontal shaded areas mark the $\ell_\mathrm{CR}$ intervals used to identify relaxed clusters.}
 \label{coh_scales}
 \end{figure*}

\begin{figure}[ht]
 \centering
 \includegraphics[width=\columnwidth]{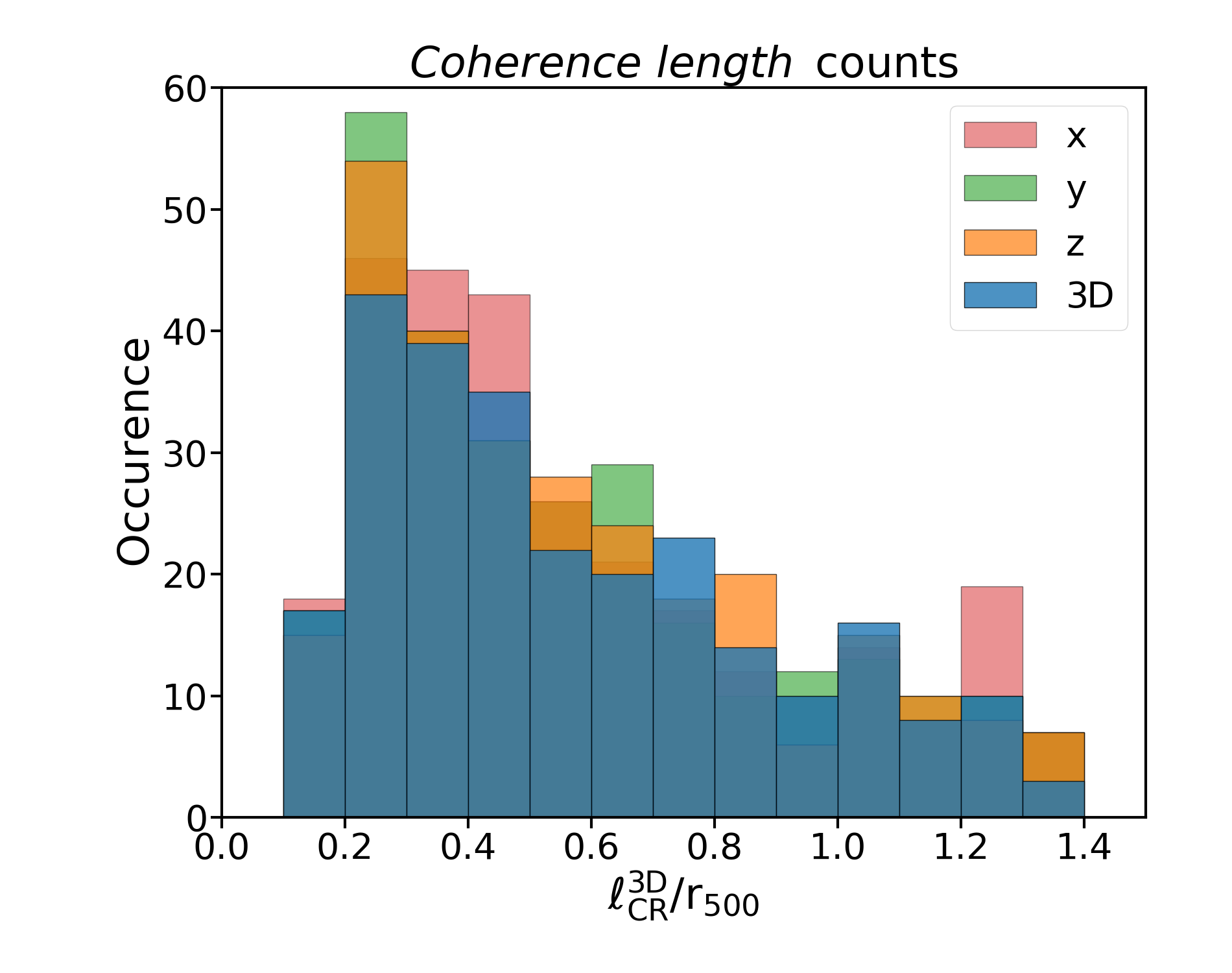} 
 \caption{\textsl{Coherence lengths} obtained from the 2D and 3D analyses.}
 \label{allc_hyst}
\end{figure}

\vspace{10pt}

\section{Gas-Mass coherence} \label{coherence}

From both the mass maps and the X-ray surface brightness maps, fluctuation fields relative to the background mean can be obtained as below:
\begin{eqnarray}
\delta F_m(x)=F_m(x)/\langle F_m \rangle -1,
\end{eqnarray}
\begin{eqnarray}
\delta F_X(x)=F_X(x)/\langle F_X \rangle -1,
\end{eqnarray}
where $\langle F_m \rangle$ and $\langle F_X \rangle$ are the average values of the two datasets. As noted in \citealt{https://Cerini22}, in previous studies based on X-ray surface brightness fluctuations of galaxy clusters (see, for instance, \citealt{Churazov_2012} and \citealt{zhuravleva17}), the global cluster emission is removed by fitting a $\beta$-model to the images and dividing the images by the best-fitting models. We do not follow this approach, since we are analyzing objects in all dynamical states, including clusters that are actively merging and are therefore far from hydrostatic equilibrium.

Through the discrete FFT, provided by the python \textsl{numpy.fft} subpackage, the following Fourier transforms were then computed:

\begin{equation}
  \Delta_m (\textbf{q})=\int \delta F_m(x)\exp(-i\textbf{x}\cdot\textbf{q})d^2x,
\end{equation}
\\
\begin{equation}
  \Delta_X (\textbf{q})=\int \delta F_X(x)\exp(-i\textbf{x}\cdot\textbf{q})d^2x, 
\end{equation}

with $\textbf{x}$ coordinate vector in the real 2D or 3D space, $\textbf{q}=2\pi \textbf{k}$ 2D or 3D wave-vector, $|\textbf{k}|=1/\theta$, and $\theta$ angular scale. For both the 2D and 3D maps, the 1D auto-power spectra can then be obtained:
 
 \begin{equation}
  P_m(q)=\langle |\Delta_m (\textbf{q})|^2 \rangle,
 \end{equation}
 \\
 \begin{equation}
  P_X(q)=\langle |\Delta_X (\textbf{q})|^2 \rangle ,
 \end{equation}
 
where the average was taken over all the independent Fourier elements that lie inside the radial interval $[q, q+dq]$. From the Fourier transforms corresponding to the two different maps, we can compute the cross-power spectrum:
 
\begin{IEEEeqnarray}{rCl}
P_{mX}(q)&=&\langle\Delta_m(q)\Delta_X^*(q)\rangle\\
  \> &=&\Re_m(q)\Re_X(q)+\Im_m(q)\Im_X(q), \nonumber
\end{IEEEeqnarray} 
 
with $\Re(q)$, $\Im(q)$ denoting the real and imaginary parts (\citealt{Kashlinsky_2012}, \citealt{Cappelluti_2013}, \citealt{Helgason_2014}, \citealt{Li_2018}, \citealt{Kashlinsky_2018}, \citealt{https://Cerini22}). Using the auto-power and cross-power spectra we can compute the coherence $C(q)$ (e.g., \citealt{2012ApJ...753...63K}, \citealt{Cappelluti_2013}, \citealt{https://Cerini22}:

\begin{equation}
   C(q)=P^2_{mX}/P_m(q)P_X(q).
   \label{c_formula}
\end{equation} 

The coherence $C$ can be quickly computed and it efficiently measures in Fourier space how well the two distributions trace each other. In this particular case, it quantifies how well the gas properties reflect and trace the overall gravitational potential. $C=1$ at a specific scale corresponds to signals that are perfectly correlated in structures at that scale, while $C=0$ indicates two totally uncorrelated signals. Therefore, for our purpose, $C=1$ at almost all scales indicates clusters in hydrostatic equilibrium. As highlighted in \citealt{https://Cerini22}, this quantity is a powerful tool to spatially resolve features that are not obviously detectable from images in real space, and it offers insights into how biased the gas is as a tracer of the potential and is a clear-cut determinant of the validity of the hydrostatic equilibrium assumption. Specifically, our method relies on the \textsl{coherence length} $\ell_{\rm CR}$, the minimum scale at which the coherence $C$ becomes larger than 0.9, as an indicator of the dynamical state of the cluster. As described in \citealt{https://Cerini22}, the lower $\ell_{\rm CR}$ is, the higher the level of equilibrium is. On the other hand, the higher $\ell_{\rm CR}$ is, the higher the level of disturbance is and the more the cluster is out-of-equilibrium. Examples of \textsl{coherence lengths} for clusters in different dynamical states are shown in Fig.~\ref{example_c} and are discussed with other results from our analysis in the next section. The mass and X-ray emissivity maps of the same clusters chosen as examples are shown in Figs.~\ref{rel_cluster},\ref{par_rel_cluster} and~\ref{unrel_cluster}.

The errors on the power spectra were obtained through the Poissonian estimators, defined as:

 \begin{equation}
  \sigma_{P_m}=P_m(q)/\sqrt{0.5N_q} 
 \end{equation}
 
 \begin{equation}
  \sigma_{P_X}=P_X(q)/\sqrt{0.5N_q} 
 \end{equation}
 
  \begin{equation}
   \sigma_{P_{mX}}=\sqrt{P_m(q)P_X(q)/N_q}
 \end{equation}

with $N_q/2$ number of independent measurements of $\Delta_i (\textbf{q})$ in a ring with $N_q$ data. Indeed, since the flux is a real quantity, only one half of the Fourier plane is independent (\citealt{Kashlinsky_2012}, \citealt{Cappelluti_2013}, \citealt{Helgason_2014}, \citealt{Li_2018}, \citealt{Kashlinsky_2018}, \citealt{https://Cerini22}). For errors on the coherence, or the square of the correlation coefficient $\textit{R}$ ($C= \textit{R}^2$), the situation is more complicated, due to the highly non-linear structure of $C$ with respect to the underlying quantities in equation ~\ref{c_formula}. This nonlinearity is especially pronounced near the boundaries, where $C\sim 0$ and $C\sim 1$. Because coherence $C$ is constrained to values between 0 and 1, these limits introduce challenges that make conventional error propagation methods less reliable. As noted by \citealt{Kashlinsky_2018}, the Fisher transformation (\citealt{Fisher1915}) analyzes the coherence (or squared correlation coefficients $\textit{R}$) by mapping $C$ ($R$) to an infinite scale, rather than being restricted to $[0,1]$ ($[-1,1]$ for $\textit{R}$). This transformation thus simplifies statistical analysis by making linear approximations valid and enabling reliable error propagation across the entire range of coherence values. Since errors are always equivalent to confidence contours, one needs to compute the $68\%$ confidence limits of \textit{R} from the errors on the auto- and cross-power spectra. The Fisher transformation technique is the standard way to obtain the probability distribution of \textit{R}, and therefore $C$, and relate these uncertainties to those of the powers. Specifically, once the central values, $\textit{R}_0=\sqrt{C}_0$, are computed from the power data mentioned above, the Fisher transformation allows the computation of the quantity

 \begin{equation}
   Z=\frac{1}{2}\ln{\frac{1+\textit{R}}{1-\textit{R}}},
 \end{equation}

which is normally distributed in most practical cases (\citealt{Fisher1915}). This transformation, and its inverse $C=(\tanh{Z})^2$, is then used to construct the corresponding confidence interval for $C$, evaluating the $68\%$ contours of $Z$ from the variances of the auto- and cross-powers. The variance in $Z$ is related to the errors on powers as 

\begin{equation}
  \sigma_Z^2=\frac{C_0}{(1-C_0)^2}(\frac{\sigma_{P_{mX}}^2}{P_{mX}^2}+\frac{1}{4}\frac{\sigma_{P_{m}}^2}{P_{m}^2}+\frac{1}{4}\frac{\sigma_{P_{X}}^2}{P_{X}^2}).
\end{equation}

The $68\%$ contours for $C$ are derived from $Z\pm \sigma_Z$. The confidence contours for $C$ are constrained to the interval of $[0,1]$ ($[-1,1]$ for \textit{R}).
 
The uncertainty on the \textsl{coherence length} can be estimated as follows:

\begin{equation}
  \sigma_{\ell_{\rm CR}}^2=(\left.\frac{dl}{dC}\right|_{C = 0.9})^2\sigma_C^2=\frac{\sigma_C^2}{(\left.\frac{dC}{dl}\right|_{l = \ell_{\rm CR}})^2},
\end{equation}

where $l=2\pi/q$ and $\sigma_C$ is the error on the coherence. The derivative of the coherence was estimated using the incremental ratio.

Finally, to mitigate edge effects in our coherence analysis, we applied the Hann window function to the data. Edge effects, arising from discontinuities at the boundaries of the finite data field, can introduce spectral leakage — the spurious spread of power across frequencies — into the Fourier transform. The Hann window, with its cosine-squared taper, effectively reduces these discontinuities by smoothly attenuating the data values to zero at the edges. This approach minimizes leakage without excessive loss of signal power, making it a balanced choice among tapering functions for spectral analysis (\citealt{harris78}). Accordingly, each 2D or 3D dataset was multiplied by the following window functions prior to Fourier transformation:

\begin{align}\label{eq:rCl}
w(n_x,n_y)= &~0.5(1-\cos{\frac{2\pi n_x}{N_x-1}})\\ 
& \times(0.5(1-\cos{\frac{2\pi n_y}{N_y-1}})], \nonumber
\end{align} 

\begin{align}
w(n_x,n_y,n_z)= &~~
0.5(1-\cos{\frac{2\pi n_x}{N_x-1}}) \\& \nonumber
\times 0.5(1-\cos{\frac{2\pi n_y}{N_y-1}})\\&\times 0.5(1-\cos{\frac{2\pi n_z}{N_z-1}}), \nonumber
\end{align} 

%\begin{equation}
%  w(n_x,n_y)=[0.5(1-\cos{\frac{2\pi n_x}{N_x-1}})]\cdot[(0.5(1-%\cos{\frac{2\pi n_y}{N_y-1}})],
%\end{equation}

%\begin{equation}
%  w(n_x,n_y,n_z)=[0.5(1-\cos{\frac{2\pi n_x}{N_x-1}})]\cdot[(0.5(1-%\cos{\frac{2\pi n_y}{N_y-1}})\cdot[(0.5(1-\cos{\frac{2\pi n_z}{N_z-1}})],
%\end{equation}

with $n_x, n_y, n_z$ pixel indices along the $x, y, z$ dimensions and $N_x, N_y, N_z$ the number of pixels along the $x, y, z$ dimensions.

 \begin{figure*}
\centering
      \subfigure[]{\includegraphics[width=0.48\textwidth]{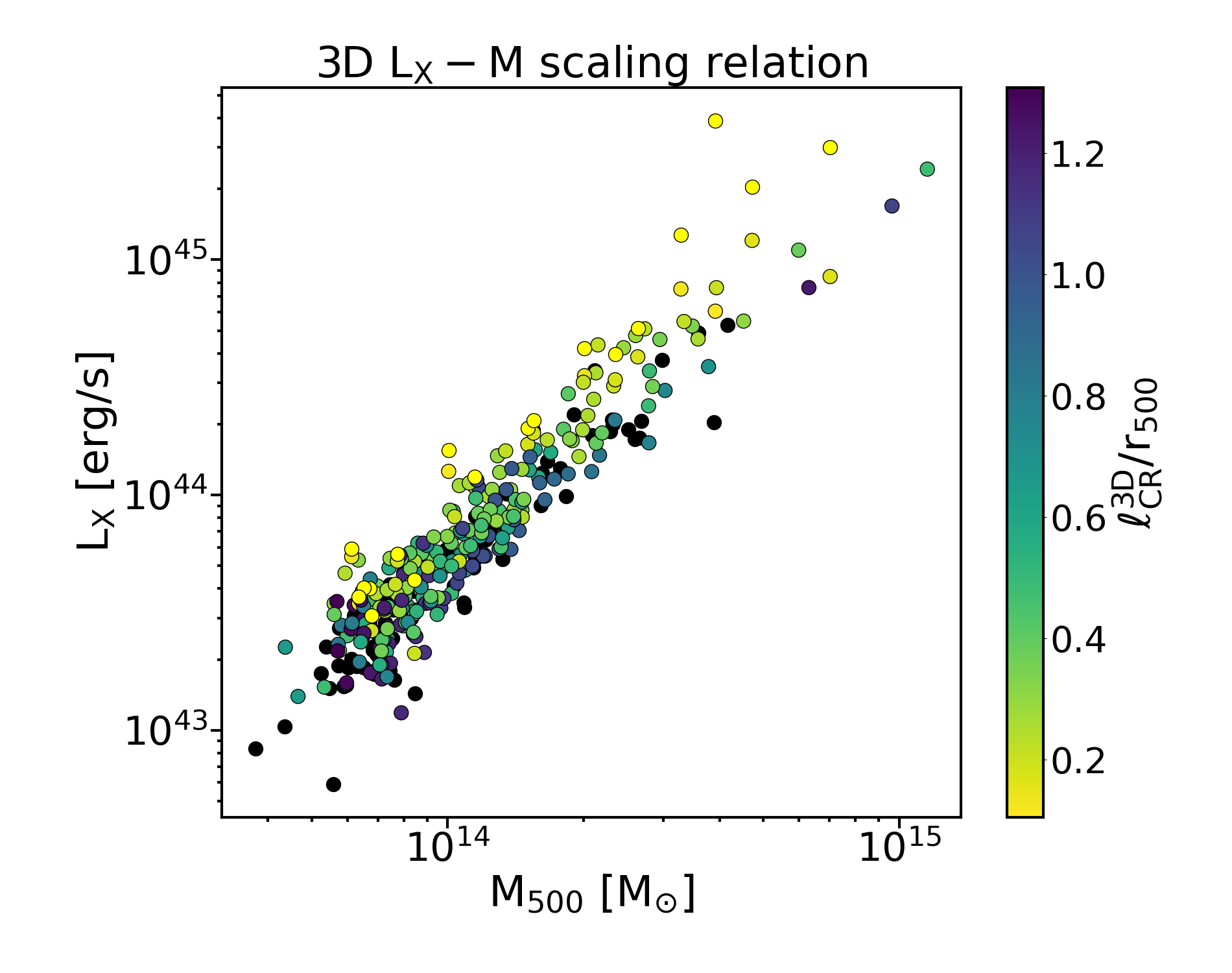}}
      \subfigure[]{\includegraphics[width=0.48\textwidth]{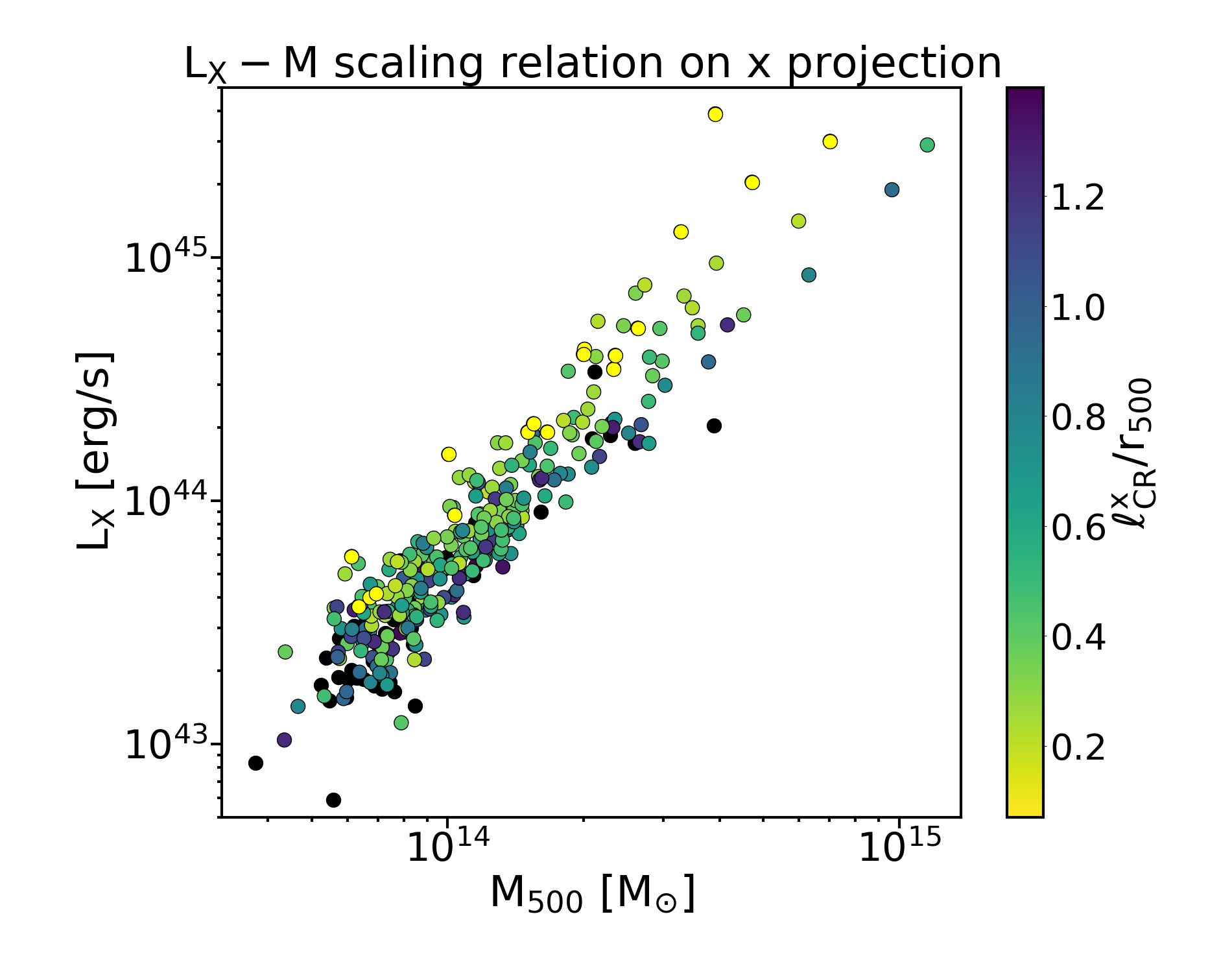}}
      \subfigure[]{\includegraphics[width=0.48\textwidth]{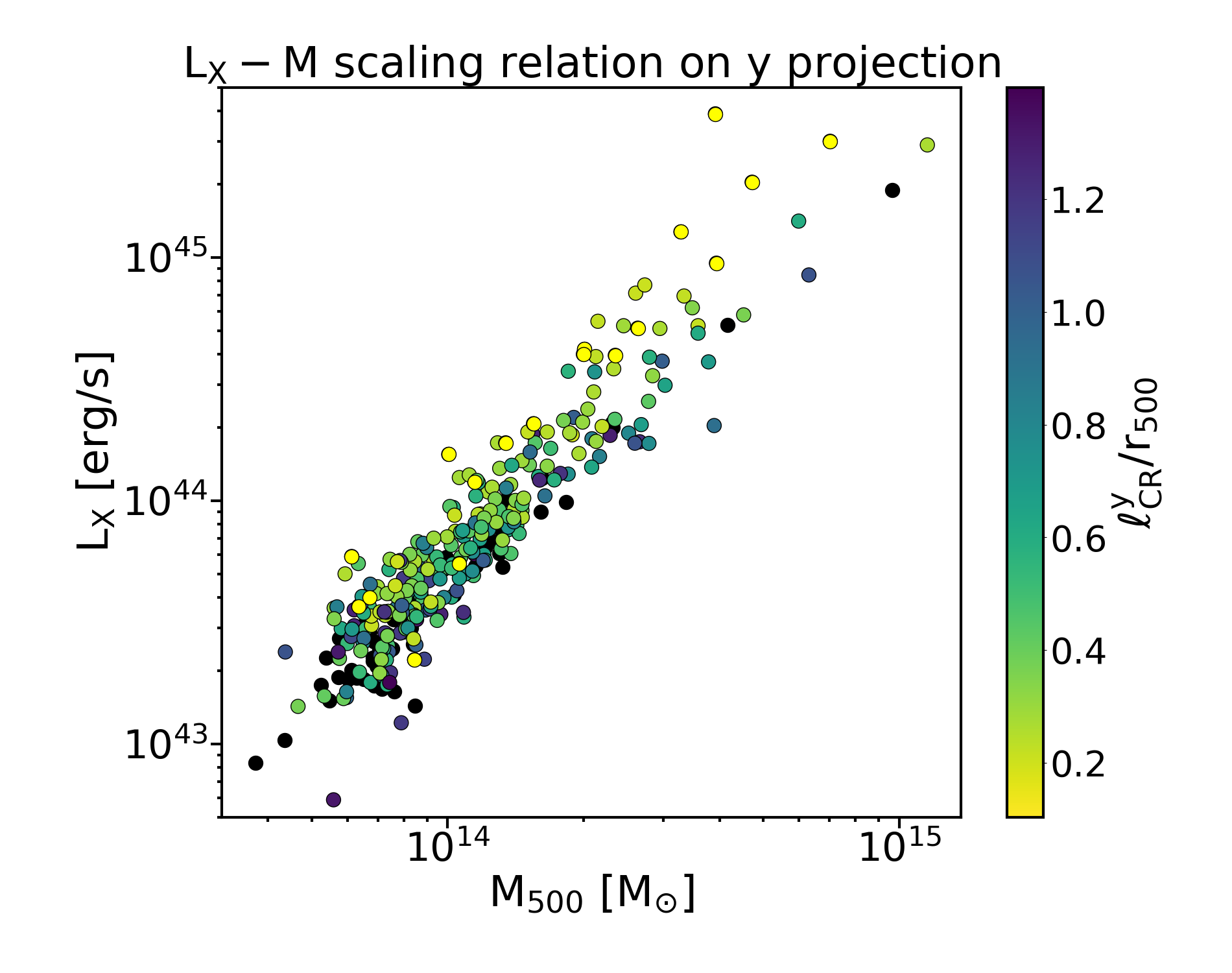}}
      \subfigure[]{\includegraphics[width=0.48\textwidth]{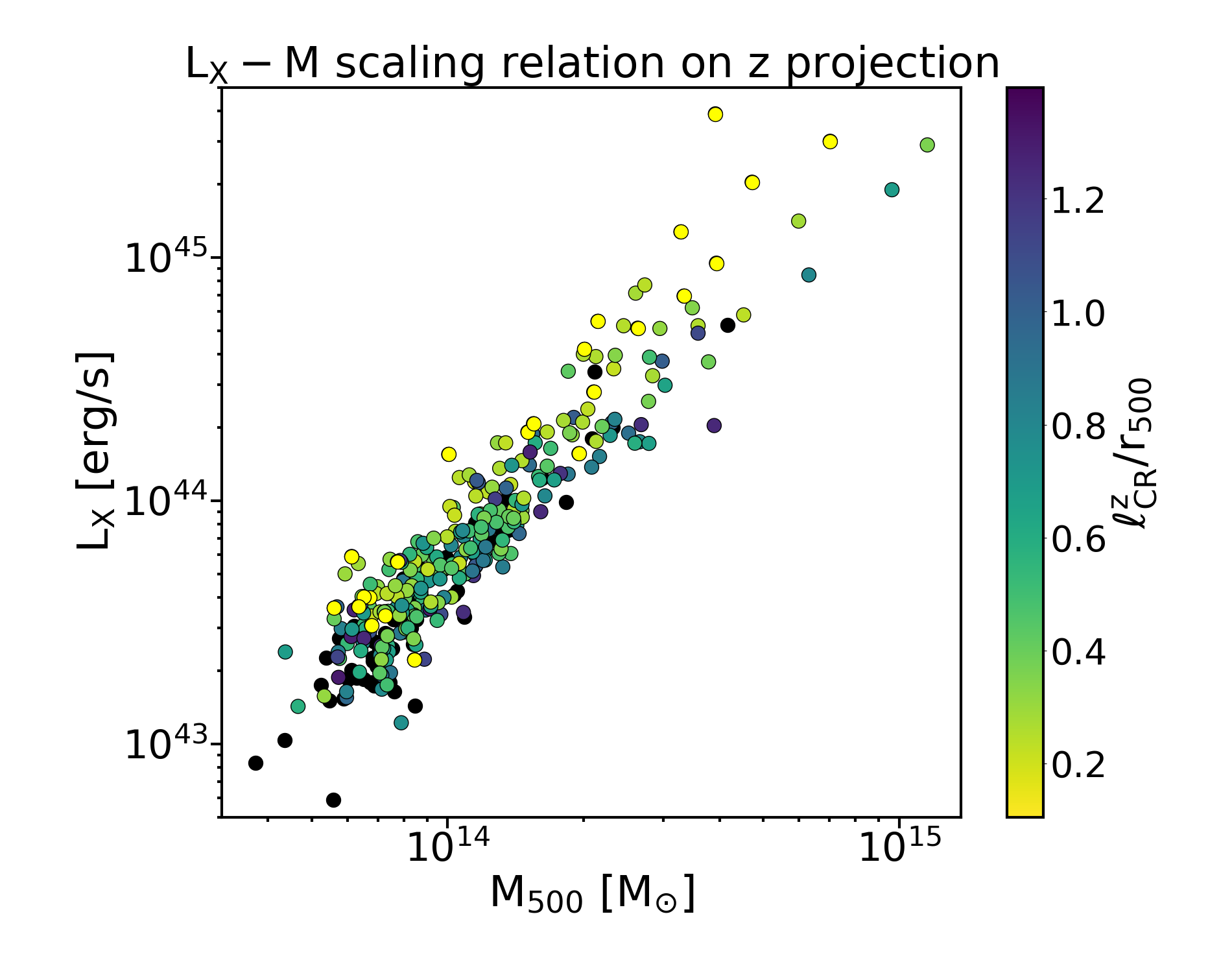}}     
    
      \caption{$L_X-M$ scaling relations for the 329 TNG-300 clusters obtained without excluding the core. The color code represents the cluster's dynamical state, depicted by the \textsl{coherence length} normalized to $r_{500}$. Brighter yellow points indicate relaxed clusters, darker blue points denote unrelaxed clusters, and intermediate colors correspond to transitional stages.}
 \label{core_lum}
 \end{figure*}

 \section{Scaling relations modeling and fitting} \label{scale_rel}
Simple forms of scaling relations for galaxy clusters were derived by \citealt{Kaiser1986} under the assumption that clusters form via a single gravitational collapse and the only source of energy input into the ICM is gravitational. These assumptions lead to a simple scenario, known as the self-similar model, in which the collapsed region that formed the cluster is virialized - i.e. the kinetic energy of its components have obtained an equilibrium with the gravitational potential, obeying the virial theorem - all clusters of the same mass and at the same redshift are identical, and their properties scale simply with mass and redshift. These self-similar relations take the general form 

\begin{equation} \label{self_rel}
  Y = AE(z)^C X^B,
\end{equation}

where $A$, $B$ and $C$ describe the normalization, slope, and evolution, respectively, of the scaling relation between some properties $X$ and $Y$, and $E(z)=\sqrt{\Omega_M(1+z)^3+\Omega_k(1+z)^2+\Omega_{\Lambda}}$, with $\Omega_M$, $\Omega_{\Lambda}$ and $\Omega_k=(1-\Omega_M-\Omega_{\Lambda})$ being the density parameters associated to the non-relativistic matter, the cosmological constant, and the curvature of the Universe, respectively. In particular, the X-ray luminosity $L_X$ and X-ray temperature $T_X$ have the following dependence on the mass:

\begin{equation}
  L_X \propto E(z)^{7/3}M^{4/3}.
\end{equation}

\begin{equation}
  T_X \propto E(z)^{2/3}M^{2/3},
\end{equation}

Nevertheless, in real galaxy clusters non-gravitational physical processes, such as radiative cooling, galactic winds, and AGN feedback, have a significant impact on the distribution of baryons in the ICM and energy budget of the system, leading to departures from self-similar relations (e.g., \citealt{Bhat2008}, \citealt{McCarthy2010}, \citealt{Fabjan2011}, \citealt{Giodini2013}, \citealt{Bulbul_2019}, \citealt{Lovisari2020}).

To fit the $L_X-M$ and $T_X-M$ scaling relations derived from the TNG300 simulations, we employed a Bayesian framework consistent with previous studies (e.g., \citealt{Mantz10} \citealt{Giles16}, \citealt{Bahar22}). Specifically, we implemented the Markov Chain Monte Carlo (MCMC) sampler \textsc{emcee} (\citealt{Foreman-Mackey_2013}) to obtain the maximum-likelihood values and posterior PDFs of the free parameters. For our sample of $z=0$ simulated clusters, we adopted the following general expression for both $L_X-M$ and $T_X-M$ scaling relations, readapting equation~\ref{self_rel}:

\begin{equation}
  Y = AX^B,
\end{equation}

where the cosmological dependence is encapsulated within the normalization factor $A$. Equivalently, this relation can be expressed as

\begin{equation}
  y=q+mx
\end{equation}

where $y=log(Y)$, $x=log(M)$, $q=log(A)$ and $m=B$.

Consistent with the widely used approach in the literature, the $Y-X$ relation assumes that the observable $Y$ is log-normally distributed around the power-law scaling relation (e.g., \citealt{Pacaud2007}, \citealt{Evrard14}, \citealt{Giles16}, \citealt{Bulbul_2019}, \citealt{Bocquet19}, \citealt{Bahar22}). The log-normal assumption implies that, for a dataset comprising $N$ clusters, where $y=\{y_1,...,y_N\}$ and $x=\{x_1,...,x_N\}$, the probability of observing 
$y$ given $x$ and the parameters $q$ and $m$ can be expressed by the likelihood function as follows:

\begin{IEEEeqnarray}{rCl}
P(Y|X,A,B) = \mathcal{L}\mathcal{N}(\mu=q+mx, \sigma=\sigma_{y|x}) \\
  \> = \prod_{i}^{N} \mathcal{N}(\mu = q+mx, \sigma = \sigma_{y|x}) \nonumber
  \\ 
  \> = \prod_{i}^{N} \frac{1}{\sqrt{2\pi\sigma^2_{y|x}}}\exp[\frac{-(y_i-q-mx_i)^2}{2\sigma^2_{y|x}}]\nonumber
\end{IEEEeqnarray} 

where $\mathcal{N}(\mu=q+mx, \sigma=\sigma_{y|x})$ represents a normal distribution with mean $\mu$ and standard deviation $\sigma_{y|x}$, $q$ is the intercept, $m$ is the slope. The quantity $\sigma_{y|x}$ represents the intrinsic scatter in $y$ at a fixed $x$. The parameters $q$, $m$ and $\sigma_{y|x}$ are treated as free parameters and are estimated from the posterior distributions obtained through the MCMC sampling.

\section{Results} \label{results}

For all 329 simulated clusters, we created 3D cubes of mass and X-ray surface brightness distributions as explained in Section~\ref{data}. 
%To produce the X-ray surface brightness, the X-ray emissivity was computed in the 0.5-7~keV band (in the observer frame) for each gas cell assuming an Astrophysical Plasma Emission Code (APEC) model \citep{smith_collisional_2001}.
Every 3D cube was then projected along the three perpendicular axes x, y, and z. We computed the coherence and \textsl{coherence length} for both the 3D distributions and the 2D projections. For example, Fig.~\ref{example_c} shows the results obtained for three clusters in different dynamical states: a relaxed cluster (panel (a)), a partially relaxed cluster (panel (b)), and an unrelaxed cluster (panel (c)). The corresponding mass and X-ray emissivity maps in the three different projections are shown in Figs.~\ref{rel_cluster}, \ref{par_rel_cluster} and \ref{unrel_cluster}, respectively. The coherences for the relaxed cluster along the three projected axes are consistent with each other, with $\ell^{3D}_{\rm CR}/r_{500}$, $\ell^x_{\rm CR}/r_{500}$, $\ell^y_{\rm CR}/r_{500}$ and $\ell^z_{\rm CR}/r_{500}$ ranging from 0.09 to 0.11. However, the farther a cluster is from equilibrium, the larger the differences between the 3D and 2D results. This demonstrates that the more a galaxy cluster deviates from equilibrium, the greater the impact of projection effects, leading to substantial variations depending on the observation angle. For all but the most relaxed clusters, these projection effects significantly alter the coherence results, underscoring the need to account for them when interpreting observational data. This behavior aligns with previous findings in the literature. For instance, \citet{2019Chen} investigated the relationship between the morphology of the X-ray emitting ICM and the mass accretion rates of OMEGA500 simulated galaxy clusters \citep{Nelson_2014}. They found a strong correlation between mass accretion rates and ICM ellipticity: clusters with lower mass accretion rates and higher relaxation levels exhibited larger 3D axis ratios, indicative of greater sphericity.

\begin{table}[h!]
  \centering
  %\caption{Priors for the parameters for $L_X-M$ and $T_X-M$ scaling relations.}
  \label{tab:scaling_relations}
  \resizebox{\columnwidth}{!}{ % Adjusts table to fit within column width
  \hspace{-3em}
  \begin{tabular}{lcc}
    \toprule
    Parameters & $L_X-M$& $T_X-M$ \\
    \midrule
    Intercept ($q$)     & $\mathcal{U}(10,20)$  & $\mathcal{U}(-10,10)$ \\
    Slope ($m$)   & $\mathcal{U}(1.4,1.9)$  & $\mathcal{U}(0.5,0.9)$  \\
    Scatter ($\sigma_{y|x}$) & $\mathcal{U}(0.05,2.0)$       & $\mathcal{U}(0.01,1.0)$      \\
    \bottomrule
  \end{tabular}
  }
    \caption{Priors for the parameters for $L_X-M$ and $T_X-M$ scaling relations.}
    \label{priors}
\end{table}

This morphological stability in relaxed clusters mitigates the influence of projection effects, making the observation angle less important for these systems. Figs.~\ref{coh_scales} and~\ref{allc_hyst} further illustrate the distribution of \textsl{coherence lengths} across all 2D projections and the 3D distributions. In particular, Fig.~\ref{coh_scales} highlights that projection effects lead to significant scatter across nearly the entire range of \textsl{coherence lengths}, with the exception of very relaxed systems, as indicated by the turquoise shaded areas. The number of clusters with $\ell_{\rm CR}<0.2r_{500}$ was found to be 18 across all analyses, with the exception of the z projection, which yielded a total of 23 clusters. Among these, only 11 clusters exhibited $\ell_{\rm CR}<0.2r_{500}$ consistently in both 3D and all projections. However, in most cases, clusters with $\ell_{\rm CR}<0.2r_{500}$ in one projection or in 3D maintained values below $0.3r_{500}$ in the other analyses, with only a few exceptions. Overall, only 6-7$\%$ of the full sample of 329 clusters are classified as relaxed using the coherence length criterion. Additionally, the number of clusters for which the coherence length could not be defined was 79 for the 3D analysis, and 44, 52, and 54 for the x, y, and z projections, respectively. This corresponds to approximately 13-24$\%$ of the total sample.

 \begin{figure*}
\centering
      \subfigure[]{\includegraphics[width=0.48\textwidth]{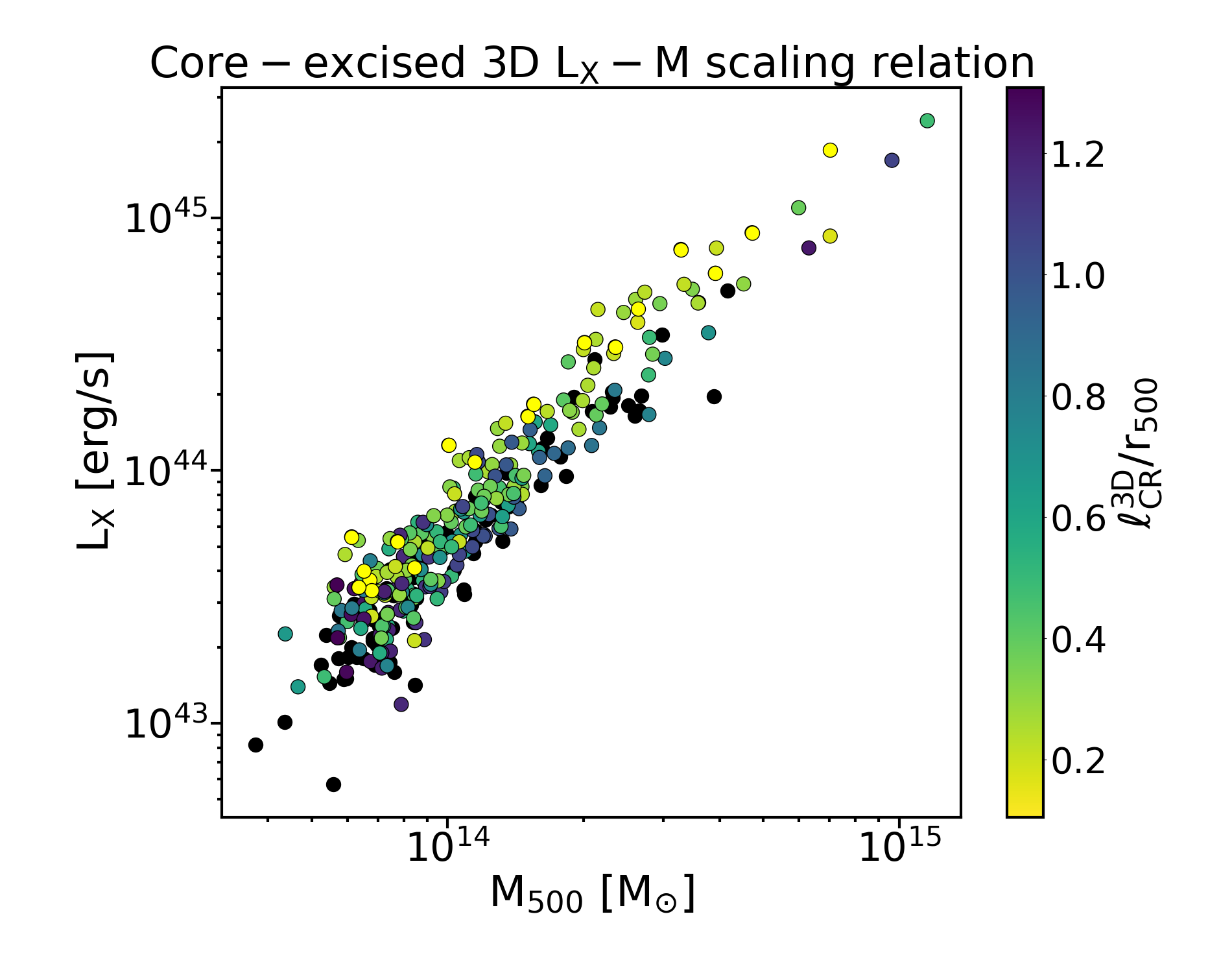}}
      \subfigure[]{\includegraphics[width=0.48\textwidth]{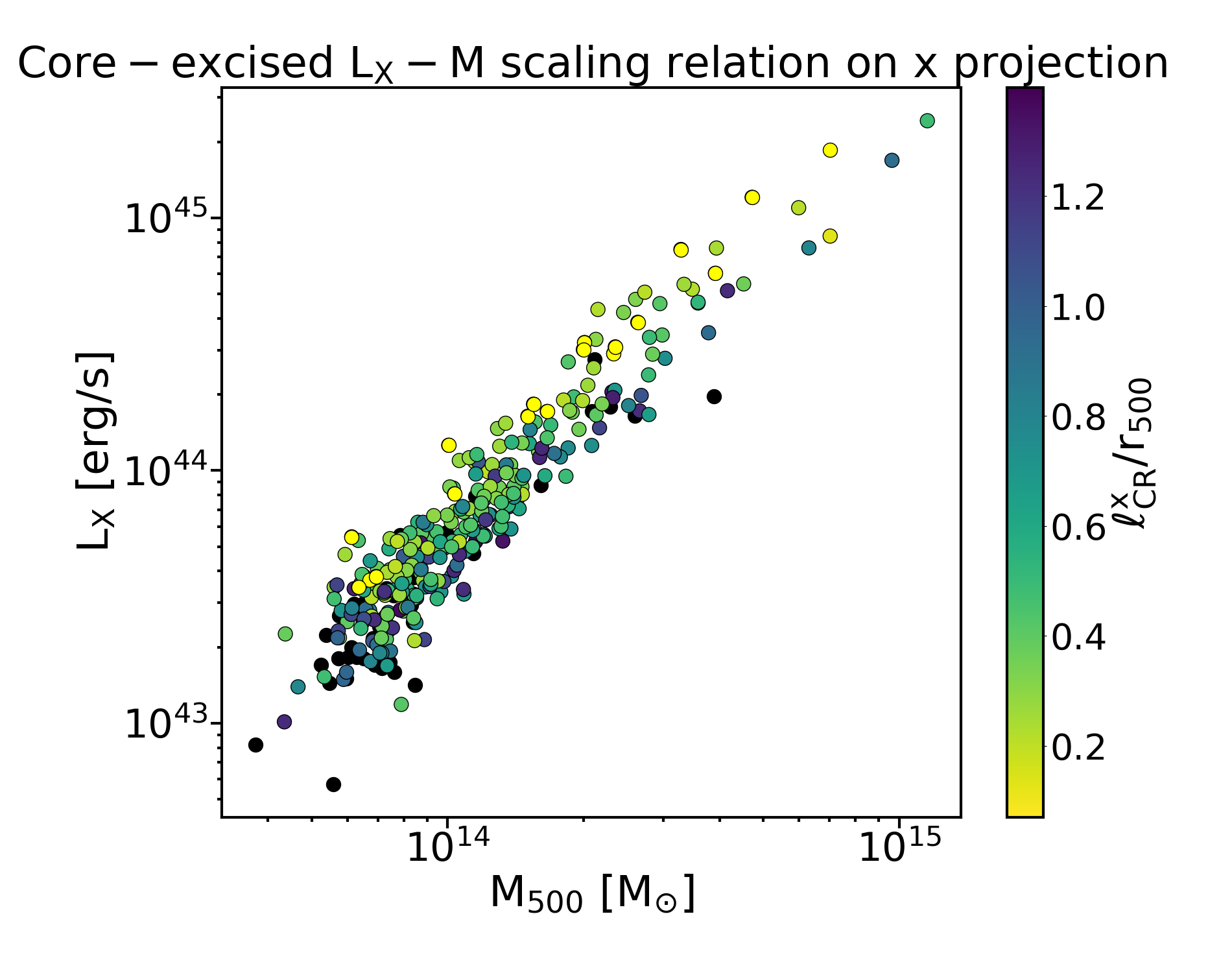}}
      \subfigure[]{\includegraphics[width=0.48\textwidth]{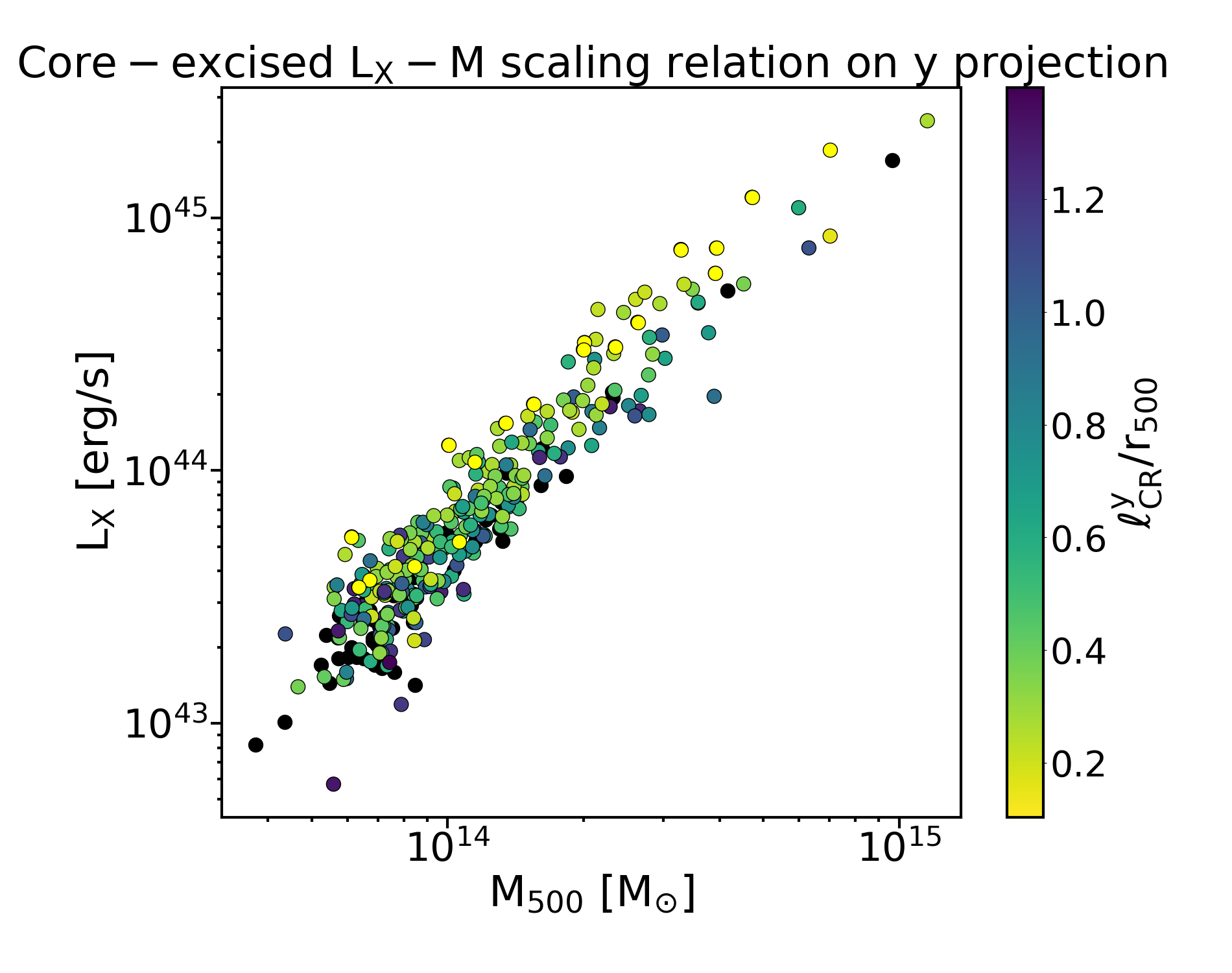}}
      \subfigure[]{\includegraphics[width=0.48\textwidth]{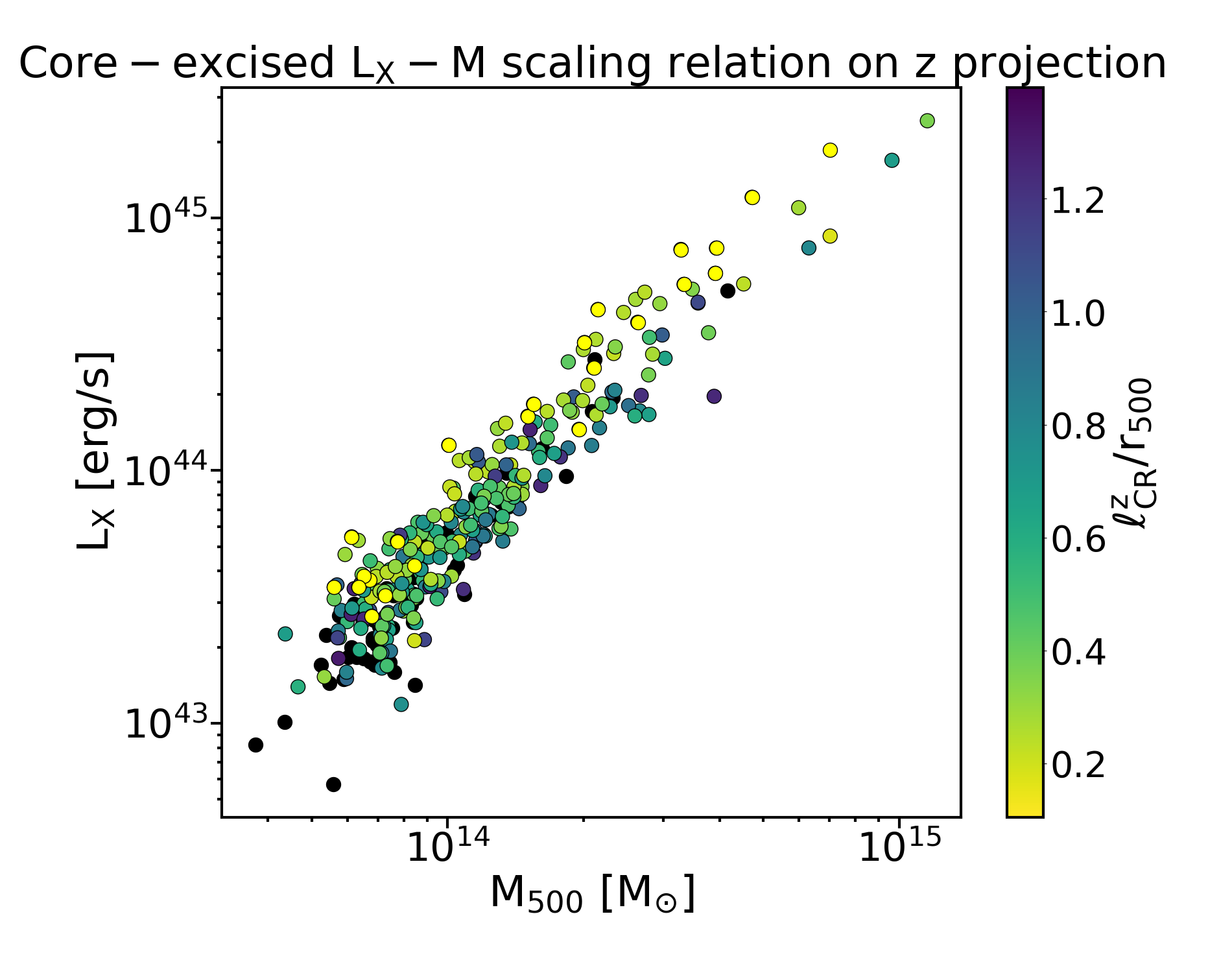}}     
    
      \caption{Core-excised $L_X-M$ scaling relations. The color code represents the cluster's dynamical state, depicted by the \textsl{coherence length} normalized to $r_{500}$. Brighter yellow points indicate relaxed clusters, darker blue points denote unrelaxed clusters, and intermediate colors correspond to transitional stages.}
 \label{nocore_lum}
 \end{figure*}

 \begin{figure*}
\centering
      \subfigure[]{\includegraphics[width=0.48\textwidth]{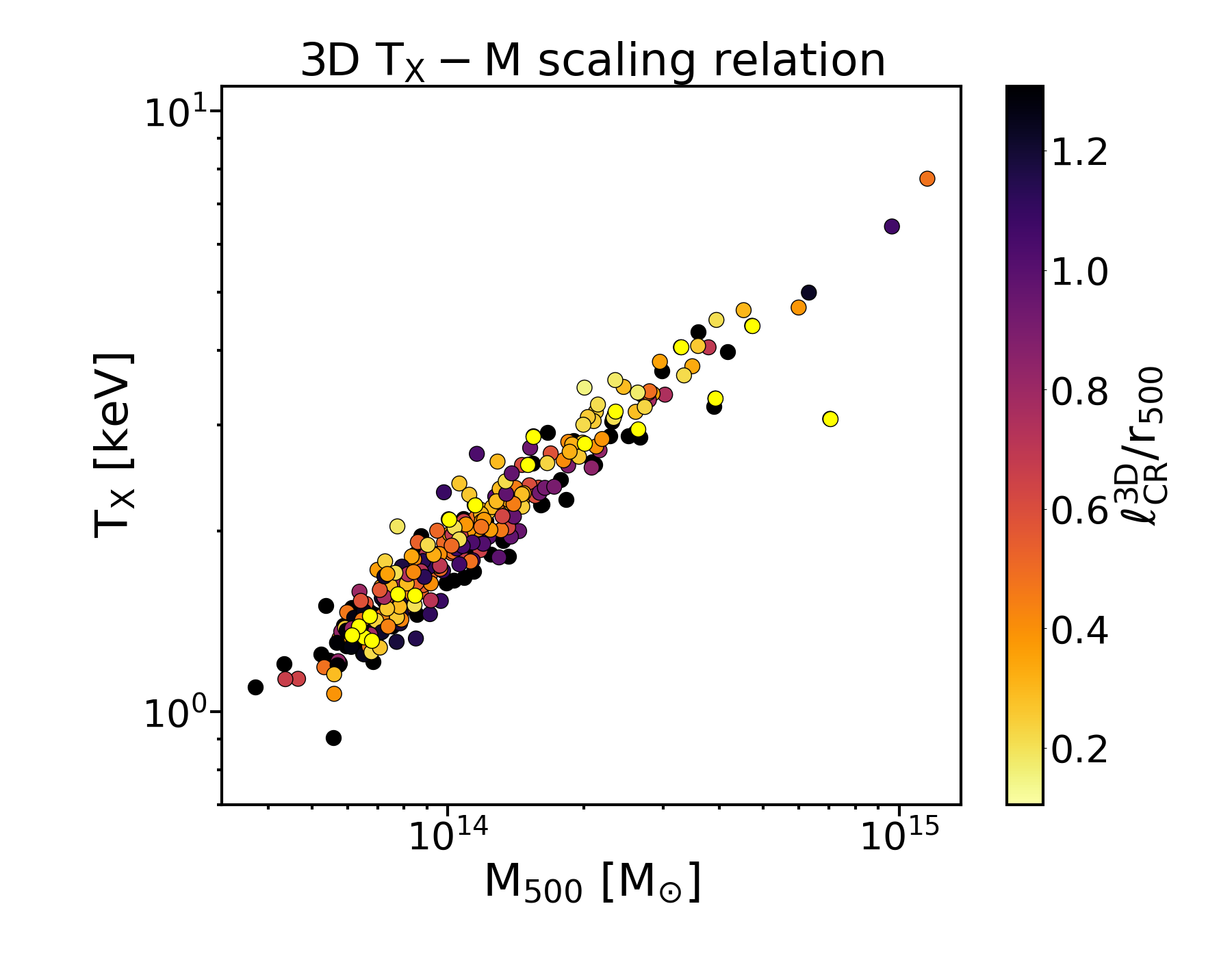}}
      \subfigure[]{\includegraphics[width=0.48\textwidth]{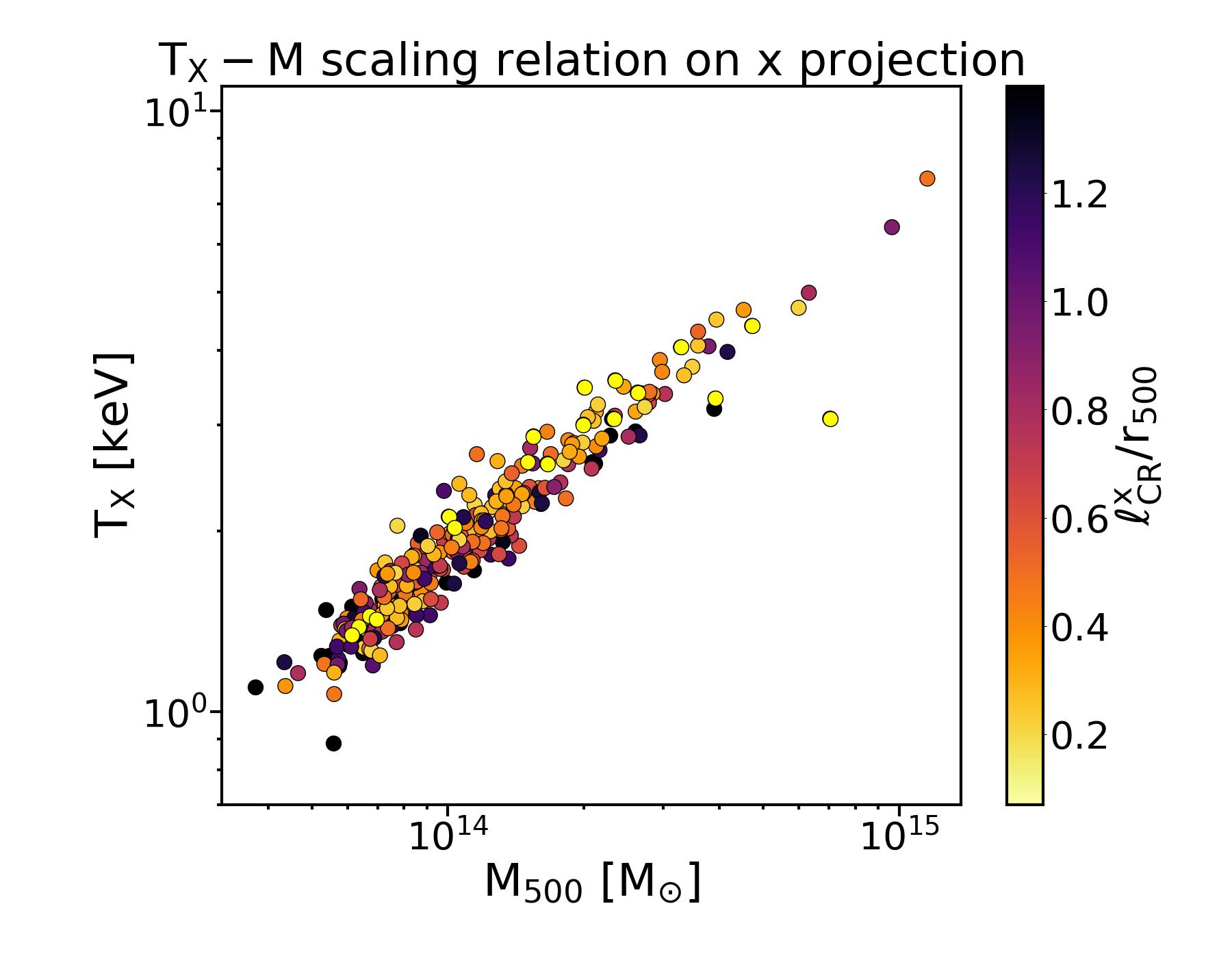}}
      \subfigure[]{\includegraphics[width=0.48\textwidth]{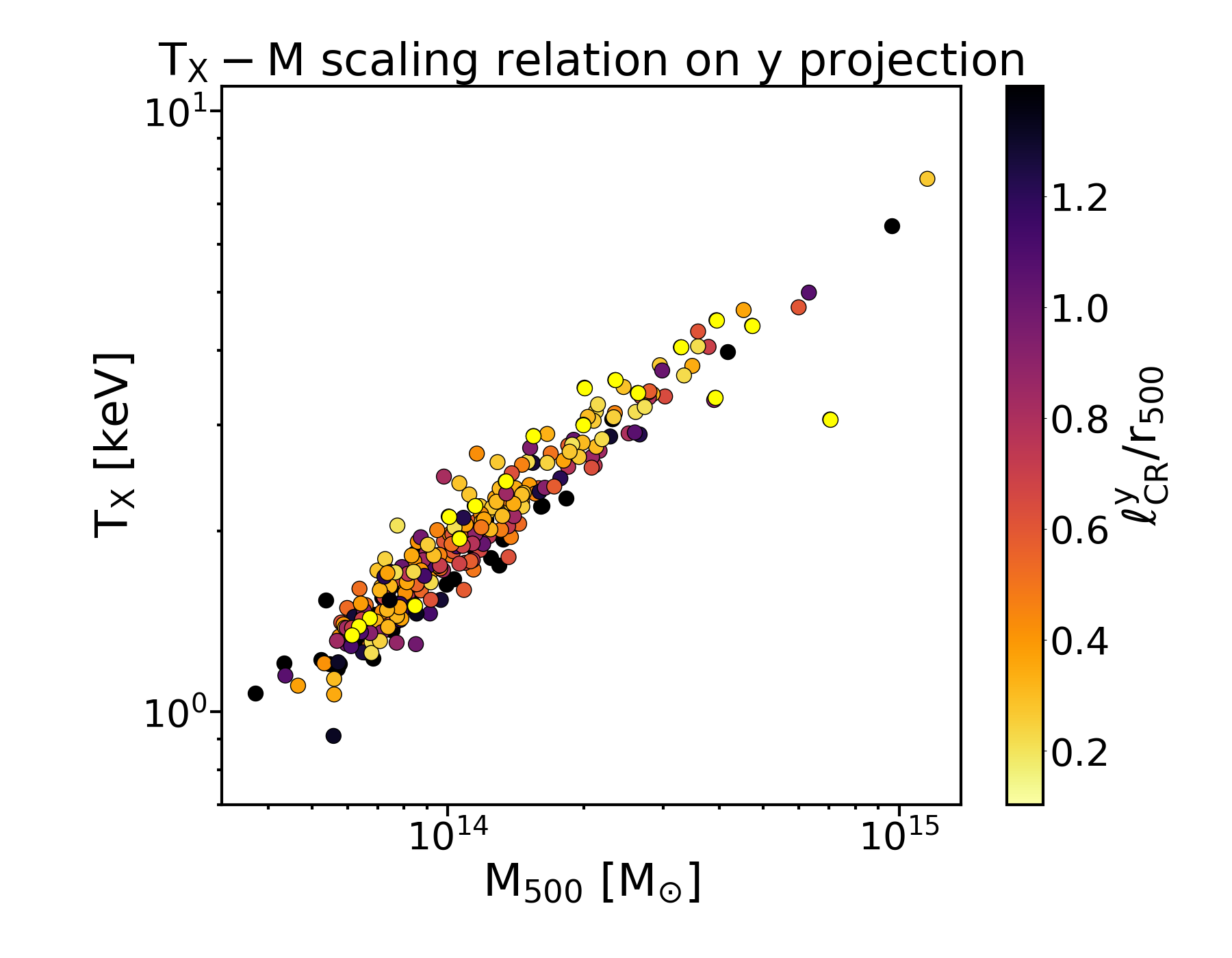}}
      \subfigure[]{\includegraphics[width=0.48\textwidth]{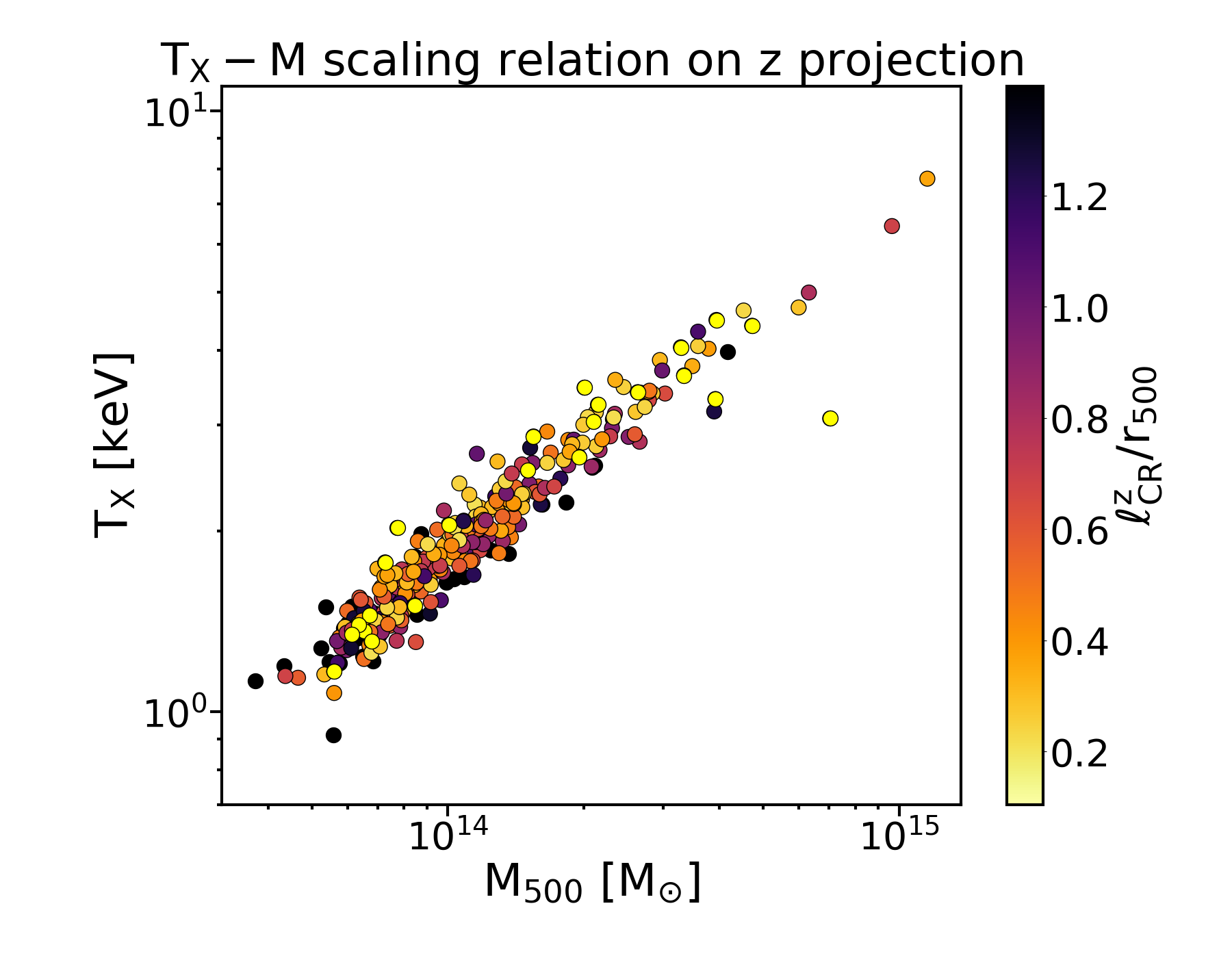}}     
    
      \caption{$T_X-M$ scaling relations obtained without excluding the core. The color code represents the cluster's dynamical state, depicted by the \textsl{coherence length} normalized to $r_{500}$. Brighter yellow points indicate relaxed clusters, darker red points denote unrelaxed clusters, and intermediate colors correspond to transitional stages.}
 \label{core_temp}
 \end{figure*}

  \begin{figure*}
\centering
      \subfigure[]{\includegraphics[width=0.48\textwidth]{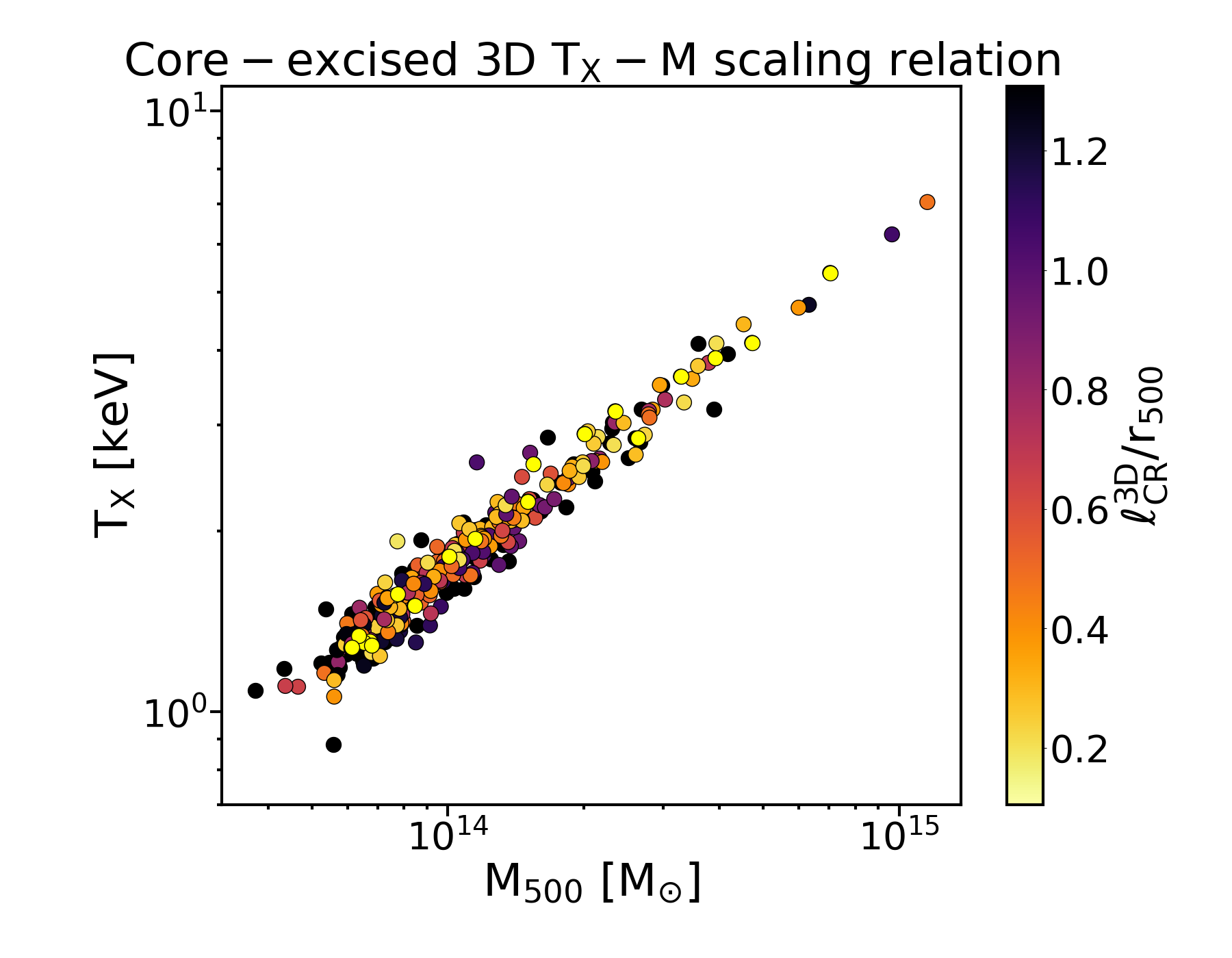}}
      \subfigure[]{\includegraphics[width=0.48\textwidth]{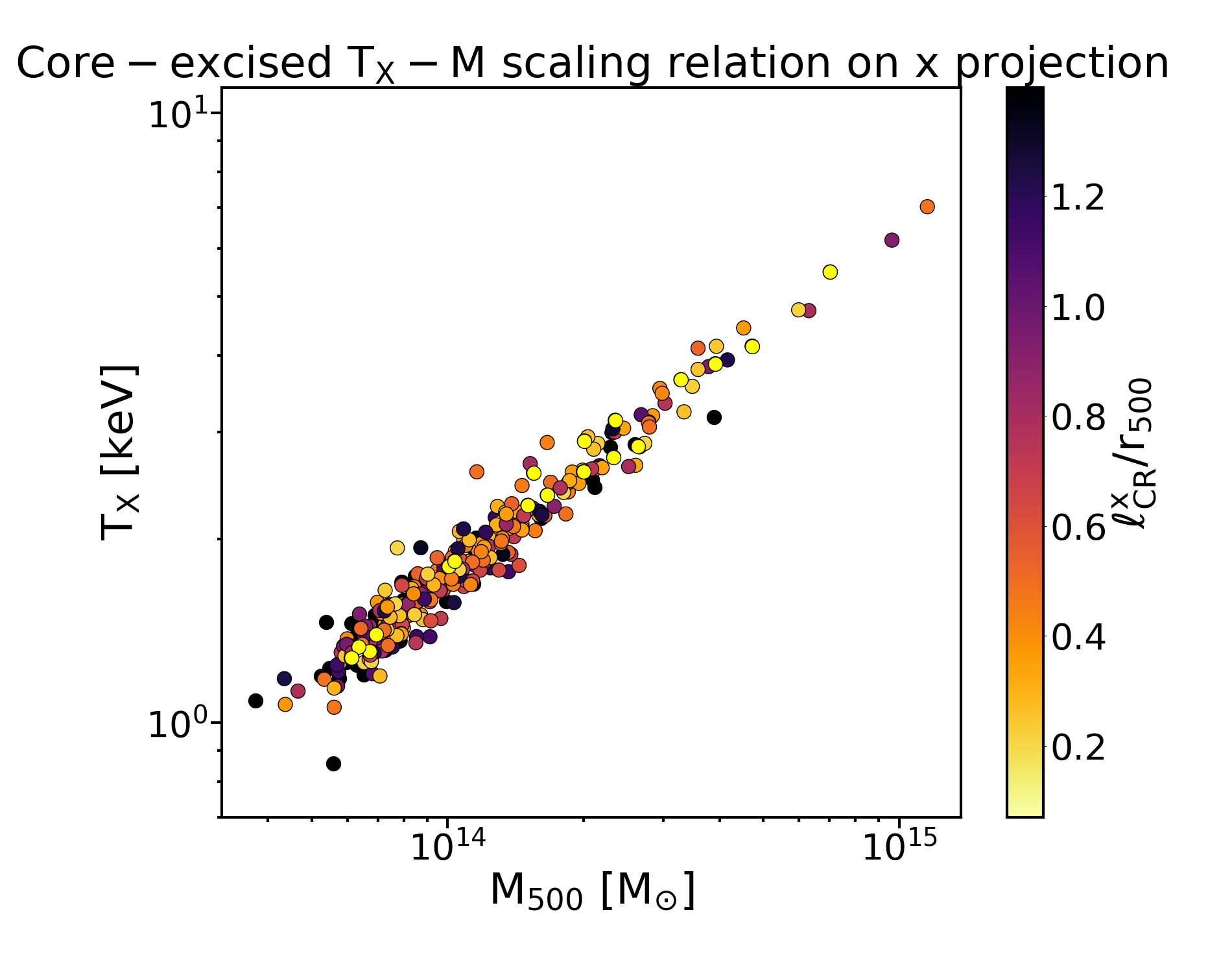}}
      \subfigure[]{\includegraphics[width=0.48\textwidth]{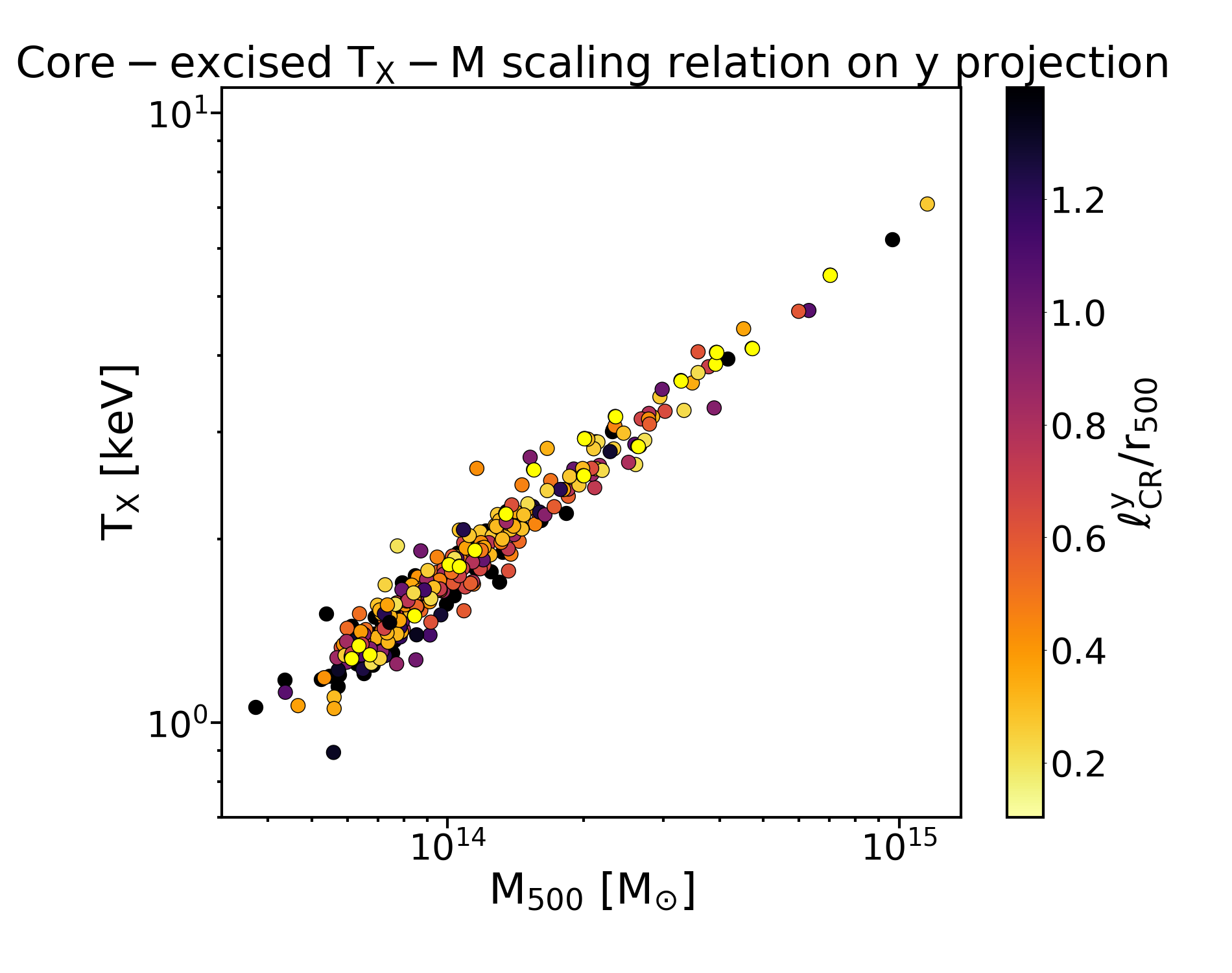}}
      \subfigure[]{\includegraphics[width=0.48\textwidth]{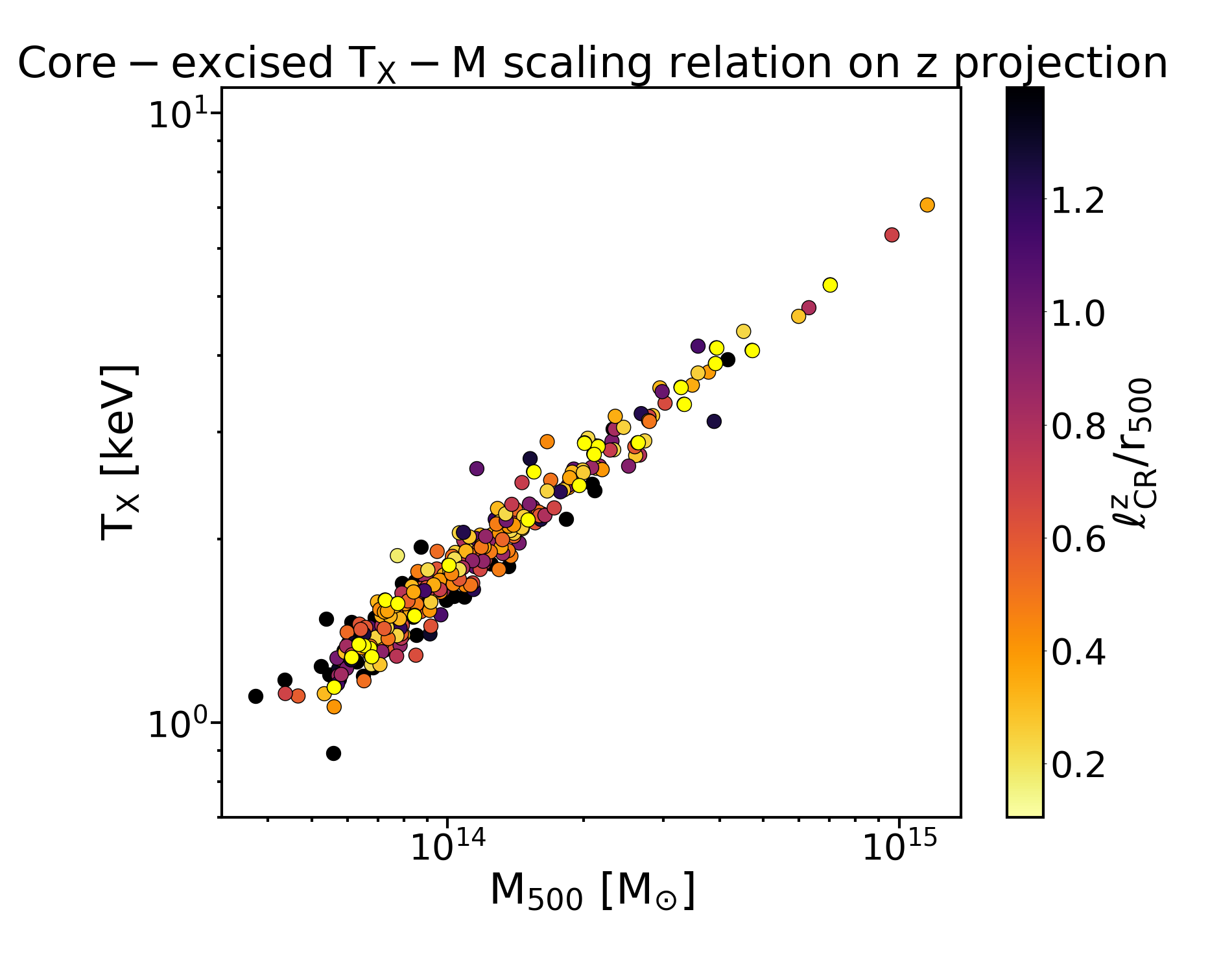}}     
    
      \caption{Core-excised $T_X-M$ scaling relations. The color code represents the cluster's dynamical state, depicted by the \textsl{coherence length} normalized to $r_{500}$. Brighter yellow points indicate relaxed clusters, darker red points denote unrelaxed clusters, and intermediate colors correspond to transitional stages}
 \label{nocore_temp}
 \end{figure*}

For all the 329 simulated clusters analyzed here, we obtained various color-coded $L_X-M$ and $T_X-M$ scaling relations, with the color code determined by the 3D and 2D \textsl{coherence lengths}, or equivalently, the dynamical state. Therefore, for each scaling relation, we produced one 3D color-coded plot and three projected ones (see Figs.~\ref{core_lum}, \ref{nocore_lum}, \ref{core_temp}, \ref{nocore_temp}). We also conducted the same analysis excluding the core region for the computation of the luminosity and temperature. Here, as usual, $L_X$, $T_X$, and $M$ represent the X-ray luminosity, the X-ray temperature, and the cluster mass, respectively. Cool-core galaxy clusters have dense, cool regions in their centers and radiate efficiently in X-rays. Studies over the past decades (e.g., \citealt{1994Fabian}, \citealt{Maughan2012}) have demonstrated that the presence of cool cores significantly impacts observations, as they substantially increase (decrease) the measured global $L_X$ ($T_X$) compared to clusters without cool cores. This represents an additional source of scatter in the galaxy cluster scaling relations. In Figs.~\ref{core_lum} and \ref{core_temp}, we present the $L_X-M$ and $T_X-M$ scaling relations obtained without excluding the core region, while Figs.~\ref{nocore_lum} and \ref{nocore_temp} illustrate the core-excised $L_X-M$ and $T_X-M$ scaling relations. In all figures, the 3D results are shown in panel (a), and the x, y, and z projections are shown in panels (b), (c), and (d), respectively. The color coding reflects the 3D \textsl{coherence length} and the \textsl{coherence lengths} from the x, y, and z projections. The difference between results with and without the core region aligns with previous studies, such as those of \citet{Maughan2012}, which reported that for $L_X-T$ scaling relations, including the core region causes even the most relaxed clusters to deviate from self-similar behavior at approximately 3.5 keV. This behavior, which we also observe, is shown in Fig.~\ref{core_temp}. Conversely, excluding the core reduces the scatter in the scaling relations. In addition, the scatter appears to be strongly correlated with the dynamical state of the cluster: the brightest yellow points represent the most relaxed clusters, while the darkest points correspond to the most unrelaxed clusters. Data for which the \textsl{coherence length} cannot be defined are shown in black, matching the color used for the most unrelaxed systems.

To thoroughly investigate the dependence of the scatter on the dynamical state, for both the 3D and projected $L_X-M$ and the $T_X-M$ relations derived from core-excised data, as shown in Figs.~\ref{nocore_lum} and ~\ref{nocore_temp}, we divided the full simulated sample of clusters into subsamples, based on different ranges of 3D and 2D \textsl{coherence length} values: $\ell_{\mathrm{CR}}<0.5/r_{500}$, $<0.4/ r_{500}$, $<0.3/ r_{500}$ and $<0.2/ r_{500}$. As shown in Fig.~\ref{coh_scales}, the 3D and 2D \textsl{coherence length} values can differ from each other, with these differences increasing as the cluster departs from equilibrium. Consequently, while the scaling relations for the full set of 329 clusters remain consistent across different projections and in 3D, sorting the clusters into subsets according to their dynamical state reveals that the classification of a given cluster can vary depending on the projection axis or the transition from 2D projections to 3D analysis. 

\begin{figure*}
\centering
      \subfigure[]{\includegraphics[width=0.48\textwidth]{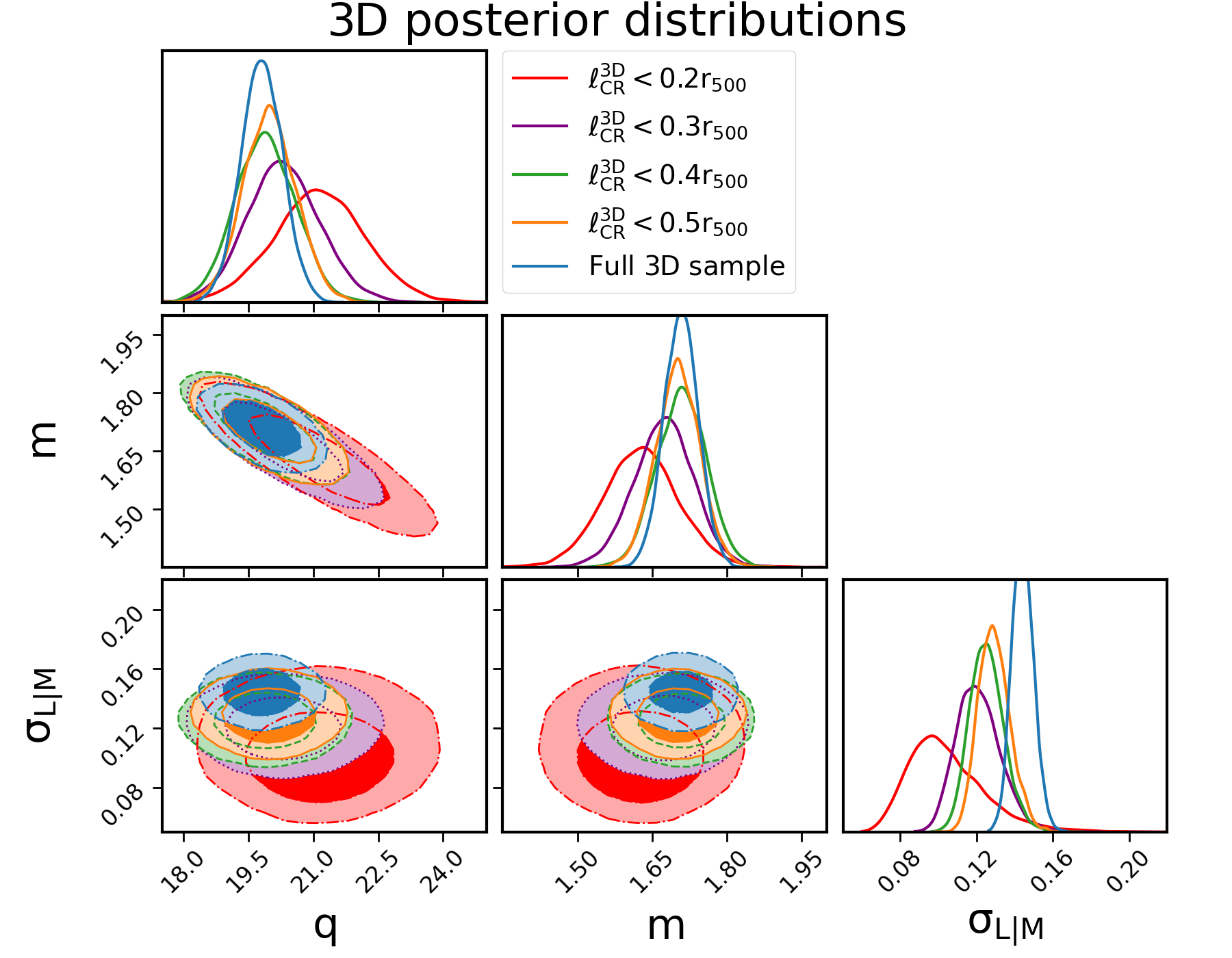}}
      \subfigure[]{\includegraphics[width=0.48\textwidth]{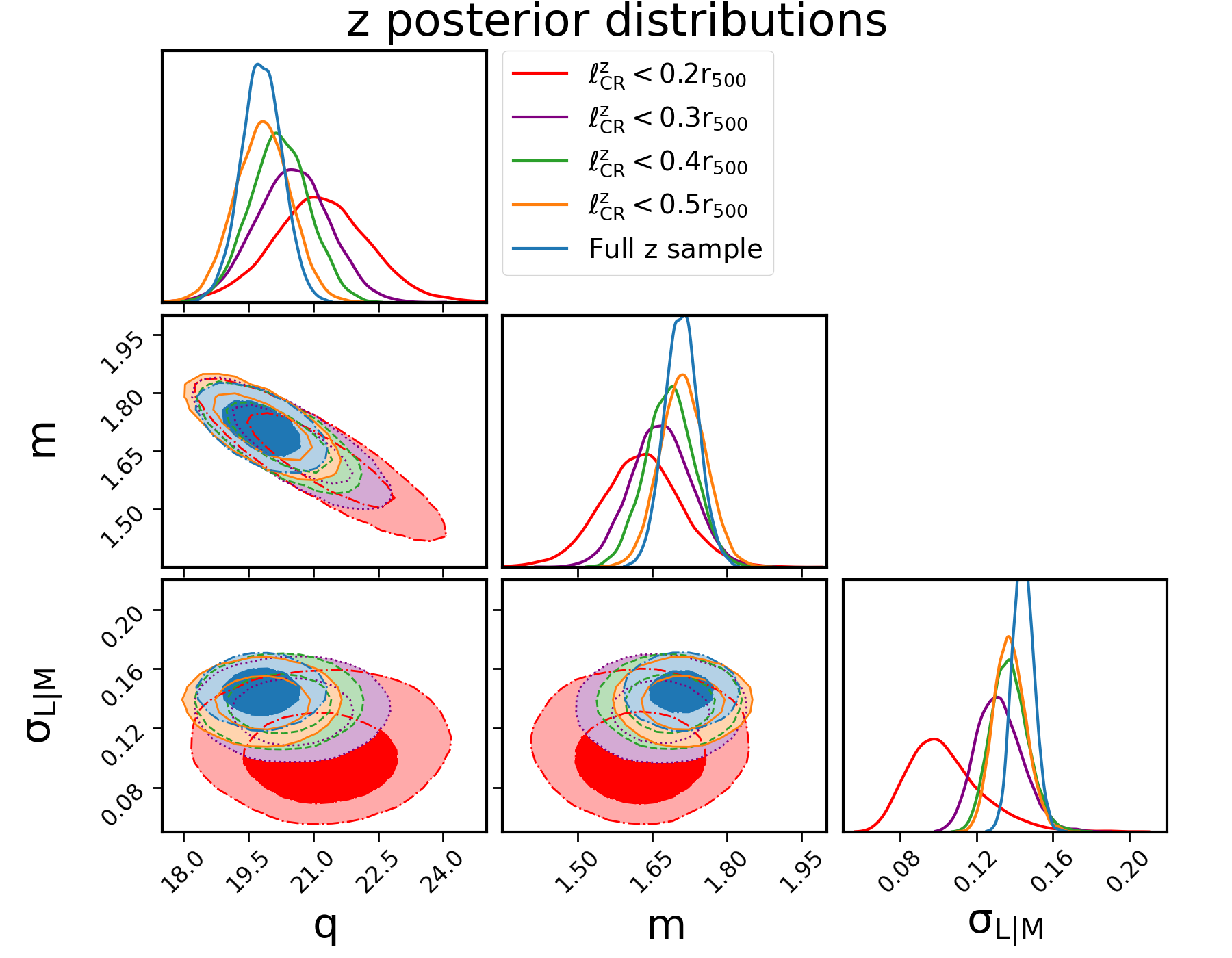}}
      \subfigure[]{\includegraphics[width=0.48\textwidth]{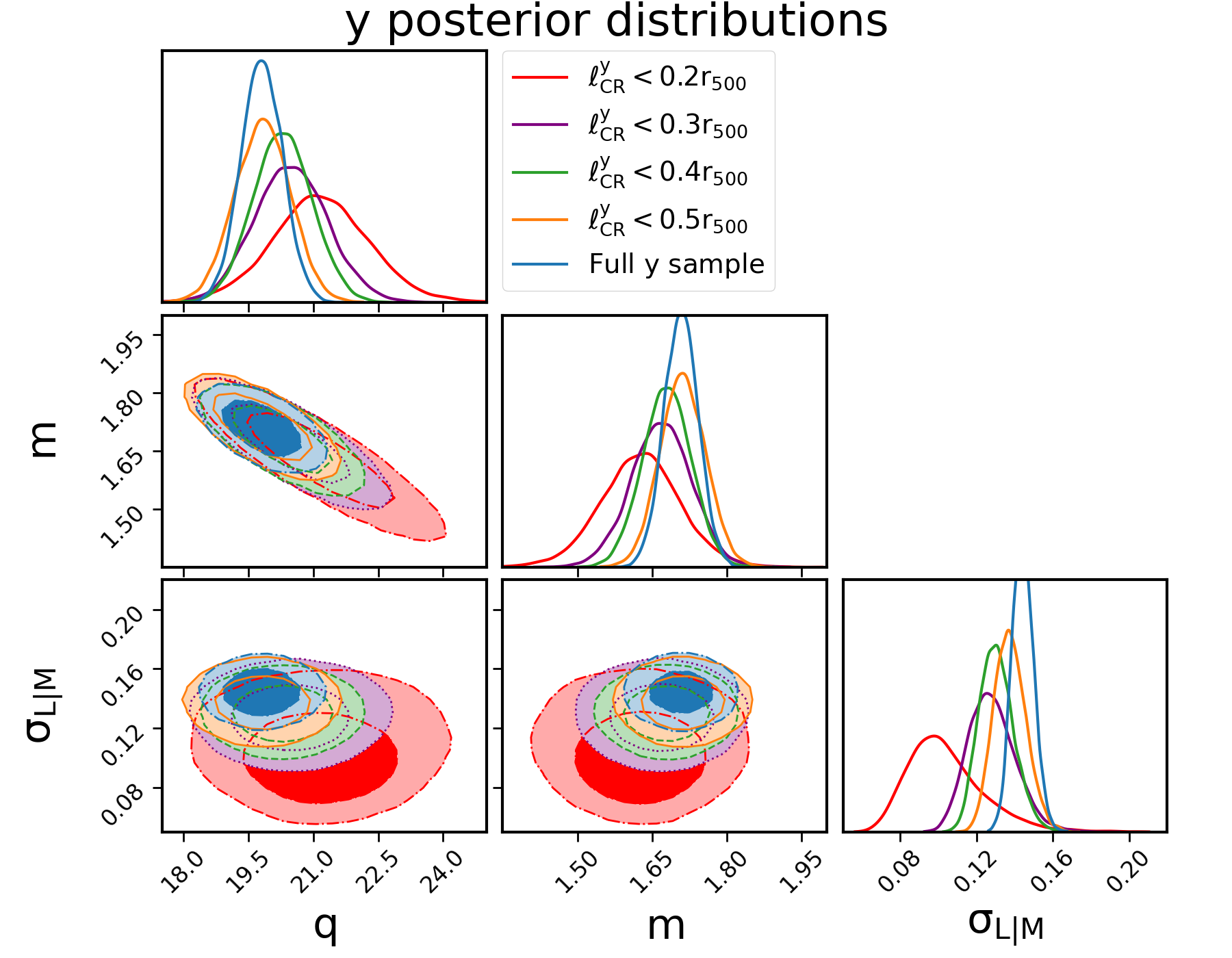}}
      \subfigure[]{\includegraphics[width=0.48\textwidth]{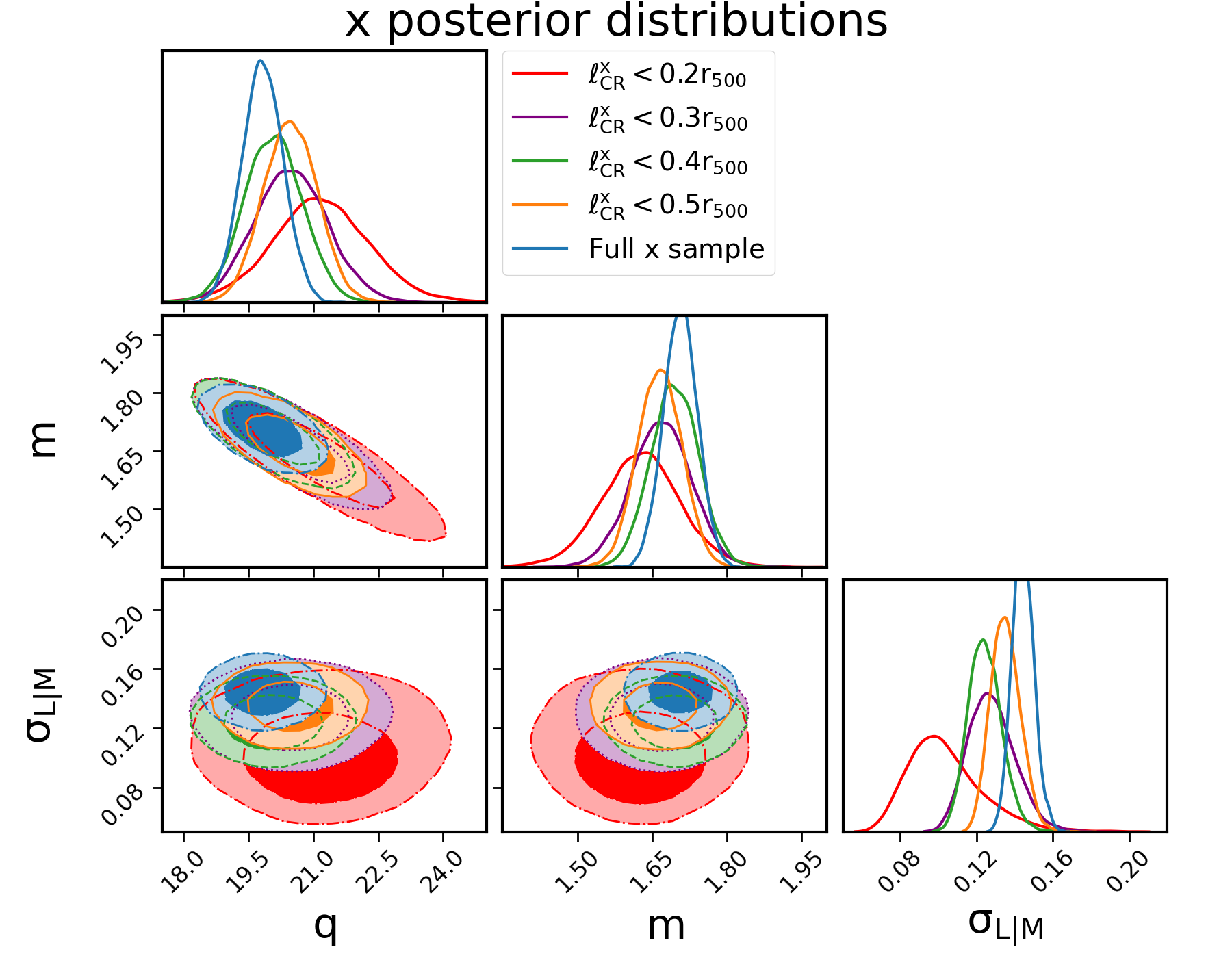}}     
    
      \caption{Posterior distributions of the $L_X-M$ scaling relations parameters obtained from the 3D analysis (upper-left corner) and x, y and z projections (upper-right, lower-left and lower-right corner, respectively). Marginalized posterior distributions are shown on the diagonal plots and the joint posterior distributions are shown on off-diagonal plots. Contours indicate 68$\%$ and 95$\%$ credibility regions. Different colors distinguish the results obtained for the various subsets.
}
 \label{L_post}
 \end{figure*}

\begin{figure*}
\centering
      \subfigure[]{\includegraphics[width=0.48\textwidth]{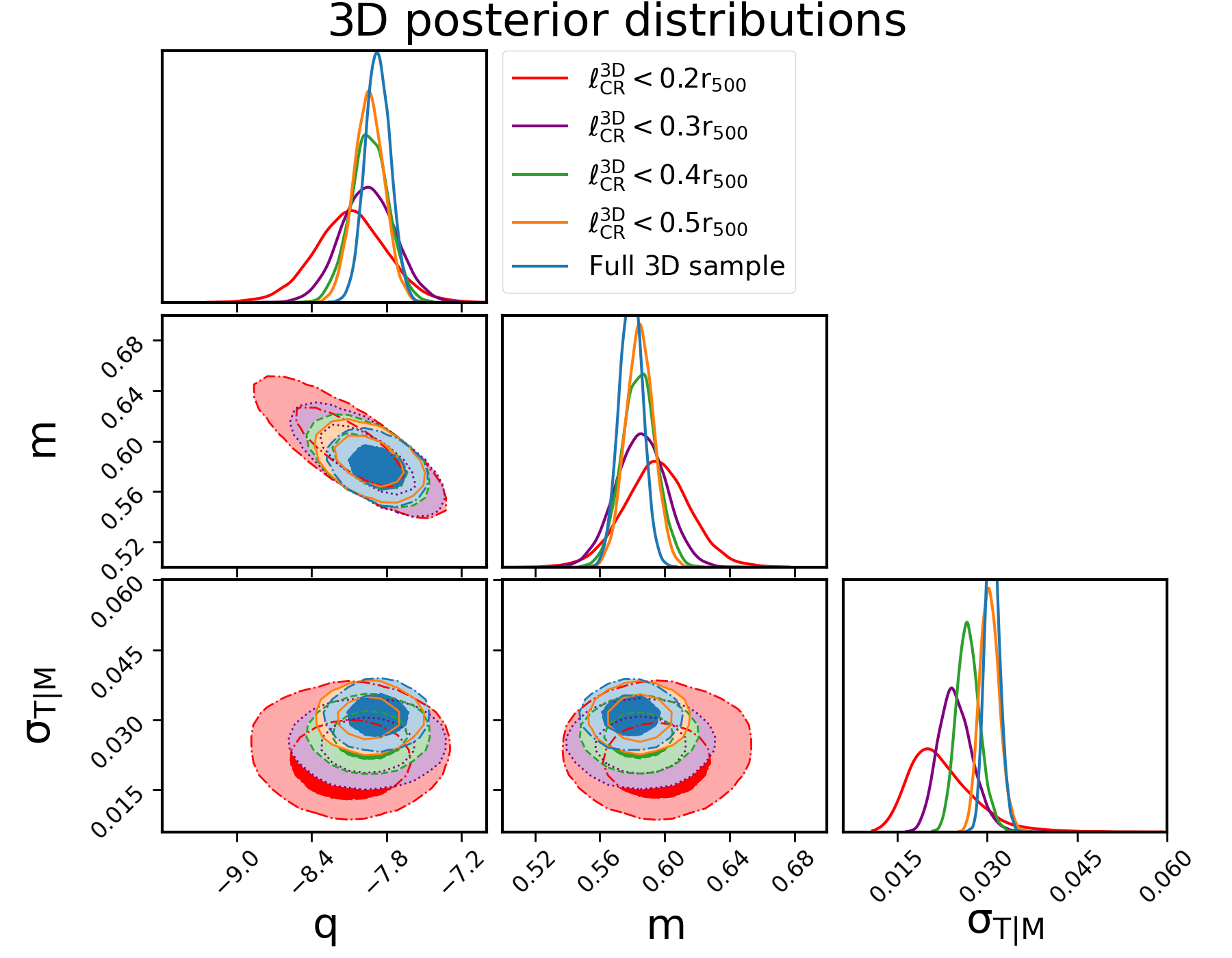}}
      \subfigure[]{\includegraphics[width=0.48\textwidth]{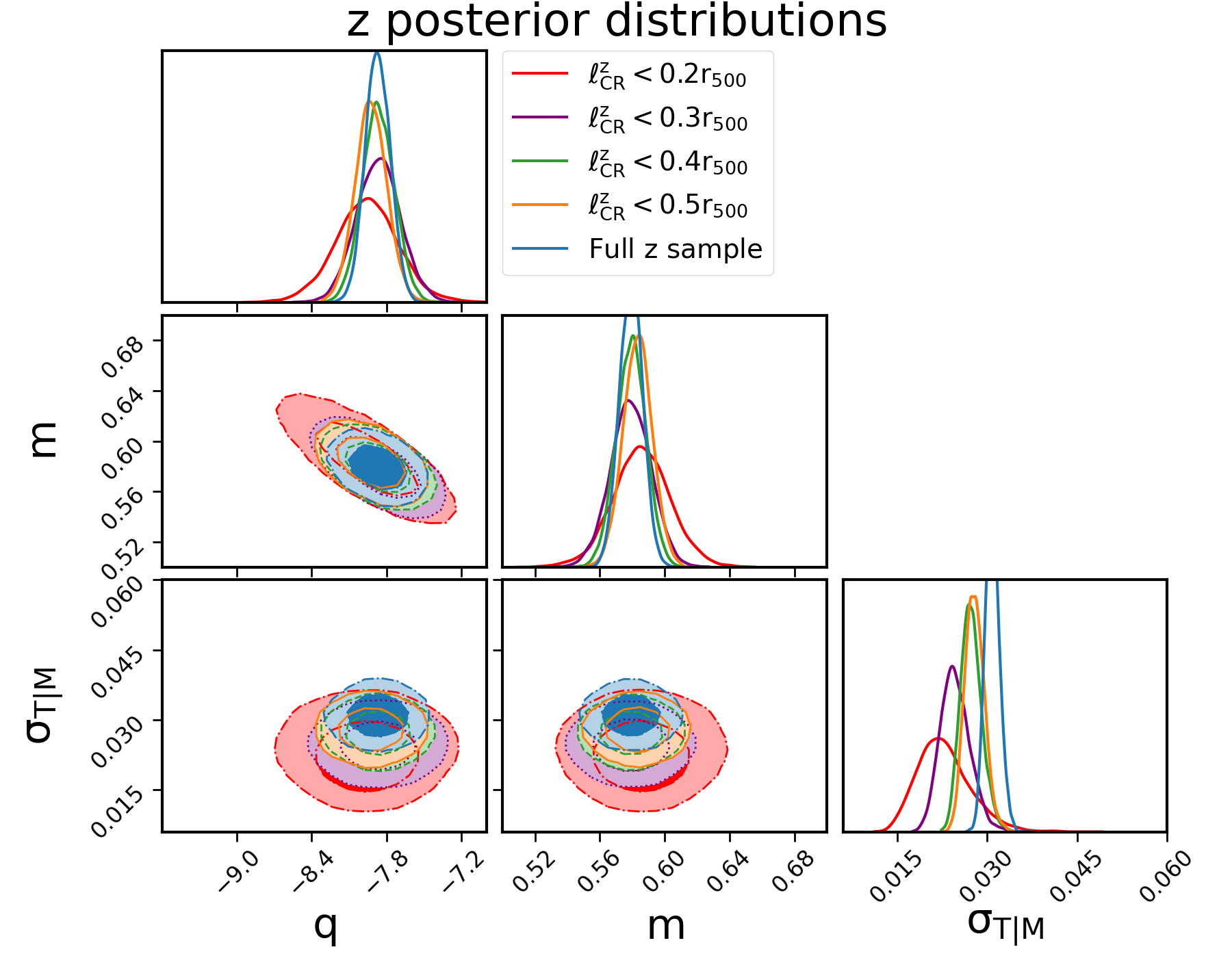}}
      \subfigure[]{\includegraphics[width=0.48\textwidth]{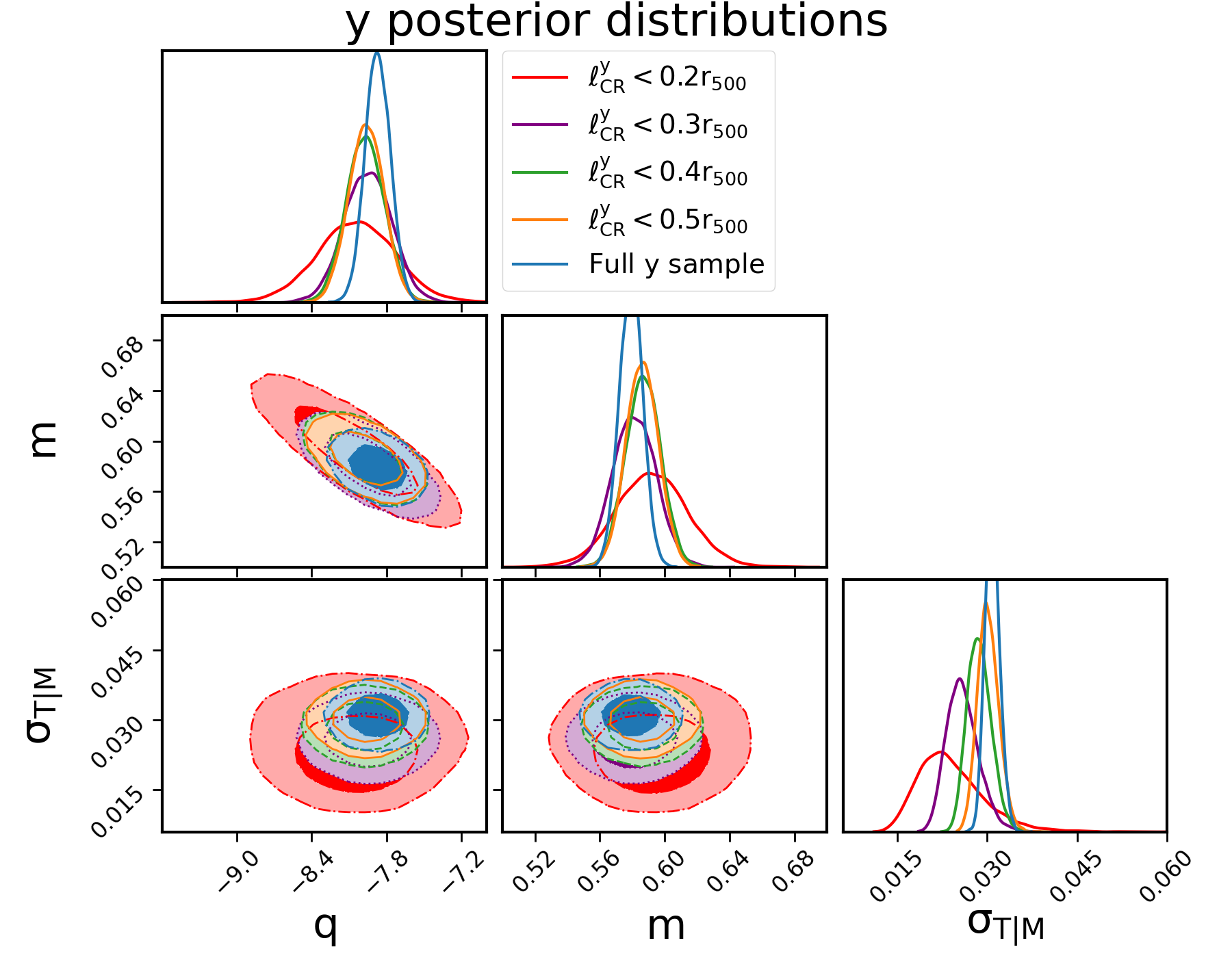}}
      \subfigure[]{\includegraphics[width=0.48\textwidth]{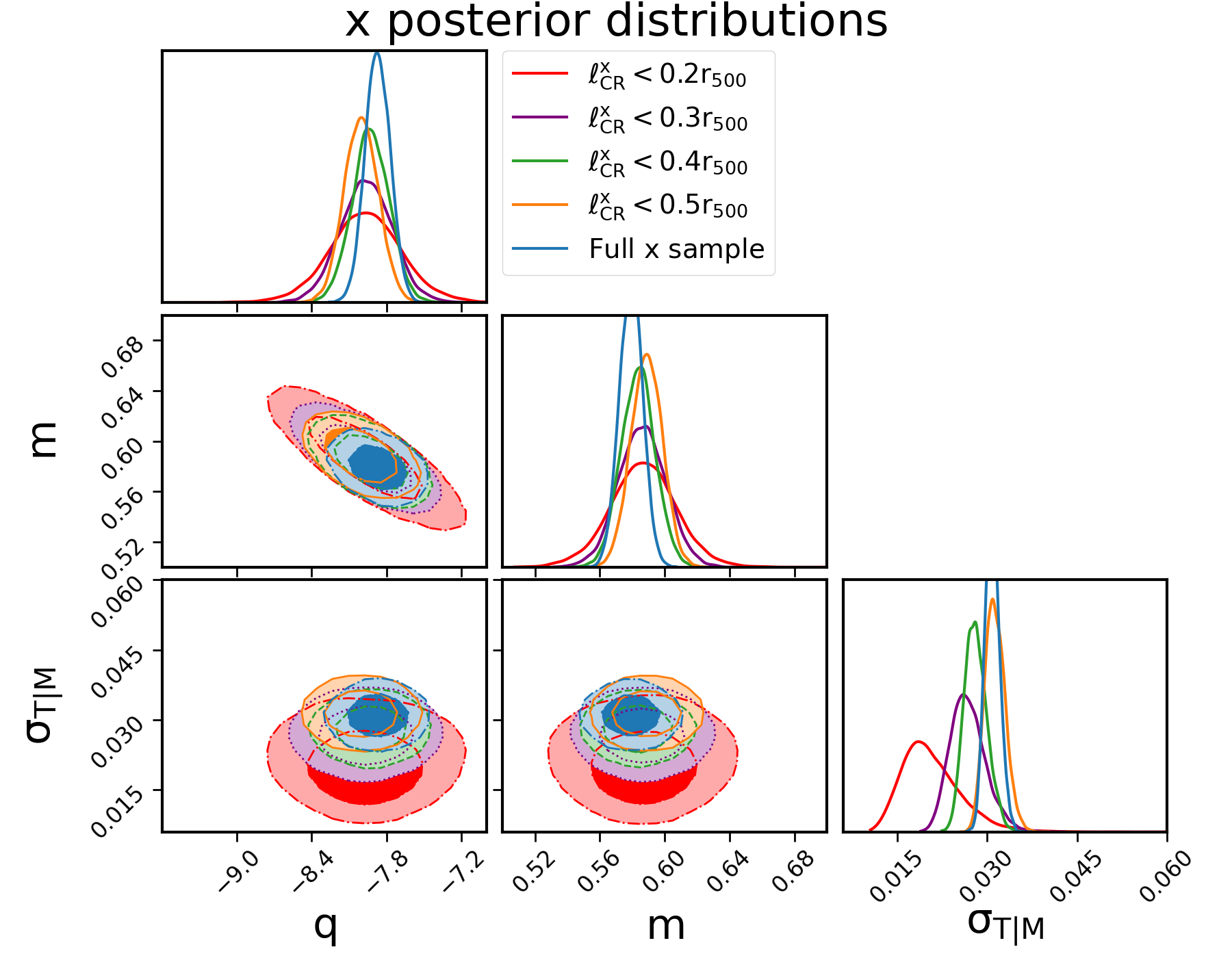}}     
    
      \caption{Posterior distributions of the $T_X-M$ scaling relations parameters obtained from the 3D analysis (upper-left corner) and x, y and z projections (upper-right, lower-left and lower-right corner, respectively). Marginalized posterior distributions are shown on the diagonal plots and the joint posterior distributions are shown on off-diagonal plots. Contours indicate 68$\%$ and 95$\%$ credibility regions. Different colors distinguish the results obtained for the various subsets.
}
 \label{T_post}
 \end{figure*}

\begin{figure*}
\centering
      \subfigure[]{\includegraphics[width=0.48\textwidth]{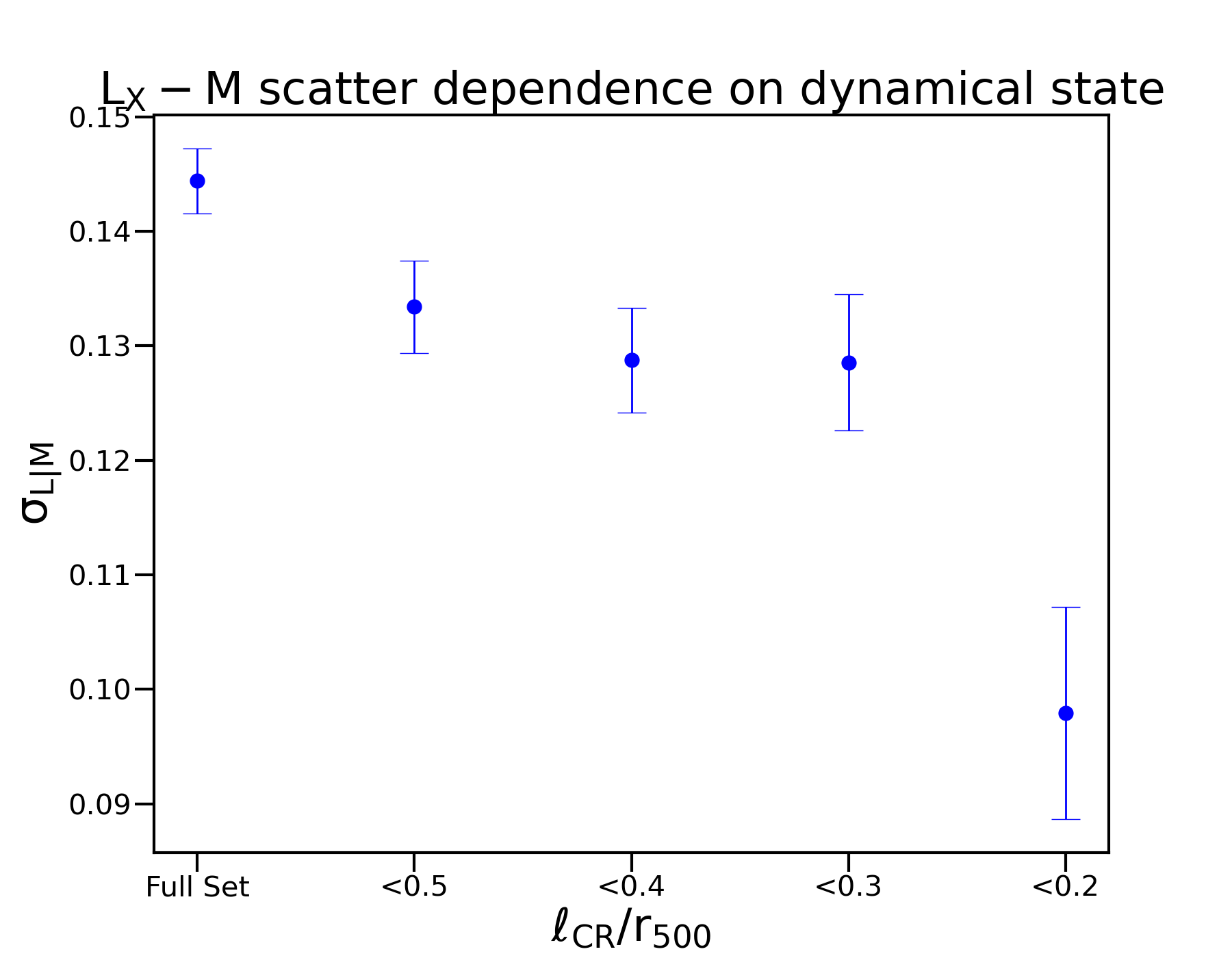}}
      \subfigure[]{\includegraphics[width=0.48\textwidth]{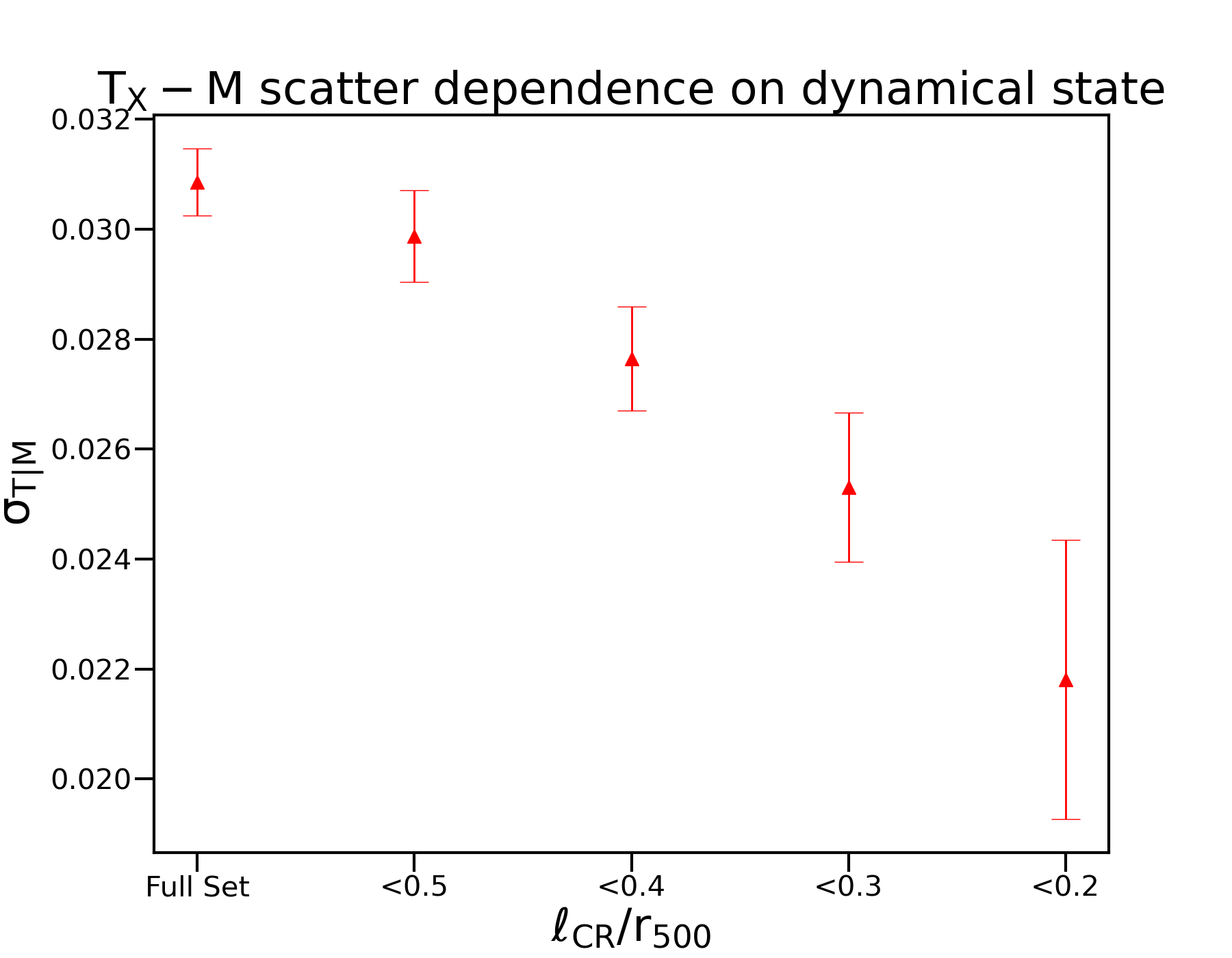}}  
    
      \caption{Average scatter dependence on the dynamical state for the $L_X-M$ (left panel) and $T_X-M$ scaling relations (right panel).}
 \label{med_scatter}
 \end{figure*}

\begin{figure}[ht]
 \centering
 \includegraphics[width=\columnwidth]{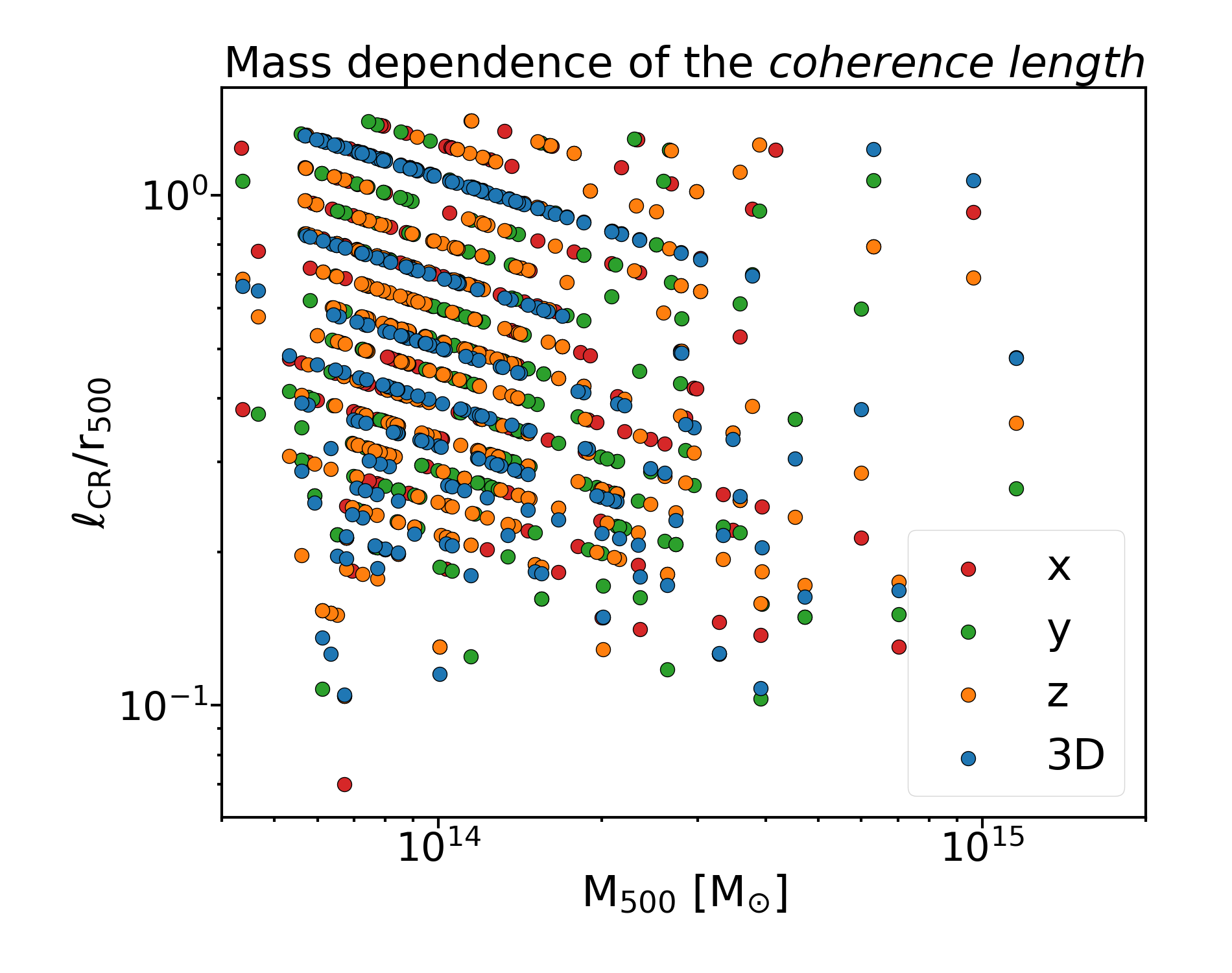} 
 \caption{Mass-\textsl{coherence length} relation.}
 \label{mass_lcr}
\end{figure}

For both the full samples and the subsets described above, we applied the likelihood method and computed posterior probability density functions (PDFs) as outlined in Section~\ref{scale_rel}, and the fitting code was run with 50 walkers for 5000 steps. Table ~\ref{priors} summarizes the priors adopted for the $L_X-M$ and $T_X-M$ scaling relations. Our choice of bounded flat priors for all scaling relations is consistent with previous investigations in the field (e.g., \citealt{2017Sharma} \citealt{2019Chen}, \citealt{Bahar22}). The results are displayed in Figs.~\ref{L_post} and ~\ref{T_post}, which compare the 3D analysis results (upper-left corners) with those from the x, y, and z projections (upper-right, lower-left, and lower-right corners, respectively). In both the 3D analysis and the projections, different colors represent the various subsets. For both the 3D analysis and the projections the different colors indicate the different subsets: full samples in blue, $\ell<0.5r_{500}$ in orange, $\ell<0.4r_{500}$ in green, $\ell<0.3r_{500}$ in purple, $\ell<0.2r_{500}$ red. Tables ~\ref{tab:LM_post} and ~\ref{tab:TM_post} summarize the results of all the best-fit parameters. The key result to emphasize is that, despite the strong influence of projection effects on the dynamical state classification and the reduced size of the subsets - which leads to weaker constraints on the parameters - we can still robustly quantify the correlation between scatter and the degree of relaxation in the system. For the most relaxed clusters, identified as those with $\ell_{\mathrm{CR}}<0.2r_{500}$, the scatter is significantly reduced - by up to $\sim40\%$ for the $L_X-M$ relation and $\sim32\%$ for the $T_X-M$ relation. To better illustrate this key result, Fig.~\ref{med_scatter} presents the average scatter as a function of dynamical state for the $L_X-M$ (left panel) and $T_X-M$ (right panel) scaling relations. For each subset, we show the average scatter, weighted by the 1$\sigma$ errors, combining results from both the 3D analysis and the projections. The average scatter decreases of $32\%$ in the $L_X-M$ relation and $29\%$ in the $T_X-M$ relation. This result is in good agreement with the findings of \citealt{damsted2023codex}, who found a relation between the cluster dynamical state and scaling relations through the analysis of the velocity substructure. Notably, the dependence of the $\mathbf{L_X-M}$ scatter on the dynamical state exhibits a sharp increase above a \textsl{coherence length} of $0.2r_{500}$, in contrast to the smoother behavior observed in the  $T_X-M$ relation. It is well established that both X-ray luminosity and temperature can deviate from self-similar expectations during mergers, driven by gas compressions and shocks (e.g., \citealt{Randall_2002}, \citealt{Pratt_2009}). However, the luminosity is particularly sensitive because it scales with the square of the electron density, making it more responsive to even modest gas density variations. The identified coherence length threshold of $0.2r_{500}$ effectively isolates the most relaxed clusters, where such merger-driven departures from the self-similar scenario are minimal, thereby avoiding the sharp rise in scatter seen in more disturbed systems. We also assessed whether there is any dependence of the \textsl{coherence length} on the clusters' mass. As shown in Fig.~\ref{mass_lcr}, our results do not indicate a significant mass dependence of the \textsl{coherence length} across the sample. This suggests that the reduction in scatter for clusters with $\ell_{\mathrm{CR}} < 0.2r_{500}$ is not driven by a preferential removal of lower-mass systems, but rather by the selection of genuinely more relaxed clusters. This is further supported by the color-coded scaling relations shown in Figs \ref{nocore_lum} and \ref{nocore_temp}, which clearly demonstrate the presence of relaxed systems across the entire mass range.

The values of the other parameters are consistent with previous studies (e.g., \citealt{Pop2022}). Specifically, the slope of the $L_X-M$ scaling relation is steeper than the self-similar prediction, while the slope of the $T_X-M$ scaling relation is shallower than expected in the self-similar scenario. These deviations from self-similar scaling in TNG300 are primarily attributed to non-gravitational physics, particularly AGN feedback. The excess energy injected by AGN activity can boost the X-ray luminosity in higher-mass clusters, resulting in a steeper $L_X-M$ slope. Conversely, AGN-driven outflows can reduce the gas temperature, particularly in lower-mass clusters with shallower gravitational potentials, leading to a flatter $T_X-M$ slope. Notably, the slopes move closer to self-similar expectations for the subset of relaxed clusters, where the influence of AGN activity and dynamical disturbances is less pronounced. Additionally, the higher normalization in the $L_X-M$ scaling relations for relaxed clusters compared to their unrelaxed counterparts was also previously found in the literature (e.g., \citealt{Zhang_2011}, \citealt{Lovisari2020}). The elevated normalization in relaxed clusters can be attributed to several factors, such as increased gas density, the presence of cool cores and reduced disturbances. Indeed, relaxed clusters often have higher central gas densities, leading to enhanced X-ray emission due to the density-squared dependence of the bremsstrahlung radiation process. As for the cool cores, even if we excised the region enclosed in $0.1r_{500}$, the cool core region might extend beyond and still contribute significantly to the total X-ray luminosity. Finally, the lack of recent mergers or significant dynamical activity allows the intracluster medium to settle into a more concentrated state, enhancing X-ray brightness.

\begin{table*}[t]
  \centering
  %\resizebox{\textwidth}{!}{
  \hspace{-10em}
  \begin{tabular}{@{}lcccccc@{}}
    \toprule
    Subset & \multicolumn{3}{c}{3D} & \multicolumn{3}{c}{z projection} \\ 
    \cmidrule(lr){2-4} \cmidrule(lr){5-7}
        & $q$ & $m$ & $\sigma_{L|M}$ & $q$ & $m$ & $\sigma_{L|M}$ \\ 
    \midrule
    Full sample & $19.803^{+0.475}_{-0.463}$ & $1.709^{+0.033}_{-0.034}$ & $0.144^{+0.006}_{-0.005}$ & $19.803^{+0.475}_{-0.463}$ & $1.709^{+0.033}_{-0.034}$ & $0.144^{+0.006}_{-0.005}$ \\
    $\ell_{\mathrm{CR}}<0.5r_{500}$ & $20.009^{+0.626}_{-0.610}$ & $1.709^{+0.043}_{-0.044}$ & $0.139^{+0.009}_{-0.008}$ & $19.921^{+0.651}_{-0.639}$ & $1.710^{+0.045}_{-0.046}$ & $0.139^{+0.009}_{-0.008}$\\
    $\ell_{\mathrm{CR}}<0.4r_{500}$ & $19.931^{+0.690}_{-0.711}$ & $1.706^{+0.050}_{-0.049}$ & $0.135^{+0.009}_{-0.008}$ & $20.220^{+0.692}_{-0.669}$ & $1.685^{+0.047}_{-0.049}$ & $0.137^{+0.010}_{-0.009}$ \\
    $\ell_{\mathrm{CR}}<0.3r_{500}$ & $20.362^{+0.849}_{-0.828}$ & $1.669^{+0.059}_{-0.060}$ & $0.120^{+0.002}_{-0.012}$ & $20.498^{+0.824}_{-0.844}$ & $1.668^{+0.060}_{-0.058}$ & $0.121^{+0.013}_{-0.010}$ \\
    $\ell_{\mathrm{CR}}<0.2r_{500}$ & $21.097^{+1.018}_{-1.009}$ & $1.630^{+0.074}_{-0.071}$ & $0.102^{+0.021}_{-0.015}$ & $20.357^{+0.943}_{-0.959}$ & $1.681^{+0.067}_{-0.067}$ & $0.107^{+0.019}_{-0.014}$ \\
    \bottomrule
  \end{tabular}
  %}
  \vspace{1em} % Adds space between blocks
  %\resizebox{\textwidth}{!}{
  \hspace{-10em}
  \begin{tabular}{@{}lcccccc@{}}
    \toprule
    Subset & \multicolumn{3}{c}{y projection} & \multicolumn{3}{c}{x projection} \\ 
    \cmidrule(lr){2-4} \cmidrule(lr){5-7}
        & $q$ & $m$ & $\sigma_{L|M}$ & $q$ & $m$ & $\sigma_{L|M}$ \\ 
    \midrule    
    Full sample & $19.803^{+0.475}_{-0.463}$ & $1.709^{+0.033}_{-0.034}$ & $0.144^{+0.006}_{-0.005}$ & $19.803^{+0.475}_{-0.463}$ & $1.709^{+0.033}_{-0.034}$ & $0.144^{+0.006}_{-0.005}$ \\
    $\ell_{\mathrm{CR}}<0.5r_{500}$ & $19.396^{+0.644}_{-0.656}$ & $1.699^{+0.047}_{-0.046}$ & $0.132^{+0.008}_{-0.008}$ & $20.086^{+0.610}_{-0.609}$ & $1.664^{+0.043}_{-0.043}$ & $0.135^{+0.008}_{-0.007}$ \\
    $\ell_{\mathrm{CR}}<0.4r_{500}$ & $19.579^{+0.687}_{-0.665}$ & $1.699^{+0.009}_{-0.009}$ & $0.027^{+0.008}_{-0.008}$ & $20.086^{+0.672}_{-0.654}$ & $1.694^{+0.046}_{-0.047}$ & $0.124^{+0.009}_{-0.008}$ \\
    $\ell_{\mathrm{CR}}<0.3r_{500}$ & $19.991^{+0.862}_{-0.842}$ & $1.718^{+0.059}_{-0.061}$ & $0.125^{+0.013}_{-0.011}$ & $20.385^{+0.839}_{-0.825}$ & $1.675^{+0.054}_{-0.060}$ & $0.116^{+0.012}_{-0.011}$ \\
    $\ell_{\mathrm{CR}}<0.2r_{500}$ & $21.144^{+1.083}_{1.068}$ & $1.627^{+0.075}_{-0.076}$ & $0.101^{+0.021}_{-0.015}$ & $22.786^{+0.931}_{-0.948}$ & $1.613^{+0.067}_{-0.065}$ & $0.087^{+0.018}_{-0.013}$ \\
    \bottomrule
  \end{tabular}
  \caption{$L_X-M$ scaling relation parameters for different subsets (left columns) and projections (second and third columns). For both the 3D analysis and all projections we indicate the best-fit parameters: the intercept $q$, the slope $m$ and the scatter $\sigma_{L|M}$.}
\label{tab:LM_post}
  %}
\end{table*}

\begin{table*}[t]
  \centering
  %\resizebox{\textwidth}{!}{
  \hspace{-10em}
  \begin{tabular}{@{}lcccccc@{}}
    \toprule
    Subset & \multicolumn{3}{c}{3D} & \multicolumn{3}{c}{z projection} \\ 
    \cmidrule(lr){2-4} \cmidrule(lr){5-7}
        & $q$ & $m$ & $\sigma_{T|M}$ & $q$ & $m$ & $\sigma_{T|M}$ \\ 
    \midrule
    Full sample & $-7.873^{+0.10}_{-0.10}$ & $0.579^{+0.007}_{-0.007}$ & $0.031^{+0.001}_{-0.001}$ & $-7.873^{+0.10}_{-0.10}$ & $0.579^{+0.007}_{-0.007}$ & $0.031^{+0.001}_{-0.001}$ \\
    $\ell_{\mathrm{CR}}<0.5r_{500}$ & $-7.941^{+0.124}_{-0.124}$ & $0.585^{+0.009}_{-0.009}$ & $0.030^{+0.002}_{-0.002}$ & $-7.928^{+0.127}_{-0.124}$ & $0.584^{+0.009}_{-0.009}$ & $0.028^{+0.002}_{-0.002}$ \\
    $\ell_{\mathrm{CR}}<0.4r_{500}$ & $-7.927^{+0.149}_{-0.147}$ & $0.584^{+0.010}_{-0.011}$ & $0.029^{+0.002}_{-0.002}$ & $-7.928^{+0.127}_{-0.124}$ & $0.584^{+0.009}_{-0.009}$ & $0.028^{+0.002}_{-0.002}$ \\
    $\ell_{\mathrm{CR}}<0.3r_{500}$ & $-7.951^{+0.208}_{-0.213}$ & $0.585^{+0.015}_{-0.015}$ & $0.026^{+0.002}_{-0.002}$ & $-7.862^{+0.172}_{-0.174}$ & $0.579^{+0.012}_{-0.012}$ & $0.025^{+0.003}_{-0.002}$ \\
    $\ell_{\mathrm{CR}}<0.2r_{500}$ & $-8.103^{+0.283}_{-0.279}$ & $0.596^{+0.020}_{-0.020}$ & $0.022^{+0.006}_{-0.004}$ & $-7.964^{+0.247}_{-0.241}$ & $0.586^{+0.017}_{-0.017}$ & $0.022^{+0.005}_{-0.004}$ \\
    \bottomrule
  \end{tabular}
  %}
  \vspace{1em} % Adds space between blocks
  %\resizebox{\textwidth}{!}{
  \hspace{-10em}
  \begin{tabular}{@{}lcccccc@{}}
    \toprule
    Subset & \multicolumn{3}{c}{y projection} & \multicolumn{3}{c}{x projection} \\ 
    \cmidrule(lr){2-4} \cmidrule(lr){5-7}
        & $q$ & $m$ & $\sigma_{T|M}$ & $q$ & $m$ & $\sigma_{T|M}$ \\ 
    \midrule    
    Full sample & $-7.873^{+0.10}_{-0.10}$ & $0.579^{+0.007}_{-0.007}$ & $0.031^{+0.001}_{-0.001}$ & $-7.873^{+0.10}_{-0.10}$ & $0.579^{+0.007}_{-0.007}$ & $0.031^{+0.001}_{-0.001}$ \\
    $\ell_{\mathrm{CR}}<0.5r_{500}$ & $-7.967^{+0.143}_{-0.146}$ & $0.587^{+0.010}_{-0.010}$ & $0.030^{+0.002}_{-0.002}$ & $-8.002^{+0.134}_{-0.135}$ & $0.589^{+0.010}_{-0.010}$ & $0.031^{+0.002}_{-0.002}$ \\
    $\ell_{\mathrm{CR}}<0.4r_{500}$ & $-7.865^{+0.131}_{-0.127}$ & $0.579^{+0.009}_{-0.009}$ & $0.027^{+0.008}_{-0.008}$ & $-7.923^{+0.150}_{-0.147}$ & $0.583^{+0.010}_{-0.011}$ & $0.028^{+0.002}_{-0.002}$ \\
    $\ell_{\mathrm{CR}}<0.3r_{500}$ & $-7.960^{+0.195}_{-0.186}$ & $0.582^{+0.013}_{-0.014}$ & $0.026^{+0.003}_{-0.002}$ & $-7.969^{+0.205}_{-0.200}$ & $0.587^{+0.014}_{-0.014}$ & $0.027^{+0.003}_{-0.003}$ \\
    $\ell_{\mathrm{CR}}<0.2r_{500}$ & $-8.047^{+0.300}_{-0.313}$ & $0.592^{+0.022}_{-0.021}$ & $0.023^{+0.006}_{-0.004}$ & $-7.974^{+0.279}_{-0.270}$ & $0.587^{+0.019}_{-0.020}$ & $0.021^{+0.005}_{-0.004}$ \\
    \bottomrule
  \end{tabular}
  \caption{$T_X-M$ scaling relation parameters for different subsets (left columns) and projections (second and third columns). For both the 3D analysis and all projections we indicate the best-fit parameters: the intercept $q$, the slope $m$ and the scatter $\sigma_{T|M}$.}
\label{tab:TM_post}
  %}
\end{table*}

\section{Discussion and Conclusions} \label{conclusions}

In this paper, we present the analysis of the luminosity-mass ($L_X-M$) and temperature-mass ($T_X-M$) scaling relations for 329 $z=0$ galaxy clusters from the TNG300 simulations as a function of the dynamical state. For all the clusters analyzed, we obtain 3D and 2D projections of mass and X-ray surface brightness distributions. We then cross-correlate the mass and X-ray maps through coherence analysis and compute the 3D and 2D \textsl{coherence lengths}, which indicates the dynamical state of the cluster from that particular projection or in 3 dimensions. The full sample of clusters was then divided into sub-samples based on different ranges of decreasing values of the 3D and 2D \textsl{coherence length}, with $\ell^{3D}_{CR}$ ($\ell^x_{CR}$/$\ell^y_{CR}$/$\ell^z_{CR}$)$<0.2\ r_{500}$ identifying the subset with the most unrelaxed clusters. The analysis of the coherence and the scatter of these different sub-samples led to the following key results:

\begin{itemize}
  \item Coherence analysis and \textsl{coherence length} show great promise in identifying and selecting a group of fully relaxed clusters and, more broadly, in assessing the dynamical state of galaxy clusters beyond the traditional binary classification of "relaxed" and "unrelaxed". This approach reveals a continuum of relaxation levels. In future work, we intend to delve deeper into improving the accuracy of 3D \textsl{coherence length} reconstruction and determining the dynamical state of clusters that are out of equilibrium.
  \item The 3D and 2D \textsl{coherence lengths} exhibit a high level of scatter, with the only exception of a few very relaxed clusters, underscoring the significant impact of projection effects. Applying this method to observed clusters in the future implies that only \textsl{coherence length} values $\lesssim 0.2/ r_{500}$ ensure that clusters are fully relaxed. Using this new criterion for selection can prevent misclassification. 
  \item While significant uncertainties can arise in the determination of the dynamical state of out-of-equilibrium clusters from a single projection, incorporating the \textsl{coherence length} into the scaling relation paradigm clearly demonstrates its power, as a strong correlation between scatter and the dynamical state of the clusters is seen. The scatter in the scaling relations of the most relaxed clusters is $\sim 30\%$ times smaller than that of the full sample comprising clusters in various dynamical states.
\end{itemize} 

It is important to emphasize that all measurements in this analysis were conducted within $r_{500}$. Using $r_{500}$ for this purpose is particularly appropriate because it restricts the analysis to the cluster region that is more dynamically stable. This stability is essential for identifying relaxed clusters, as our goal is to correlate their dynamical state with the scatter in scaling relations. An extension of this study to larger areas and volumes, such as $r_{200}$ and beyond, could uncover additional aspects, including the influence of the outer regions during the transition to the infall zone, the impact of the surrounding environment, insights into accretion processes, and the presence of gas clumping. These factors will be explored in future investigations. 

Although our results indicate that incorporating the \textsl{coherence length} as an additional criterion for identifying relaxed cluster samples is a powerful tool for quantifying the contribution of dynamical state to the scatter in scaling relations, we acknowledge the current limitations of this approach. As previously mentioned, our results on the dependence of scatter on dynamical state are in agreement with recent quantitative findings, such as \citealt{damsted2023codex}. While it is well known that dynamical state is an important source of scatter, these quantitative analyses further emphasize its impact and motivate more refined approaches. Rather than advocating for strict cuts to exclude non-relaxed systems - an approach that would inevitably reduce the cluster sample size and affect the uncertainty on $\sigma_8$ - we are actively exploring methods to incorporate a dynamical state proxy such as \textsl{coherence length} directly into cosmological parameter estimation frameworks. This strategy aims to fully utilize all available clusters and can be adapted to other dynamical state indicators as well. A follow-up investigation addressing these aspects is currently underway and beyond the scope of the present manuscript.

As explained in Section~\ref{data}, in this analysis, for projected maps we adopted a grid of $512\times 512$ pixels over a $3\times 3\ Mpc$ field of view, resulting in a pixel scale of approximately 5.9 kpc. This resolution allows us to resolve substructures in the simulated clusters and compute the coherence at fine spatial scales. As part of the same follow-up investigation mentioned above, we are preparing a dedicated paper that will quantitatively forecast the expected behavior of the coherence and \textsl{coherence length} using realistic mock observations that incorporate instrument Point Spread Functions (PSFs), noise, and survey characteristics, including Euclid, Nancy Grace Roman, Vera C. Rubin Observatory Legacy Survey of Space and Time (LSST), eROSITA, XMM-Newton, and Chandra. However, based on well-established instrumental properties and prior studies, we are able to provide realistic estimates already at this stage.

In future applications aimed at cosmological studies, we plan to employ weak-lensing convergence maps, as they enable large statistical samples of clusters to be analyzed. For wide-field weak-lensing surveys like Euclid, LSST and Roman, expected source galaxy densities are approximately $30-40\ gal/arcmin^2$ (e.g. \citealt{LSSTbook}, \citealt{laureijs2011eucliddefinitionstudyreport}, \citealt{Chang2013}, \citealt{spergel2015widefieldinfrarredsurveytelescopeastrophysics},  \citealt{Euclid_Scar}, \citealt{Euclid_Blanc}). When constructing weak-lensing convergence maps, which represent the projected mass density reconstructed from shear measurements using the Kaiser $\&$ Squires inversion method (\citealt{Kaiser1986}), the reduced shear field derived from galaxy ellipticities is typically sampled on a grid with a chosen pixel size. This pixel size must balance spatial resolution with the noise introduced by the finite number of background galaxies. In practice, the smoothing scale can be set so that there are about 10 galaxies per pixel, providing a compromise that retains spatial detail while minimizing shape noise (e.g. \citealt{Jullo2014}). Based on the expected galaxy densities in current and upcoming wide-field lensing-oriented surveys, we estimate that a pixel scale of approximately $30"$ is required to meet this criterion consistently across different datasets. This weak-lensing resolution becomes the dominant limiting factor when combined with X-ray maps from Chandra or XMM-Newton, which have sharper PSFs. Indeed, Chandra and XMM-Newton have an on-axis Half Power Diameter (HPD) of approximately $0.5"$ and $15.0"$, respectively; therefore, their contribution to resolution degradation is less critical in this regime.

In our simulations, relaxed clusters maintain high coherence ($>0.9$) down to $< 200\ kpc$. Both the pixel resolution and the PSF act to suppress power spectra and coherence at small scales by effectively smoothing or blurring fine structures (e.g.\citealt{Cappelluti_2012}), \citealt{Kashlinsky_2018}, \citealt{https://Cerini22}. To provide concrete estimates, with an angular resolution of $30"$ (typical of weak-lensing convergence maps with ~10 galaxies per pixel), and in the adopted cosmology, $30"$ corresponds to a physical scale of approximately 221 kpc at redshift $z = 0.7$. Thus, to reliably probe scales below 200 kpc, the redshift limit is $z = 0.6$. A similar situation holds for eROSITA, which has an on-axis PSF of approximately $28"$, resulting in a comparable threshold. Even within these redshift constraints, the wide-area surveys mentioned, covering thousands to tens of thousands of square degrees, will still provide access to large statistical samples, comprising thousands of clusters. We emphasize that rigorously quantifying these limits - including noise, realistic PSF convolution, and survey-specific systematics — requires dedicated mock simulations, which will be addressed in detail in a separate forthcoming study.

\hspace{5pt}
\section*{Acknowledgements}
This research has been made use of data obtained from IllustrisTNG simulations (https://www.tng-project.org). G.C. and N.C. acknowledge the University of Miami and the Center for Astrophysics Harvard $\&$ Smithsonian for the support. G.C., N.C. and P.N. acknowledge Chandra X-ray Observatory grant \#AR3-24011X and NASA Astrophysics Data Analysis Program (ADAP) grant \#80NSSC24K1025. G.C. acknowledges the Jet Propulsion Laboratory, operated by the California Institute of Technology under a contract with NASA (80NM0018D0004), Jason Rhodes, Eric Huff and Diana Scognamiglio for providing additional resources and support during the completion of this work. This research was sponsored by the Jet Propulsion Laboratory, California Institute of Technology through a contract with ORAU. The views and conclusions contained in this document are those of the authors and should not be interpreted as representing the official policies, either expressed or implied, of the Jet Propulsion Laboratory, California Institute of Technology or the U.S. Government. The U.S.Government is authorized to reproduce and distribute reprints for Government
purposes notwithstanding any copyright notation herein. P.N. acknowledges support from DOE grant \#DE-SC0017660. J.A.Z. is funded by the Chandra X-ray Center, which is operated by the Smithsonian Astrophysical Observatory for and on behalf of NASA under contract NAS8-03060.
\newpage

%\end{acknowledgments}

%\appendix \label{appen}

\clearpage
\newpage

\bibliography{cluster_project}{}
\bibliographystyle{aasjournal}

\end{document}